\documentclass[onecolumn]{elsart3p}

\usepackage{ifpdf}
\usepackage{natbib}
\usepackage{graphicx}
\usepackage{amsmath}
\usepackage{amssymb}
\usepackage{color}

\newcommand{\hete}{\textit{HETE-2} }

\newcommand{\hst}{\textit{HST} }

\newcommand{\Swift}{{\it Swift} }
\newcommand{\swift}{\textit{Swift} }

\newcommand{\chandra}{\textit{Chandra} }

\newcommand{\batse}{\textit{BATSE} }
\newcommand{\BATSE}{\textit{BATSE} }
\newcommand{\glast}{\textit{GLAST} }

\newcommand{\konus}{\textit{Konus-Wind} }

\def\simlt{\mathrel{\hbox{\rlap{\hbox{\lower4pt\hbox{$\sim$}}}\hbox{$<$}}}}
\def\simgt{\mathrel{\hbox{\rlap{\hbox{\lower4pt\hbox{$\sim$}}}\hbox{$>$}}}}

\newcommand{\Msunyr}{\mbox{$M_\odot$ yr$^{-1}$}}

\newcommand{\peryear}{yr$^{-1}$}

\newcommand{\mnras}{MNRAS }
\newcommand{\apjl}{ApJ }
\newcommand{\apj}{ApJ }
\newcommand{\apjs}{ApJ }
\newcommand{\aj}{AJ }
\newcommand{\nat}{Nature }
\newcommand{\aap}{A\&A }
\newcommand{\aaps}{A\&AS }
\newcommand{\pre}{Phys. Rev. E }
\newcommand{\prd}{Phys. Rev. D }
\newcommand{\prc}{Phys. Rev. C }
\newcommand{\physrep}{Physics Reports }
\newcommand{\aplett}{Astrophysical Letters }
\newcommand{\apss}{Astrophysics and Space Science }
\newcommand{\araa}{Annual Review of Astronomy and Astrophysics}

\definecolor{webgreen}{rgb}{0,.5,0}
\definecolor{webbrown}{rgb}{.6,0,0}

\ifpdf
\usepackage[%
  pdftitle={Short/Hard Gamma-Ray Burst},%
  pdfauthor={Ehud Nakar},%
  pdfsubject={},%
  pdfkeywords={},%
  pdfstartview=FitH,%
  bookmarks=true,%
  bookmarksopen=true,%
  breaklinks=true,%
  colorlinks=true,%
  linkcolor=blue,anchorcolor=blue,%
  citecolor=blue,filecolor=green,%
  menucolor=blue,pagecolor=blue,%
  urlcolor=blue]{hyperref}
\else
\usepackage[%
  breaklinks=true,%
  colorlinks=true,%
  linkcolor=blue,anchorcolor=blue,%
  citecolor=blue,filecolor=green,%
  menucolor=blue,pagecolor=blue,%
  urlcolor=blue]{hyperref}
\fi

\makeatletter
\def\elsartstyle{%
    \def\normalsize{\@setfontsize\normalsize\@xiipt{14.5}}
    \def\small{\@setfontsize\small\@xipt{13.6}}
    \let\footnotesize=\small
    \def\large{\@setfontsize\large\@xivpt{18}}
    \def\Large{\@setfontsize\Large\@xviipt{22}}
    \skip\@mpfootins = 18\p@ \@plus 2\p@
    \normalsize
} \@ifundefined{square}{}{\let\Box\square} \makeatother

\begin{document}


\begin{frontmatter}

 \title{Short-Hard Gamma-Ray Bursts}
 \author{Ehud Nakar}
 \ead{udini@tapir.caltech.edu}
 \address{Theoretical Astrophysics, California Institute of Technology, MC 130-33, Pasadena, CA 91125, USA\\
  }




\begin{abstract}
Two types of Gamma-ray bursts (GRBs) are observed: short duration
hard spectrum GRBs and long duration soft spectrum GRBs. For many
years long GRBs were the focus of intense research while the lack of
observational data limited the study of short-hard GRBs (SHBs). In
2005 a breakthrough occurred following the first detections of SHB
afterglows, longer wavelength emission that follows the burst of
gamma-rays. Similarly to long GRBs, afterglow detections led to the
identification of SHB host galaxies and measurement of their
redshifts. These observations established that SHBs are cosmological
relativistic sources that, unlike long GRBs, do not originate from
the collapse of massive stars, and therefore constitute a distinct
physical phenomenon. One viable model for SHB origin is the
coalescence of compact binary systems (double neutron stars or a
neutron star and a black hole), in which case SHBs are the
electromagnetic counterparts of strong gravitational-wave sources.
The theoretical and observational study of SHBs following the recent
pivotal discoveries is reviewed, along with new theoretical results
that are presented here for the first time.

\end{abstract}

\end{frontmatter}
{\bf Contents}
\newline
\contentsline {section}{\numberline {1}Introduction}{1}{section.1}
\contentsline {section}{\numberline {2}Observations}{3}{section.2}
\begin{small}
\contentsline {subsection}{\numberline {\hspace{0.9cm} 2.1}\hspace{0.9cm}Prompt emission}{3}{subsection.2.1}
\contentsline {subsection}{\numberline {\hspace{1.6cm}2.1.1}\hspace{1.6cm}Duration}{3}{subsubsection.2.1.1}
\contentsline {subsection}{\numberline {\hspace{1.6cm}2.1.2}\hspace{1.6cm}Temporal structure}{5}{subsubsection.2.1.2}
\contentsline {subsection}{\numberline {\hspace{1.6cm}2.1.3}\hspace{1.6cm}Spectrum}{5}{subsubsection.2.1.3}
\contentsline {subsection}{\numberline {\hspace{1.6cm}2.1.4}\hspace{1.6cm}Isotropic equivalent energy}{7}{subsubsection.2.1.4}
\contentsline {subsection}{\numberline {\hspace{0.9cm} 2.2}\hspace{0.9cm}Afterglow}{7}{subsection.2.2}
\contentsline {subsection}{\numberline {\hspace{1.6cm}2.2.1}\hspace{1.6cm}The late time afterglow}{7}{subsubsection.2.2.1}
\contentsline {subsection}{\numberline {\hspace{1.6cm}2.2.2}\hspace{1.6cm}Soft X-ray ``tails'' of the prompt emission and the early afterglow}{10}{subsubsection.2.2.2}
\contentsline {subsection}{\numberline {\hspace{0.9cm} 2.3}\hspace{0.9cm}Observed redshift distribution}{11}{subsection.2.3}
\contentsline {subsection}{\numberline {\hspace{0.9cm} 2.4}\hspace{0.9cm}Host galaxies and cluster associations}{13}{subsection.2.4}
\contentsline {subsection}{\numberline {\hspace{0.9cm} 2.5}\hspace{0.9cm}Limits on a supernova component}{13}{subsection.2.5}
\contentsline {subsection}{\numberline {\hspace{0.9cm} 2.6}\hspace{0.9cm}Observed rate}{13}{subsection.2.6}
\contentsline {subsection}{\numberline {\hspace{0.9cm} 2.7}\hspace{0.9cm}Identifying an SHB}{14}{subsection.2.7}
\contentsline {subsection}{\numberline {\hspace{0.9cm} 2.8}\hspace{0.9cm}Additional putative SHBs}{15}{subsection.2.8}
\contentsline {subsection}{\numberline {\hspace{0.9cm} 2.9}\hspace{0.9cm}SHB interlopers}{16}{subsection.2.9}
\contentsline {subsection}{\numberline {\hspace{1.6cm}2.9.1}\hspace{1.6cm}Extra-galactic SGR giant flares}{16}{subsubsection.2.9.1}
\contentsline {subsection}{\numberline {\hspace{1.6cm}2.9.2}\hspace{1.6cm}Very short GRBs}{17}{subsubsection.2.9.2}
{\normalsize \contentsline {section}{\numberline {3}Relativistic outflows and the prompt emission}{17}{section.3}}
\contentsline {subsection}{\numberline {\hspace{0.9cm} 3.1}\hspace{0.9cm}Relativistic effects}{17}{subsection.3.1}
\contentsline {subsection}{\numberline {\hspace{1.6cm}3.1.1}\hspace{1.6cm}Time scales}{17}{subsubsection.3.1.1}
\contentsline {subsection}{\numberline {\hspace{1.6cm}3.1.2}\hspace{1.6cm}Causal connection and quasi-sphericity}{18}{subsubsection.3.1.2}
\contentsline {subsection}{\numberline {\hspace{1.6cm}3.1.3}\hspace{1.6cm}High latitude emission}{18}{subsubsection.3.1.3}
\contentsline {subsection}{\numberline {\hspace{0.9cm} 3.2}\hspace{0.9cm}The Lorentz factor of the outflow}{18}{subsection.3.2}
\contentsline {subsection}{\numberline {\hspace{1.6cm}3.2.1}\hspace{1.6cm}Opacity constraints}{19}{subsubsection.3.2.1}
\contentsline {subsection}{\numberline {\hspace{1.6cm}3.2.2}\hspace{1.6cm}Constraints from the onset of the afterglow}{20}{subsubsection.3.2.2}
\contentsline {subsection}{\numberline {\hspace{0.9cm} 3.3}\hspace{0.9cm}The composition of the relativistic outflow}{20}{subsection.3.3}
\contentsline {subsection}{\numberline {\hspace{1.6cm}3.3.1}\hspace{1.6cm}Baryonic flow}{20}{subsubsection.3.3.1}
\contentsline {subsection}{\numberline {\hspace{1.6cm}3.3.2}\hspace{1.6cm}Magnetized flow}{21}{subsubsection.3.3.2}
\contentsline {subsection}{\numberline {\hspace{0.9cm} 3.4}\hspace{0.9cm}The prompt emission}{21}{subsection.3.4}
\contentsline {subsection}{\numberline {\hspace{1.6cm}3.4.1}\hspace{1.6cm}Internal or external dissipation?}{22}{subsubsection.3.4.1}
\contentsline {subsection}{\numberline {\hspace{1.6cm}3.4.2}\hspace{1.6cm}Internal shocks}{22}{subsubsection.3.4.2}
\contentsline {subsection}{\numberline {\hspace{1.6cm}3.4.3}\hspace{1.6cm}Prompt emission from a magnetized flow}{24}{subsubsection.3.4.3}
\contentsline {subsection}{\numberline {\hspace{0.9cm} 3.5}\hspace{0.9cm}High energy Cosmic-rays and neutrinos}{24}{subsection.3.5}
\contentsline {subsection}{\numberline {\hspace{1.6cm}3.5.1}\hspace{1.6cm}High energy Cosmic-rays}{24}{subsubsection.3.5.1}
\contentsline {subsection}{\numberline {\hspace{1.6cm}3.5.2}\hspace{1.6cm}High energy neutrinos}{25}{subsubsection.3.5.2}
{\normalsize \contentsline {section}{\numberline {4}The afterglow - Theory}{25}{section.4}}
\contentsline {subsection}{\numberline {\hspace{0.9cm} 4.1}\hspace{0.9cm}Standard model - synchrotron radiation from a spherically symmetric adiabatic blast wave}{25}{subsection.4.1}
\contentsline {subsection}{\numberline {\hspace{1.6cm}4.1.1}\hspace{1.6cm}Comparison to the observations}{27}{subsubsection.4.1.1}
\contentsline {subsection}{\numberline {\hspace{0.9cm} 4.2}\hspace{0.9cm}Synchrotron self-Compton}{27}{subsection.4.2}
\contentsline {subsection}{\numberline {\hspace{0.9cm} 4.3}\hspace{0.9cm}Early afterglow and the reverse shock}{28}{subsection.4.3}
\contentsline {subsection}{\numberline {\hspace{0.9cm} 4.4}\hspace{0.9cm}``Naked'' afterglow}{29}{subsection.4.4}
\contentsline {subsection}{\numberline {\hspace{0.9cm} 4.5}\hspace{0.9cm}X-ray dark afterglows and $\gamma $-ray efficiency}{30}{subsection.4.5}
\contentsline {subsection}{\numberline {\hspace{0.9cm} 4.6}\hspace{0.9cm}Angular structure of the outflow}{30}{subsection.4.6}
\contentsline {subsection}{\numberline {\hspace{0.9cm} 4.7}\hspace{0.9cm}Afterglow variability}{32}{subsection.4.7}
\contentsline {subsection}{\numberline {\hspace{0.9cm} 4.8}\hspace{0.9cm}The early X-ray ``tail''}{33}{subsection.4.8}
\contentsline {subsection}{\numberline {\hspace{0.9cm} 4.9}\hspace{0.9cm}Macronova}{34}{subsection.4.9}
\contentsline {subsection}{\numberline {\hspace{0.9cm} 4.10}\hspace{0.9cm}Relativistic collisionless shocks}{34}{subsection.4.10}
{\normalsize \contentsline {section}{\numberline {5}Progenitors and the central engine}{35}{section.5}}
\contentsline {subsection}{\numberline {\hspace{0.9cm} 5.1}\hspace{0.9cm}Progenitors lifetime and intrinsic rate}{36}{subsection.5.1}
\contentsline {subsection}{\numberline {\hspace{1.6cm}5.1.1}\hspace{1.6cm}The sample}{36}{subsubsection.5.1.1}
\contentsline {subsection}{\numberline {\hspace{1.6cm}5.1.2}\hspace{1.6cm}Constraints from SHB redshift distribution}{36}{subsubsection.5.1.2}
\contentsline {subsection}{\numberline {\hspace{1.6cm}5.1.3}\hspace{1.6cm}Based on host galaxy types}{38}{subsubsection.5.1.3}
\contentsline {subsection}{\numberline {\hspace{1.6cm}5.1.4}\hspace{1.6cm}Intrinsic local rate}{39}{subsubsection.5.1.4}
\contentsline {subsection}{\numberline {\hspace{0.9cm} 5.2}\hspace{0.9cm}Coalescence of a compact binary}{39}{subsection.5.2}
\contentsline {subsection}{\numberline {\hspace{1.6cm}5.2.1}\hspace{1.6cm}The ``central engine''}{39}{subsubsection.5.2.1}
\contentsline {subsection}{\numberline {\hspace{1.6cm}5.2.2}\hspace{1.6cm}The lifetime of compact binaries and their merger rate}{44}{subsubsection.5.2.2}
\contentsline {subsection}{\numberline {\hspace{1.6cm}5.2.3}\hspace{1.6cm}Offsets from host galaxies and external medium densities}{45}{subsubsection.5.2.3}
\contentsline {subsection}{\numberline {\hspace{1.6cm}5.2.4}\hspace{1.6cm}Comparison with the observations}{46}{subsubsection.5.2.4}
\contentsline {subsection}{\numberline {\hspace{0.9cm} 5.3}\hspace{0.9cm}Other progenitor models}{46}{subsection.5.3}
\contentsline {subsection}{\numberline {\hspace{1.6cm}5.3.1}\hspace{1.6cm}Accretion induced collapse}{46}{subsubsection.5.3.1}
\contentsline {subsection}{\numberline {\hspace{1.6cm}5.3.2}\hspace{1.6cm}Magnetars}{47}{subsubsection.5.3.2}
\contentsline {subsection}{\numberline {\hspace{1.6cm}5.3.3}\hspace{1.6cm}Quark stars}{48}{subsubsection.5.3.3}
\contentsline {subsection}{\numberline {\hspace{1.6cm}5.3.4}\hspace{1.6cm}Type Ia SN}{48}{subsubsection.5.3.4}
{\normalsize \contentsline {section}{\numberline {6}Gravitational waves from SHBs}{48}{section.6}}
\contentsline {subsection}{\numberline {\hspace{0.9cm} 6.1}\hspace{0.9cm}NS-NS or NS-BH coalescence}{48}{subsection.6.1}
\contentsline {subsection}{\numberline {\hspace{0.9cm} 6.2}\hspace{0.9cm}Other processes that radiate gravitational waves }{49}{subsection.6.2}
\end{small}

\listoftables
\newpage

\section{Introduction}
Gamma-ray bursts (GRBs) are short intense flashes of soft
($\sim$MeV) $\gamma$-rays that are detectable once or twice a day.
The observed bursts arrive from apparently random directions in the
sky and they last between tens of milliseconds and thousands of
seconds. Their physical origin remains a focal point of research and
debate ever since they were first detected, more than three decades
ago.

More than twenty years ago hints that the GRB duration distribution
is bimodal have emerged \citep{Mazets81,Norris84} , suggesting that
the GRB population might not be monolithic. The idea that GRBs are
most likely composed out of two major distinctive sub-populations
became the common view following the influential work of
\cite{Kouveliotou93}. Based on 222 GRBs detected by the Burst And
Transient Source Experiment ({\it
BATSE}\footnote{\href{http://www.batse.msfc.nasa.gov/batse/}{http://www.batse.msfc.nasa.gov/batse/}})
on board the Compton Gamma-Ray Observatory, \cite{Kouveliotou93}
confirmed that the GRB duration distribution is indeed bimodal, with
a minimum around $2$ s. They have further shown that  bursts with
durations shorter than 2 s are composed, on average, of higher
energy (harder) photons than longer bursts. It took a dozen
additional years before recent observations confirmed that the two
sub-populations, defined in duration-hardness space, indeed
represent two distinctive physical phenomena. Throughout this review
I use the term short-hard GRBs (SHBs), or simply short GRBs, for
bursts that share the physical properties of the population that
constitutes the majority of the short-hard events in the
duration-hardness space\footnote{Note that it is not necessary that
the duration of a specific burst corresponds one-to-one to its
physical properties (e.g., the nature of the progenitor). It is
possible for example that there is a small fraction of bursts that
are associated physically with long GRBs but their duration is short
and vice versa.}. Bursts with physical properties common to the
population that constitutes the majority of long-soft events are
denoted simply as long GRBs. The term GRBs is used to describe the
whole population of the busters, short and long.

A major revolution in GRB research came with the launch of \BATSE in
1991. \BATSE detected bursts at a rate of one per day, a quarter of
which were short. The isotropic distribution of bursts on the sky
(both long and short) and the paucity of weak bursts indicated that
the origin of GRBs is extra-Galactic and most likely cosmological.
However, because of the inability to identify the distance of any
specific burst based on the observed gamma-rays alone, the final
confirmation of the cosmological origin of GRBs had to wait
\cite[for a review of \BATSE results see][]{Fishman95}. The final
validation that long GRBs arrive from remote locations in the
Universe came following the discovery by the Dutch-Italian satellite
BeppoSAX that the prompt $\gamma$-ray emission of long GRBs, lasting
minutes or less, is followed by X-ray emission that can be detected
for hours and days \citep{Costa97}. Observations of this X-ray
emission enabled accurate localizations of bursts and led to the
detection of associated optical \citep{vanParadijs97} and radio
\citep{Frail97} emission, that can be observed, in some cases, for
weeks and years, respectively. Longer wave-length emission that
follows the prompt GRB is called the ``afterglow'' and it provides a
wealth of information about the physics of the burst. Most
importantly, sub-arcsecond localization of optical afterglows
enabled secure identification of galaxies that host long GRBs,
leading to the measurement of their redshifts, and finally
establishing an unambiguous distance scale for these events
\citep[e.g.,][]{Kulkarni98}. In some cases, the burst redshift was
measured directly by detection of absorption lines in the spectra of
optical afterglows  \citep[e.g.,][]{Metzger97}.

These observations revolutionized the study of long GRBs. The
confirmation of their cosmological origin and the detailed
observations of a growing sample of afterglows supported a physical
picture according to which long GRBs are produced by a catastrophic
event involving a stellar-mass object or system that releases a vast
amount of energy ($\gtrsim 0.001 M_\odot c^2$) in a compact region
($< 100$ km) on time scales of seconds to minutes. This energy
source, referred to as the ``central engine'', accelerates an
ultra-relativistic outflow to Lorentz factor $\gtrsim 100$ and it is
this outflow that generates the observed $\gamma$-ray prompt
emission, and later the afterglow, at large distances from the
source. This model is now generally accepted.

The nature of the stellar progenitor of long GRBs remained a matter
of debate for many years. Until 2003, when a detection of a long GRB
at a relatively low redshift (z=0.168) by {\it HETE-2}\footnote{The
High Energy Transient Explorer;
\href{http://heasarc.gsfc.nasa.gov/docs/hete2/hete2.html}{http://heasarc.gsfc.nasa.gov/docs/hete2/hete2.html}}
led to the identification of a type Ic supernova (SN) spectrum
superposed on the afterglow of this burst \citep{Stanek03,Hjorth03}.
This observation confirmed previous suggestions that at least some
long GRBs are associated with SNe \citep[e.g.][]{galama98,Bloom99b}.
Following this association and several additional evidence
\citep[e.g.,][]{Fruchter06} the consensus today is that most, and
probably all, long GRBs are produced by the collapse of very massive
stars \cite[e.g.,][]{Woosley93,Paczynski98,MacFadyen99}.

Comparable studies of SHBs were not conducted during this time.
Being shorter and harder they eluded accurate localization and no
SHB afterglow was detected despite the effort. The breakthrough in
the study of SHBs occurred, finally, during the spring-summer of
2005 when \swift and \hete succeeded in localizing several SHBs,
leading to afterglow detections and to the determination of their
redshifts. Following these discoveries, and fueled by a continuous
flow of SHB detections by {\it Swift}, the study of SHBs progressed
rapidly. The first conclusions from these observations were that
SHBs are cosmological, but unlike long GRBs, their progenitors are
not massive stars, thereby confirming that these two observationally
defined classes are distinctive physical phenomena. On the other
hand, the comparable luminosities and roughly similar afterglows of
long and short GRBs, suggested that similar physical processes are
involved in both types of explosions.

The aim of this paper is to review and summarize the theoretical
study of SHBs following the recent observations. These observations
are described in \S\ref{SEC: observations} and aspects that have
direct implications for the current theoretical understanding of
SHBs are emphasized. Special attention is given to the observational
challenge of identifying a burst as a SHB.

Relativistic effects and the prompt emission theory model are
discussed in \S\ref{SEC: Rel+propmpt} while the afterglow theory is
discussed in \S\ref{SEC: afterglow theory}. The theory of both the
prompt and the afterglow emission is discussed in the framework of
models in which the emission is produced by a relativistic outflow
that first expands quasi-spherically (see definition in \S\ref{SEC:
Rel qausi-spher}) and that later its interaction with the ambient
medium leads to the formation of a blast wave that propagates into
this medium. These models include the ``classic'' fireball model of
radiation accelerated baryonic plasma as well as other types of
relativistic outflows (e.g., Poynting-flux-dominated)\footnote{Other
GRB models, such as the cannonball model \citep[for review
see][]{Dar04} and the precessing Jet model
\citep[e.g.,][]{Fargion99,Fargion06}, are not discussed in detail in
this review.}. The physics of GRB emission (prompt and afterglow)
was intensively studied in the context of long GRBs, while there is
little work specific to SHBs. As both the prompt emission and the
afterglows of long and short GRBs are quite similar, long GRB models
are typically applied to SHBs as well. The fireball and related
models were comprehensively covered by several excellent reviews,
and I refer the reader to these reviews for a detailed description
of these models
\citep{Piran99,Piran00,ZhangMeszaros04,Piran05,Meszaros06,Lyutikov06}.
In order for this review to be self contained, I include a brief
overview of the basics of the fireball and related  models,
emphasizing aspects that are relevant to SHBs. I focus on the
interpretation of SHB observations in the framework of these models.
In several cases I extend existing models and re-derive predictions
and constraints that are relevant to SHB parameter space. These
derivations and some novel conclusions that follow appear for the
first time in this
review (see highlighted sections in Table 1). 

I go on to review proposed SHB progenitor models and the formation
of the central engine \S\ref{SEC: progenitors} \citep[for an
excelent comprehansive review of the topic see][]{LeeRuiz07}. The
leading progenitor candidate is the coalescence of a neutron star
(NS) with another neutron star or with a black hole (BH). This idea
was commented upon some twenty years ago
\citep{Blinnikov84,Paczynski86,Goodman86,Goodman87} and was first
explored in detail by \cite{Eichler89}. It successfully survived
twenty years of observations and got some support from the
properties of SHB afterglows and of the identified host galaxies. I
therefore dedicate most of the discussion on SHB progenitors to the
aspects of these mergers that are relevant to SHBs.  Nevertheless, a
merger origin for SHBs is not confirmed, and other progenitor models
are still viable, and are discussed here as well.

Mergers of NS-NS or NS-BH binaries are also the most promising
sources of gravitational-waves that may be detected by ground-based
observatories. This of course makes SHBs very interesting, as the
possible electromagnetic counterparts of gravitational-wave signals
that are expected to be detected within a decade. The prospects of
gravitational wave detection from SHBs are discussed in \S\ref{SEC:
GW}.

For convenience, Table 1 
summarizes and compares observational properties of short and long
GRBs, and their theoretical interpretations in the fireball model
framework. This table can also serve as a quick reference guide for
the different topics covered in this review. Through the review
$\Omega_m = 0.3$, $\Omega_\Lambda = 0.7$ and $H_0=70 ~\rm km s^{-1}~
Mpc^{-1}$ cosmology is used.

\begin{table}
{\bf \caption{Short vs. Long GRBs}} \label{Table: short Vs. long}

\begin{tabular}{@{}cccc@{}}
&&&\\
 & {\bf Short GRBs} & {\bf Long GRBs} & {\bf Section$^*$}\\\hline\hline
{\bf General}&&\\\hline
\BATSE observed all sky rate  & $\approx 170\rm yr^{-1}$  &   $\approx 500 \rm ~yr^{-1}$ & \S\ref{SEC: obs rate} \\
\BATSE observed local rate density  & $\sim 10 \rm ~Gpc^{-3}~yr^{-1}$ &$\sim 0.5 \rm ~Gpc^{-3}~yr^{-1}$ & \S\ref{SEC: obs rate}\\
Host Galaxy types & Early \& Late & Late & \S\ref{SEC: host obs} \\
Host specific SFR & $\lesssim 1 \rm ~M_\odot/yr/(L/L_*)$& $\sim 10 \rm ~M_\odot/yr/(L/L_*)$ & \S\ref{SEC: host obs}\\
Median observed redshift & $\approx 0.25$ & $\approx 2.5$ & \S\ref{SEC: z_obs}\\
Supernovae association&  No & Yes (at least some) & \S\ref{SEC: SN limit}\\
Progenitor& NS-BH/NS-NS/? & Massive star & \S\ref{SEC: progenitors}\\
\hline {\bf Prompt emission} &&&\\\hline
Typical \BATSE duration & $\approx 0.8$ s & $\approx 30$ s & \S\ref{SEC: Duration}\\
Best fit spectral model$^{\text a}$ & Power-law + exp cutoff & Band function  & \S\ref{SEC: spectrum prompt} \\
$E_{\gamma,iso} ^{\text b}$ & $10^{49}-10^{51}$ erg  & $10^{52}-10^{54}$ erg & \S\ref{SEC: prompt energy}\\
$L_{\gamma,iso} ^{\text b}$ & $10^{50}-10^{52}$ erg/s  & $10^{50}-10^{52}$ erg/s  & \S\ref{SEC: prompt energy}\\
Average observed flux $^{\text c}$& $\sim 5\cdot10^{-10} \rm ~GeV/cm^2/s/sr $ & $\sim 10^{-8} \rm ~GeV/cm^2/s/sr$  \\
Local energy output rate $^{\text d}$& $\sim 10^{51} \rm~erg~Gpc^{-3}~yr^{-1}$ & $\sim 10^{53} \rm~erg~Gpc^{-3}~yr^{-1} $ \\
Total energy output $^{\text e}$& $\sim 10^{63}$ erg &  $\sim 10^{66}$ erg  \\
\hline {\bf X-Ray afterglow} &&\\\hline
X-ray dark & Some bursts & None & \S\ref{SEC: obs late aft} \& \S\ref{SEC: gamma eff}\\
Typical decay phase ($\sim t^{-1}$)& Yes & Yes & \S\ref{SEC: obs late aft} \& \S\ref{SEC: afte standard theory}\\
Shallow decay phase ($\sim t^{-0.25}$)& Not observed yet & Yes & \S\ref{SEC: obs late aft}\\
X-ray flares & Yes & Yes & \S\ref{SEC: obs late aft} \& \S\ref{SEC: aft variability}\\
\hline {\bf Fireball model$^{\text f}$ } &&&\\\hline
Lorentz factor &  $\gtrsim 30$ &$\gtrsim 100$ &  {\bf \S\ref{SEC: LF}}\\
Prompt emission energy source& Internal processes$^{\text g}$ & Internal processes$^{\text g}$ & \S\ref{SEC: prompt internal VS. external} \\
Afterglow energy source & External shock & External shock & \S\ref{SEC: afte standard theory} \\
Synchrotron self-Compton & $Y << 1$ & $Y \gtrsim 1$ & {\bf \S\ref{SEC: SSC}}\\
Early afterglow \& prompt emission & Well separated & May overlap & \S\ref{SEC: early aft theory} \\
Early afterglow (baryonic flow) & Optically faint & Optically bright & \S\ref{SEC: early aft theory}\\
Gamma-ray efficiency & $\sim 10\%$ & $\gtrsim 10 \%$ &{\bf \S\ref{SEC: gamma eff}} \\
X-ray dark afterglows$^{\text h}$ & $n \lesssim 10^{-5}\rm ~cm^{-3}$  & - & {\bf \S\ref{SEC: gamma eff}}  \\
Average beaming & $1 < f_b^{-1} \lesssim 100$ & $f_b^{-1} \sim 75$ & \S\ref{SEC: beaming} \\\hline
\end{tabular}
\newline\newline\newline

$^*$ The section in which the SHB table entry is discussed. Highlighted sections present new results that appear in this review for the first time.  \\
$^{\text a}$ The spectral model that provides the best fit to most of the bursts (may be affected by observational selection effects).\\
$^{\text b}$ Isotropic equivalent quantities per burst.\\
$^{\text c}$ The total observed $\sim$MeV $\gamma$-ray flux from GRBs, averaged over long time ($\gg 1$ day).\\
$^{\text d}$ The energy output density rate in $\sim$MeV $\gamma$-rays of the entire local GRB population.\\
$^{\text e}$ The total energy output in $\sim$MeV $\gamma$-rays of the entire GRB population in the observed universe.\\
$^{\text f}$ Theoretical interpretations in a framework of models where outflow of any type
(e.g., baryonic, Poynting-flux-dominated) expands quasi-spherically and interacts with the ambient medium by driving a blast wave into it.\\
$^{\text g}$ In Poynting-flux-dominated outflow models the dissipation of the internal magnetic field that leads to the prompt emission may be induced by interaction with the external medium.\\
$^{\text h}$ The most likely, but not the only, explanation of X-ray
dark afterglows (see \S\ref{SEC: gamma eff} for details).
\end{table}

\section{Observations}\label{SEC: observations}
Gamma-Ray burst observations are grossly divided into two main
phases - the prompt gamma-ray emission and the afterglow. The prompt
gamma-rays are intense bursts of $\sim MeV$ $\gamma$-rays that are
detected by gamma-ray space observatories and are localized by
current instruments to within several arcminutes. Afterglows are
X-ray, optical and radio emission that follow the prompt gamma-rays
and can be observed in some cases weeks, months and years after the
bursts, respectively.

{\it BATSE} detected the prompt emission from  $\sim 3000$ GRBs,
about 1/4 of which are SHBs. Additionally, prompt emission from more
than a hundred SHBs were observed by {\it
Konus-Wind}\footnote{\href{http://lheawww.gsfc.nasa.gov/docs/gamcosray/legr/konus/}{http://lheawww.gsfc.nasa.gov/docs/gamcosray/legr/konus/}}
and other spacecraft that were, or still are, part of the
Interplanetary
Network\footnote{\href{http://www.ssl.berkeley.edu/ipn3/}{http://www.ssl.berkeley.edu/ipn3/}}
(IPN). This is compared to a handful of SHB afterglows that were
observed following accurate localization of the prompt emission by
{\it Swift} and {\it HETE-2}. In some of these cases the
sub-arcsecond localization of the afterglow led to an unambiguous
identification of the burst redshift and host galaxy. These
observations are briefly reviewed below.

\subsection{Prompt emission}
Most of the prompt emission properties are derived using \BATSE
bursts. Typically, the SHB sample is drawn out of the complete
\BATSE GRB sample using the criterion $T_{90}<2$ s (defined below).
This is clearly a rough cut (see \S\ref{SEC: SHB identification}),
but it is sufficient for the purpose of statistical studies which
are weakly affected by a small contamination of the sample.

\subsubsection{Duration}\label{SEC: Duration}
\begin{figure}[t]
\includegraphics[width=12cm]{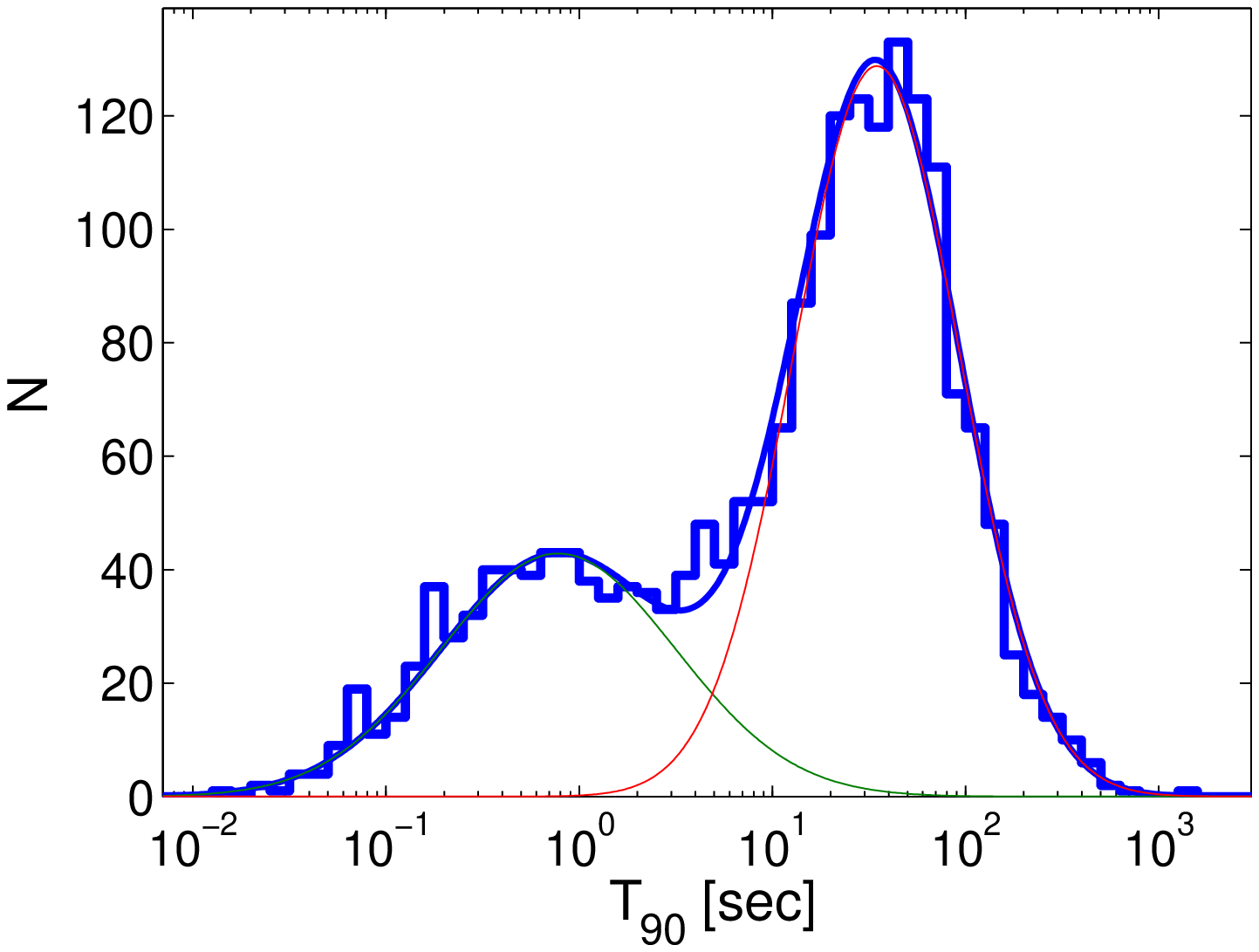}
\caption{\label{FIG: T90} The bimodal duration distribution of GRBs.
The observations (2041 bursts in the current \BATSE catalog) are
marked by the thick stairs. The decomposition of the distribution
into two lognormal distributions, as determined by \cite{Horvath02},
({\it thin solid lines}) and the sum of these components ({\it thick
solid line}) are superposed.}
\end{figure}

The true (intrinsic) duration distribution of SHBs is unknown. At
the short end, the observed distribution is affected by the minimal
\BATSE trigger time ($64$ ms) while the long end the distribution
blends into that of long GRBs. Figure \ref{FIG: T90} depicts the
$T_{90}$ distribution of the entire \BATSE  GRB sample, where
$T_{90}$ is the duration encompassing the 5'th to the 95'th
percentiles of the total counts in the energy range $20-2000$ keV.
The  duration distribution is bimodal with a minimum around $2$ s
\citep{Kouveliotou93}. For this reason the dividing line between
short and long GRBs is usually drawn at $T_{90}=2$ s. This
distinction suffices for the purpose of statistical population
analysis, but one should bear in mind that there are short GRBs with
$T_{90}>2$ s, and long GRBs with $T_{90}<2$ s. \cite{Horvath02}
finds that this bimodal distribution can be decomposed into two
lognormal distributions, as presented in figure \ref{FIG: T90}. The
SHB distribution peaks  at $T_{90} \approx 0.8$ s, and the
full-width-half-maximum of the distribution is 1.4 dex. If this
decomposition represents the physical one then $25\%[3.5\%]$ of
\BATSE SHBs last longer than $2[10]$ s.

It was recently realized that at least in some SHBs the initial
short and hard $\gamma$-ray prompt emission is followed by a much
longer X-ray ``tail'' that lasts tens to hundreds of seconds.
Typically,  this tail is too faint to affect \BATSE $T_{90}$,
although in some cases it probably does \citep{Norris06}. The
properties of this soft tail and its implications for the
identification of SHBs are discussed later (\S\ref{SEC: X-ray tail}
and \S\ref{SEC: SHB identification}).

\subsubsection{Temporal structure}\label{SEC: prompt temporal}
\begin{figure}[t]
\includegraphics[width=12cm]{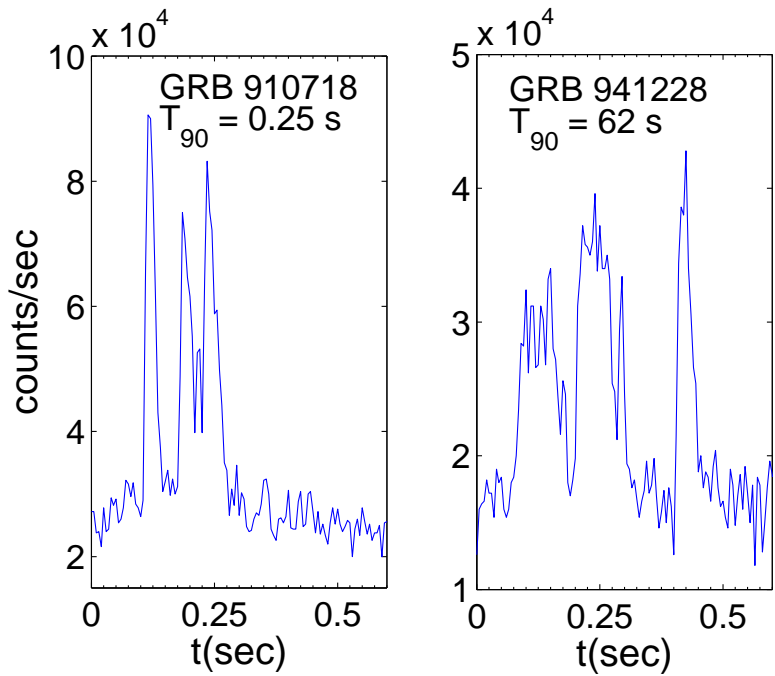}
\caption{\label{FIG: LightCurve} \textbf{Left}: The light curve of
GRB 910718, a bright SHB with $T_{90}=0.25$ s.  \textbf{Right}: The
first 0.7 s of GRB 941228, a  bright long GRB with $T_{90}=62$ s.
The figure demonstrates  the similarity in short-time-scale
structure in long and short GRBs. The resolution of both light
curves is 5 ms. Both bursts show variability down to the resolution
limit.}
\end{figure}

Only a small fraction of the \batse SHB sample have sufficient
signal to noise ratio (S/N) to conduct analysis of the fine temporal
structure. \cite{Nakar02} analyze a sample of 33 bright SHBs at a
resolution of 2 ms, looking for variability (separate pulses) within
the bursts. The low S/N and the limited time resolution imply that
the observed variability is only a lower limit of the true one. Yet,
they find that most of the bursts in the sample exhibit variability
on time scales that are shorter than the bursts' durations (see Fig.
\ref{FIG: LightCurve} for a typical SHB light curve). More than half
of the bursts in their sample show at least two well-separated
pulses and more than a third show rapid variability in the sense
that the shortest pulse is shorter by more than an order of
magnitude than the burst duration. No correlation is found between
the duration of a burst and the duration of its sub-pulses (given
that the burst is not single pulsed). The duration distribution of
single pulses ranges from 5 ms to 300 ms with a broad peak around 50
ms. Thus, the lower limit on the shortest time scale observed in
these SHBs is of the order of 10 ms, and is set by the resolution
limit. Shorter time scales are probably present, as evident from a
single case in which a very bright $<1$ ms pulse is observed in a
SHB \citep[][fainter pulses than this one cannot be resolved on ms
time scale]{Scargle98}. \cite{McBreen01} analyzed the distribution
of various temporal properties of pulses in 100 bright \batse SHBs.
They find that the rise times, fall times, FWHM, pulse amplitudes
and areas are all consistent with lognormal distributions and that
time intervals between pulses and pulse amplitudes are highly
correlated with each other.

A comparison of the temporal structure of bright SHBs to the initial
$2$ s of a sample of long GRBs\footnote{The sample includes only
long GRBs that have high resolution light curve and an initial pulse
that is shorter than $2$ s} shows similar time scales and similar
distributions of pulse durations \citep{Nakar02}. This similarity is
demonstrated in figure \ref{FIG: LightCurve}. Similarly,
\cite{McBreen01} find a great similarity between the lognormal
distributions and correlations in the temporal structure of short
and long GRBs. On the other hand, an examination of the temporal
evolution of pulses as a function of frequency shows a different
behavior in long and short bursts. \cite{Norris01} compare the
spectral lags of short and long GRBs. Spectral lag is a measurement
of the spectral evolution timescale of the pulse structure, where a
positive value indicates a hard-to-soft evolution \citep[see][for an
exact definition]{Norris00}. They find that long bursts show
positive spectral lags that extend up to $\sim 2$ s with a core
around $50$ ms. SHBs, however, show a symmetric distribution of lags
that ranges between $\pm 30$ ms.

Thus the temporal structure of SHBs shows both similarities and
dissimilarities to that of long GRBs. Unfortunately additional
comparisons of the temporal structure of short and long bursts were
not carried out so far, mostly because of the difficulty in the
analysis of SHB light curves. While the temporal structure of long
GRBs was explored in detail \citep[e.g.,][]{Norris96}, only the
general properties of the temporal structure of SHBs are known.

\subsubsection{Spectrum}\label{SEC: spectrum prompt}
\begin{figure}[t]
\includegraphics[width=12cm]{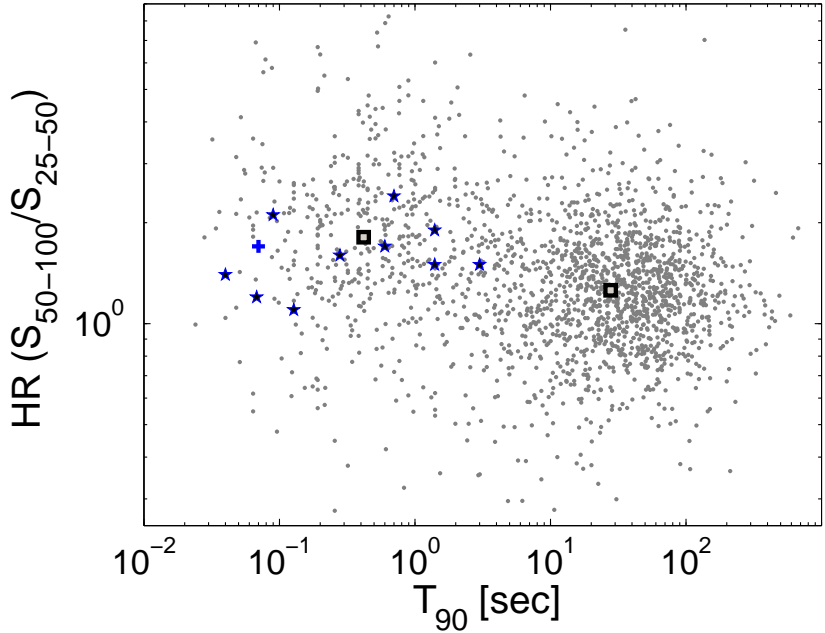}
\caption{\label{FIG: T90_HR} The duration $T_{90}$ and the hardness
ratio of all GRBs with available data from the \BATSE catalog ({\it
dots}). \Swift ({\it stars}) and \hete ({\it plus}) SHBs are marked
as well. The average logarithmic values of \BATSE bursts with
$T_{90}>2$ s and $T_{90}<2$ s are marked by the two squares. The
hardness ratio used here is the ratio of the fluence in the 50-100
keV and 25-50 keV energy bands. The \BATSE data is taken from the
current \BATSE catalog and \Swift and \hete data are taken from
Table 2, 
where the hardness ratio is calculated
using the photon index. Note that the fluence ratio that is used
here is different than the typical hardness ratio, that uses
counts.}\end{figure}

Comparison of the hardness ratio of SHBs to that of long GRBs shows
that SHBs are on average harder. This result was used by
\cite{Kouveliotou93}, together with the bimodal duration
distribution, to suggest that SHBs and long GRBs are two distinctive
populations. Figure \ref{FIG: T90_HR} depicts $T_{90}$ and the
hardness (ratio between the $50-100$ keV and the $25-50$ keV
fluence) of \BATSE GRBs. It is evident that while SHBs are on
average harder, the variance in the hardness ratio is such that the
two populations overlap and the distinction between them is not well
determined.

\cite{Ghirlanda04} explore the time integrated spectra of $28$
bright \BATSE SHBs. They attempt to fit the spectrum of each burst
with four different spectral models $-$ a single power-law, a broken
power-law, a Band function \citep[a smoothly broken power-law;
][]{Band93} and a power-law with an exponential cut-off (PLE):
\begin{equation}\label{EQ: prompt spectrum}
    \frac{dN}{dE}=N_0 E^{\alpha} {\rm exp}\left[-E/E_0\right]
\end{equation}
where $N$ is the photon count and $E_p = (2+\alpha)E_0$ is the peak
of  $\nu F_\nu$. They find that PLE model (Eq. \ref{EQ: prompt
spectrum}) provides the best fit to the data. The values of $E_p$
that \cite{Ghirlanda04} find are in the range $50-1000$ keV (mean
value $355 \pm 30$ keV), which are comparable to values observed in
long GRBs. They find values of $\alpha$ between $-2$ and $0.5$ (mean
value $-0.58 \pm 0.1$), which are significantly harder than those of
long GRBs. These higher values of the low energy power-law index are
the main driver of the higher hardness ratio of SHBs.

The \konus SHB catalog \citep{Mazets04} contains $140$ spectra of
$98$ SHBs (for some SHBs the spectrum is time integrated and for
some it is given in two or three different time intervals).
\cite{Mazets04} use the same spectral models as \cite{Ghirlanda04}
to fit the spectra of \konus SHBs. They fit $60\%$ of the spectra
with a PLE, $21\%$ with a single power-law and 16\% with a Band
function. The average $\alpha$ value of the PLE spectra is $-0.78$,
rather similar to the one found by \cite{Ghirlanda04} for \BATSE
SHBs. The average $E_p$ of PLE spectra is $837$ keV, significantly
higher than the value found for \BATSE events. This difference is
probably a result of the wider energy window of {\it Konus-Wind},
$15-10000$ keV compared to $25-2000$ keV for {\it BATSE}. The
highest $E_p$ value in the \konus catalog is $3.1$ MeV. In about
$30\%$ of the sample photons are observed up to $5$ MeV but only in
$\approx 5\%$ of the bursts harder photons are observed. Only a
single burst shows $10$ MeV photons (the one with a PLE spectrum
with $E_p=3.1$ MeV).

An important point is that while most spectra are consistent with
exponential cut-offs, the low energy spectrum of most SHBs is too
soft to be consistent with a blackbody spectrum. \cite{Lazzati05}
studied the spectra of $76$ \batse SHBs and find that in more than
$75\%$ of the bursts the spectrum is inconsistent with a blackbody
spectrum.

The fact that PLE provides the best fit to most SHB spectra is in
contrast to long GRBs, where a Band function, with a rather shallow
high energy power-law ($\sim E^{-2.25}$), provides the best spectral
fit for most of the bursts \citep{Preece00,Ghirlanda02}. This
dissimilarity may reflect a genuine intrinsic difference between the
spectra of short and long GRBs, in which case long GRBs are actually
harder than SHBs at high energies ($\gg$MeV). However, this
difference may also be a result of observational selection effects.
A PLE fit may be preferred over a Band function fit in cases of low
signal-to-noise ratio at high energies, as suggested by the results
of \cite{Kaneko06} that analyze the spectra of the 17 brightest
\batse SHBs and find that most of them are well fitted by a Band
function or a broken power-law.


Among SHBs with known redshifts there are only two bursts with good
spectral data (the spectral range of \swift alone is too narrow to
constrain the broad band spectrum). The prompt emission
time-integrated spectrum of \hete SHB 050709 is best fitted by a PLE
with $E_p=86.5^{+16}_{-11}$ keV and $\alpha = -0.82^{+0.13}_{-0.14}$
\citep[$E_0 \approx 73$ keV;][]{Villasenor05}. The \konus spectrum
of SHB 051221A is fitted by a PLE with $E_p=402^{+93}_{-72}$ keV and
$\alpha = -1.08^{+0.13}_{-0.14}$ \citep[$E_0 \approx 436$
keV;][]{Golenetskii05_GCN4394}.

As I discuss below, the hardest observed non-thermal photons in the
prompt emission play an important role in constraining the Lorentz
factor of the prompt emission source (\S\ref{SEC: LF}). The
observations described above show that there are many SHB spectra
that are consistent with having no non-thermal photons above $\sim
1$ MeV, while other bursts show non-thermal emission at least up to
$\sim 5$ MeV. There is no published evidence that SHBs emit
non-thermal photons above $10$ MeV. This is in contrast to long GRBs
were there are bursts that show $>100$ MeV photons
\citep[e.g.,][]{Schneid92,Sommer94,Hurley94}. The lack of observed
$\gtrsim 100$ MeV photons from SHBs may be in part an observational
selection effect as well. At very high energies the background is
negligible on prompt emission time scales and therefore the
detectability depends on the fluence and not on the flux. Hence,
SHBs, which have a significantly lower fluence than long GRBs, are
much harder to detect at these energies.

\subsubsection{Isotropic equivalent energy}\label{SEC: prompt energy}
In the few cases where SHB redshifts are known, the isotropic
equivalent gamma-ray energy, $E_{\gamma,iso}$, and luminosity,
$L_{\gamma,iso}$, can be calculated (i.e., assuming that the source
emission is constant over the whole $4\pi$ sr solid angle). These
values are listed in Table 2. 
The isotropic equivalent energy of SHBs ranges between
$10^{49}-10^{51}$ erg, which is $2-3$ orders of magnitude smaller
than the energy range of long GRBs. The peak isotropic equivalent
luminosity in most SHBs with known redshift is $10^{50}$ erg/s and
in a single case (SHB 051221A) it is even larger than $10^{52}$
erg/s. These values are comparable to those observed in long GRBs
(see Table 2). 

\cite{Amati02} have shown that there is a tight relation  between
$E_{\gamma,iso}$ and the redshift correct peak spectral energy,
$E_{p}(1+z)$, of long GRBs with known redshifts \citep[see also
][]{Lloyd00a,Lloyd02,Lamb04}. Even before redshifts of some SHBs
were determined, it was shown that SHBs do not follow the Amati
relation\footnote{\cite{NakarPiran05} and \cite{Band05} have used
the same method to show that many \batse long GRBs do not follow the
Amati relation as well.} \citep{Ghirlanda04,NakarPiran05}. The
recent detection of SHB redshifts has shown that SHBs with known
redshifts do not follow the Amati relation as well \citep{Amati06}.

\subsection{Afterglow}\label{SEC: afterglow observations}
The breakthrough in the study of SHBs came in the spring-summer of
2005 with the first detections of X-ray, optical and radio afterglow
emission  \citep{Gehrels05,
Castro05b,Prochaska05,Fox05,Hjorth05,Hjorth05b,Bloom06,Covino06,Berger05}.
This discovery was facilitated by timely and accurate localizations
by \swift and \hete, and came 8 years after the first afterglow
detection of a long GRB. The difficulty in detecting SHB afterglows
 is to achieve precise localization from a small number of photons
(compared to long GRBs) and then to quickly point a sensitive X-ray
instrument in order to detect the afterglow, which is  significantly
fainter relative to long GRB afterglows.

The importance of afterglow detection is twofold. First, afterglow
detection enables sub-arcsecond localization and, possibly,
unambiguous determination of the host galaxy and its redshift.
Second, the afterglow itself can teach us about the processes that
take place after the explosion and to provide clues about the
physics of the central engine and about the properties and
environment of the progenitor. Here I discuss the observational
properties of the afterglows themselves, while the environmental
properties are discussed later. The discussion below is based on the
small sample of the SHBs with observed afterglows, and even within
this small sample a large variance exists. It is likely that as more
SHB afterglows will be observed, the discussion below will have to
be updated or revised.

\begin{table} \label{Table: prompt}

{\bf \caption{Prompt emission and afterglow properties}}
 \begin{tabular}{lc@{~~~~~~}lc@{~~~~~~}lc@{~~~~~~}lc@{~~~~~~}lc@{~~~~~~}lc@{~~~~~~}lc@{~~~~~~}lc@{~~~~~~}l}
 \hline
 SHB & $T_{90}^a$ & z & $S_\gamma^b$ & $E_{\gamma,iso}^c$ & $L_{\gamma,peak}^d$ &$f_b^{-1}~^e$ & $E_\gamma~^f$ & $F_{XE}^g$ & $\rm Aft.^h$ & ref.\\
     &   [s]    &   & $\times 10^{-7}$& $\times 10^{49}$  & $\times 10^{50}$  &     & $\times 10^{49}$ &$\times 10^{-11}$&           &     \\
  \hline
  \hline
050509B& 0.04 & 0.225 & $0.23 \pm 0.09$ & 0.25      &0.7[35ms]&       &            &       0.06          &   XE        &  ~1\\ \hline
$050709^*$& 0.07 & 0.16  & $3\pm0.38$    & 1.6      &3[60ms] &        &            &       800           &   {\tiny XE/L,O}   & ~2 \\ \hline
050724 &  3   &0.258    & $6.3\pm1$     & 9.1       &1[0.8s] &$<13$ &  $>0.7$    &       1200          &    XE/L   &  ~3\\
$^{**}$& $[1.3]$&      &                &           &        &[$<500$]&$[>0.02]$   &                     &  O,R        &  \\ \hline
050813 & 0.6  & 0.7 or&$1.24\pm0.46$    & 11        &3[0.3s] &        &            &       0.3           &   XE        &  ~4\\
       &      & 1.8   &                 & 48        &20[0.2s]&        &            &                     &             &  \\ \hline
050906 & 0.13 &       &$0.84\pm0.46$    &           &        &        &        &       $<0.007 $     &    None     &  ~5\\\hline
050925$^\dagger$ & 0.07 &       &$0.92\pm0.18$      &        &        &       &            &      $<0.003$       &None&  ~6\\\hline
051105A& 0.28 &       &$0.4\pm0.09$     &           &        &        &            &                     &    None     &  ~7\\\hline
051210 & 1.4  &       &$1.9\pm0.3$      &           &        &        &            &         40          &   XE        &  ~8\\\hline
051221 & 1.3  & 0.546 &$22.2\pm0.8$     & 130       &        &   80   &    1.5     &         10          &  XE/L &  ~9\\
       &      &       &$(32.2^{+1}_{-17})$& 250     &(550[3ms])&        &    3       &                     &  O,R         &       \\\hline
060313 & 0.7  &$<1.7$&$32.1\pm1.4$      &           &        &        &            &         30          &   XE/L   &  ~10\\
       &      &       &$(110\pm 20)$    &           &        &        &            &                     &   O         &       \\\hline
060502B& 0.09 &  0.287(?)&$1\pm0.13$    &  0.8(?)   &        &        &            &         0.1         &   XE        &  ~11     \\\hline
060801&  0.04 &          &$0.8\pm0.1$   &           &        &        &            &         0.1         &   XE        &  ~12 \\\hline
061201$^\ddag$& 0.8 &    &$3.3\pm0.3$   &           &        &        &            &         $\sim$10    &{\tiny XE/L,O}&  ~13 \\\hline
061217$^\ddag$& 0.3 &    &$0.46\pm0.08$ &           &        &        &            &         $\sim$0.1   &   XE/L    &  ~14 \\\hline
 \hline

\end{tabular}
\newline\newline
Various properties of SHB prompt and afterglow emission as detected
by \Swift and \hete. All the quantities are in c.g.s units.\\
$^a$ $T_{90}$ as measured from \Swift observations in the $15-350$
keV energy band. This value is not strictly equivalent to the \BATSE
$T_{90}$, where the $20-2000$ keV energy band is used.\\
$^b$ Fluence [$\times 10^{-7} \rm erg/cm^2$] in the $15-350$ keV
energy band (except for 060801, 061201 \& 061217 - $15-150$ keV). The fluence in the
20-2000 keV energy band is given in parentheses for bursts that are
also detected by
\konus.\\
$^c$ Energy [$\times 10^{49}$ erg] in the source frame $15-350~
(20-2000)$ keV energy band.\\
$^d$ Peak isotropic equivalent luminosity [$\times 10^{50}$ erg/s] in the source frame $15-350~
(20-2000)$ keV energy band. This luminosity is measured over the source
frame time interval that is indicated in the brackets.\\
$^e$ The inverse beaming factor, $f_b=\theta_j^2/2$, assuming a
two sided 'top hat' jet with a half opening angle $\theta_j$.\\
$^f$ The beaming corrected $\gamma$-ray energy: $E_\gamma = f_b E_{\gamma,iso}$. \\
$^g$ The early X-ray flux [$\times 10^{-11} \rm erg/cm^2/s$] in the
$0.3-10$ keV energy band at $t \approx 100$ s (observer frame).\\
$^h$ Detection of afterglow in: XE - early ($t \sim 100$ s) X-ray,
XL - late ($t \sim 1$ d) X-ray, O - optical, R-radio.\\
$^*$ This burst was detected by \hete, so $T_{90}$, the fluence and
the energy are given in the $30-400$ keV energy band.\\
$^{**}$ $T_{90}$ when calculated by simulating the \BATSE algorithm.
The maximal beaming factor and the minimal energy in the parenthesis
are derived with the more conservative limit on the jet opening
angle of SHB 050724(see \S\ref{SEC: beaming}).\\
$^\dagger$ The galactic latitude ($0.1$) and the spectrum of this
burst are consistent with an origin of a Galactic SGR. However,
there is no known Galactic SGR at this location. \\
$^\ddag$ These bursts were observed while the manuscript was in the refereeing
process and are therefore not discussed in the text.\\
(?) Based on a low significance association ($\approx 90\%$
confidence).\\
\newline\newline References - 1. \cite{Gehrels05,Bloom06} 2.
\cite{Villasenor05,Fox05,Hjorth05} 3.
\cite{Barthelmy05,Berger05,Krimm05,Grupe06} 4.
\cite{Sato05,Prochaska05,Berger06a} 5. \cite{Parsons05} 6.
\cite{Markwardt05} 7. \cite{Barbier05} 8. \cite{Parola06} 9.
\cite{Cummings05,Soderberg06a,Burrows06} 10. \cite{Roming06} 11.
\cite{Bloom07,Sato06c} 12. \cite{Sato06_GCN5381} 13.
\cite{Markwardt06_GCN5882} 14. \cite{Parsons06_GCN5930}


\end{table}

\subsubsection{The late time afterglow}\label{SEC: obs late aft}

\begin{figure}[!t]
\includegraphics[width=11cm]{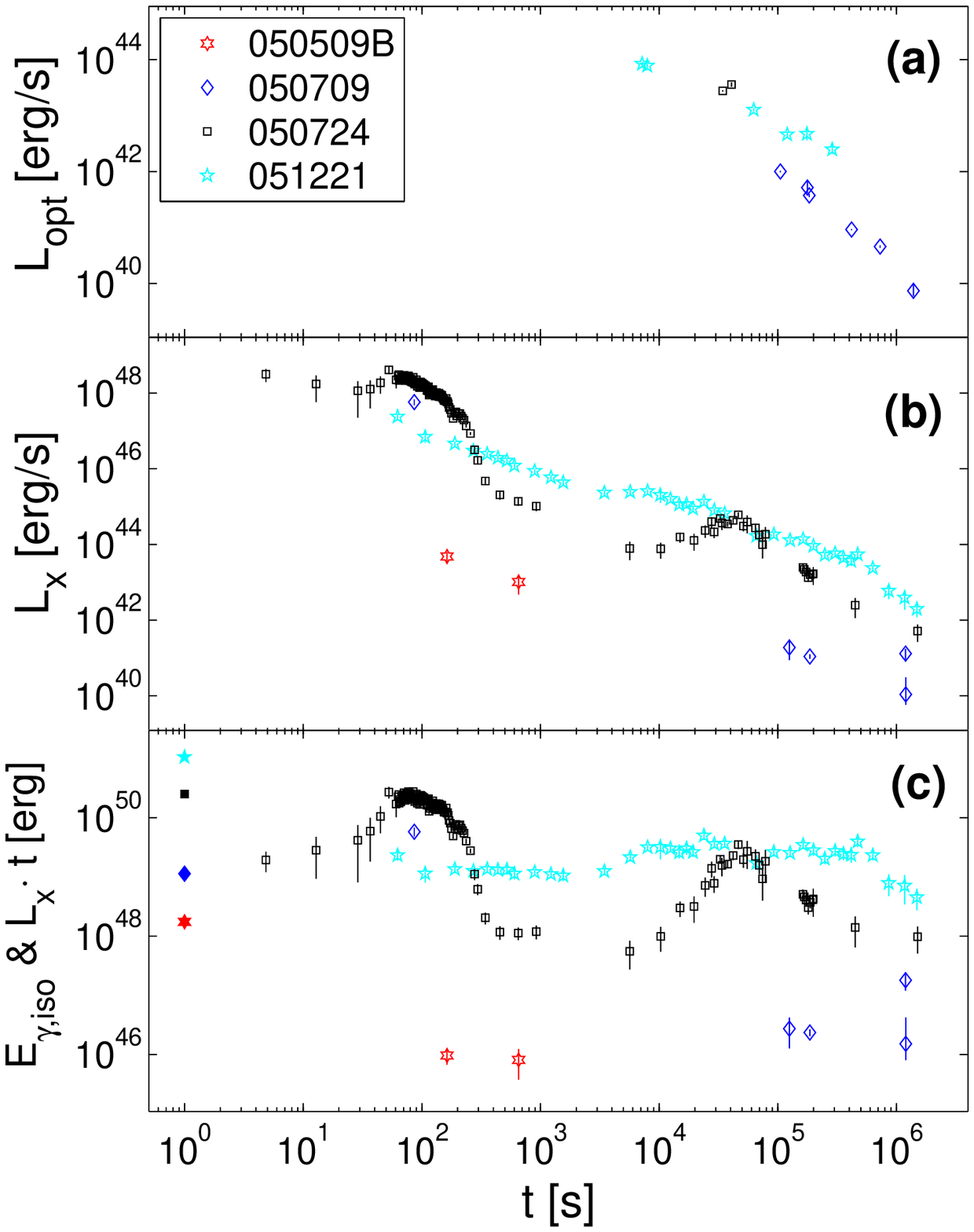}
\caption{\label{FIG: Aft_lc} The isotropic equivalent optical and
X-ray luminosity as a function of the source frame time. Plotted are
all SHB afterglows with a known redshift. {\bf (a)} Optical
luminosity (there was no optical detection of SHB 050509B). {\bf
(b)} X-ray luminosity. {\bf (c)} X-ray luminosity multiplied by the
time since the burst. This quantity illustrates the total energy
emitted in the X-ray afterglow at a given time. The isotropic
equivalent prompt emission fluence is plotted at $t=1$ s ({\it solid
symbols}) . References: SHB 050509B: \cite{Gehrels05}; SHB 050709:
\cite{Villasenor05,Fox05,Hjorth05,Watson06}; SHB 050724:
\cite{Barthelmy05,Grupe06,Berger05}; SHB 051221A:
\cite{Soderberg06a,Burrows06}}
\end{figure}

\begin{figure}[!t]
\includegraphics[width=13cm]{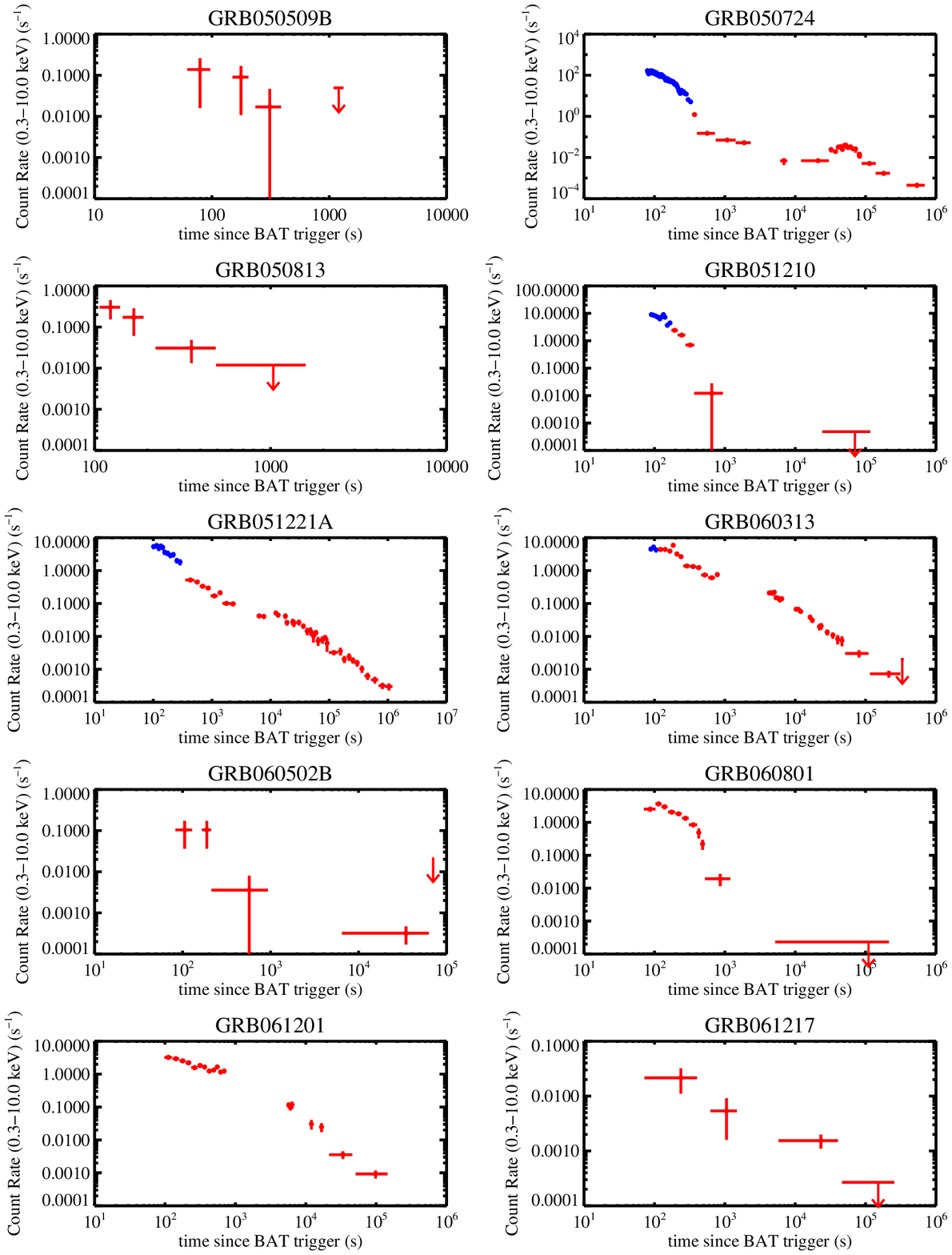}
\caption{\label{FIG: XRT_lightcurves} SHB X-ray afterglow light
curves observed by the \Swift X-ray telescope (XRT; Windowed Timing
mode in blue and Photon Counting mode in red). The light curves are
given in units of count/s. The conversion factor to energy flux
units is roughly $1$ count/s $\approx$ $5 \times 10^{-11}
\rm{erg/cm^2/s}$, where the exact conversion factor depends on the
X-ray spectrum of each afterglow. {\it Courtesy of  Judith Racusin
and David Burrows}.}
\end{figure}

Figures \ref{FIG: Aft_lc}a \& \ref{FIG: Aft_lc}b depict the X-ray
and optical afterglow isotropic equivalent luminosities as a
function of the source time of all SHBs with known redshift. The
observed X-ray flux as a function of observer time  is presented if
Fig. \ref{FIG: XRT_lightcurves}. The light curves roughly show a
power-law decay on scales of hours to days, with significant
additional variability in some cases. After several hours, the X-ray
and optical temporal power-law decay appear to be rather similar in
all observed SHBs with indices in the range of $\alpha \sim 1-1.5$
($F_{\nu} \propto t^{-\alpha}\nu^{-\beta}$). An exception is the
X-ray afterglow of SHB 051210 \citep{Parola06}, which decays as
$t^{-2.57}$ during at least the first several hours (the redshift of
this burst is unknown and therefore its afterglow is not in the
figure). The optical and X-ray afterglows show power-law spectra
with indices in the range $\beta \sim 0.5-1.5$. These general
properties of late SHB afterglows are similar to those observed in
long bursts.

Whenever a dense temporal sampling of the afterglow (mostly in
X-rays) exists, significant superposed variability is evident, with
diverse characteristics. The afterglow of SHB 051221 shows an X-ray
brightening after $\sim 2$ hr, with a similar power-law decay before
and after the brightening. A bright X-ray bump is observed,
superimposed on a power-law decay, between $3$ hr and $5$ days after
SHB 050724 occurred. In both of these cases no high-amplitude, rapid
variability ($\Delta F/F \gg 1$ and $\Delta t / t \ll 1$) is seen.
SHB 050709 does show rapid variability; after $\approx 15$ days, a
bright X-ray spike is observed ($\Delta F/F \approx 10$) that decays
within $2$ hr ($\Delta t / t \approx 0.01$). However, the detection
of this spike is based on a single \chandra observation of nine
photons \citep{Fox05}. This kind of variability, if confirmed, has
far-reaching implications on theoretical models, as discussed in
\S\ref{SEC: afterglow theory}.

\begin{figure}[!t]
\includegraphics[width=12cm]{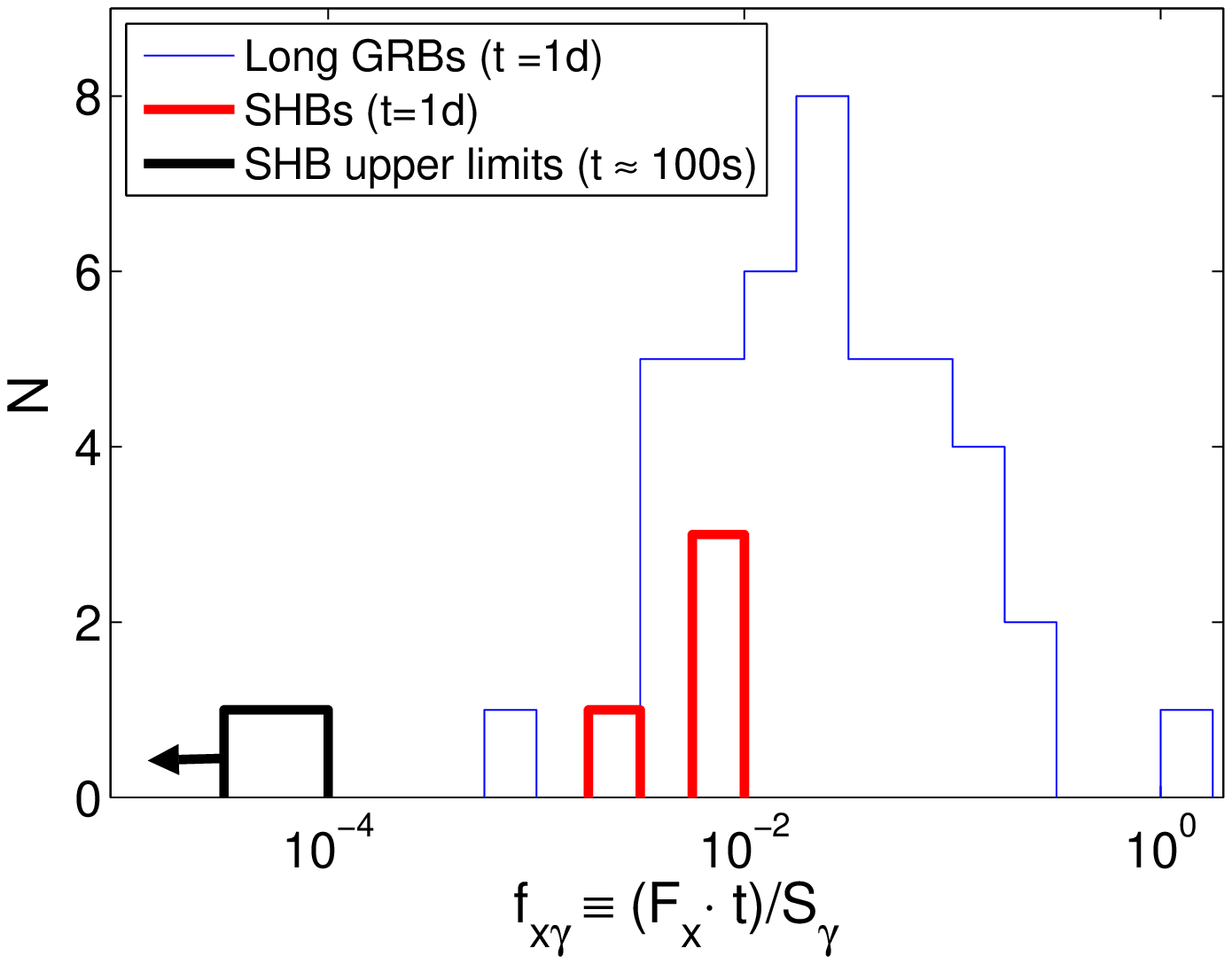}
\caption{\label{FIG: ratio_gx} A histogram of the ratio between the
X-ray energy flux at time $t$ multiplied by $t$ and the prompt
gamma-ray fluence. This is an estimate of the ratio between the
energy emitted in the late X-ray afterglow and in the prompt
emission. The ratio is given for \Swift long bursts ({\it thin
line}) and for SHBs with X-ray afterglow observed after $\sim 1$ day
({\it thick line}). The upper limit at $t \sim 100$ s for two \swift
SHBs without detected X-ray afterglow is marked with arrow ({\it
thick line + arrow}). The theoretical implications of this figure
are discussed in \S\ref{SEC: gamma eff}. Reference: \swift archive,
\href{http://swift.gsfc.nasa.gov/docs/swift/archive/grb\_table}{http://swift.gsfc.nasa.gov/docs/swift/archive/grb\_table}.
}
\end{figure}

Observed SHB afterglows are significantly fainter than those of long
GRBs. For example, after 1 day the SHB X-ray flux is fainter by more
than an order of magnitude ($\sim 5 \cdot 10^{-14} ~\rm erg/cm^2/s$
in the four detected SHB late afterglows, compared to $\sim 5 \cdot
10^{-13} ~\rm erg/cm^2/s$ in long GRBs\footnote{Values are taken
from the \swift archive, \rm
\href{http://swift.gsfc.nasa.gov/docs/swift/archive/grb\_table}{http://swift.gsfc.nasa.gov/docs/swift/archive/grb\_table}.
}). Moreover, early ($t \sim 100$ s) X-ray afterglow emission was
detected in every long GRB that was observed at early times by
\Swift, while three out of the 11 SHBs observed by the \Swift X-ray
telescope (XRT) within $\sim 100$ s, lack detectable X-ray emission.
If SHB afterglow flux is related to the flux of the prompt emission,
as observed in long GRBs, then SHB afterglows are expected to be
fainter (see \S\ref{SEC: prompt energy}). Fig. \ref{FIG: Aft_lc}c
shows that indeed this is part of the reason for the faintness of
SHB afterglows. The figure presents SHB X-ray light curves
multiplied by the time since the burst (illustrating the total
energy emitted in the X-ray afterglow), as well as the prompt
emission fluence (plotted at $t=1$ s). This figure clearly shows
that the X-ray afterglow is strongly correlated with the gamma-ray
fluence and that the energy emitted in the X-ray afterglow is
smaller than the gamma-ray energy by no more than a factor of $\sim
10$. However, Fig. \ref{FIG: ratio_gx} shows that in SHBs in which
the X-ray afterglow is not detected, the upper limits on the
afterglows indicate that in these cases the emission is very faint,
even compared to the gamma-ray fluence. See \S\ref{SEC: gamma eff}
for a discussion of the implications of this result.

We conclude that late SHB afterglows (hours-weeks) show roughly
similar temporal and spectral properties to those observed in long
GRBs, while being fainter on average. The afterglows of some SHBs
are considerably fainter than these of long GRBs, even after
correcting for the lower $\gamma$-ray energy output of these SHBs.
The small number of SHB afterglows prevents a more detailed
comparison. These observations are discussed in the context of the
external shock model for GRB afterglows in \S\ref{SEC: afterglow
theory}.

\subsubsection{Soft X-ray ``tails'' of the prompt emission and the early
afterglow}\label{SEC: X-ray tail}

Several SHBs show a bright X-ray tail that follows the short prompt
gamma-ray emission and lasts for $\sim 100$ s. This X-ray component
is most prominent in SHBs 050709 and 050724, where its energy is
comparable to (SHB 050724) or even larger by a factor of $\sim 3$
(SHB 050709) than the energy in the prompt gamma-rays
\citep{Barthelmy05,Villasenor05}. This early luminous X-ray emission
seems to be a common feature in SHBs, although there are cases in
which the X-ray tail is either much weaker or does not exist at all
(e.g. SHBs 050906 and 050925). An early hint to the existence of
this component was found by \cite{Lazzati01} that summed the
gamma-ray light curves of 76 \BATSE SHBs and detected an X-ray
signal starting $30$ s after the trigger and lasting for $\sim 100$
s \citep[see also ][]{Connaughton02,Frederiks04}. The longer time
scale and the softer spectrum of the X-ray tail, as well as the
temporal gap from the short-hard prompt emission, clearly
distinguishes it from the prompt emission. However, extrapolation of
the late afterglow back to early times suggests that the X-ray tail
is not the onset of the late afterglow. It is therefore unclear if
the physical origin of this component is related to the prompt
emission, to the afterglow or to a third process. In those cases
where the energy in the X-ray tail is large, it may be detected also
in gamma-rays, causing a SHB to look like a long GRB (see
\S\ref{SEC: SHB identification}). \cite{Norris06} find  eight \BATSE
GRBs with $T_{90} > 2$ s that show an initial short duration
emission that is followed by a long episode of softer emission. The
spectral lags of all these bursts are short, suggesting that these
are SHBs. The extended emission in some of these bursts is highly
variable and its energy is comparable to that of the initial spike.
Based on comparison between their sample and the results of
\cite{Lazzati01},  they suggest that the peak flux ratio between
the initial spike and the extended tail can vary by a factor
$10^{4}$ between different bursts.

\begin{figure}[t]
\includegraphics[width=12cm]{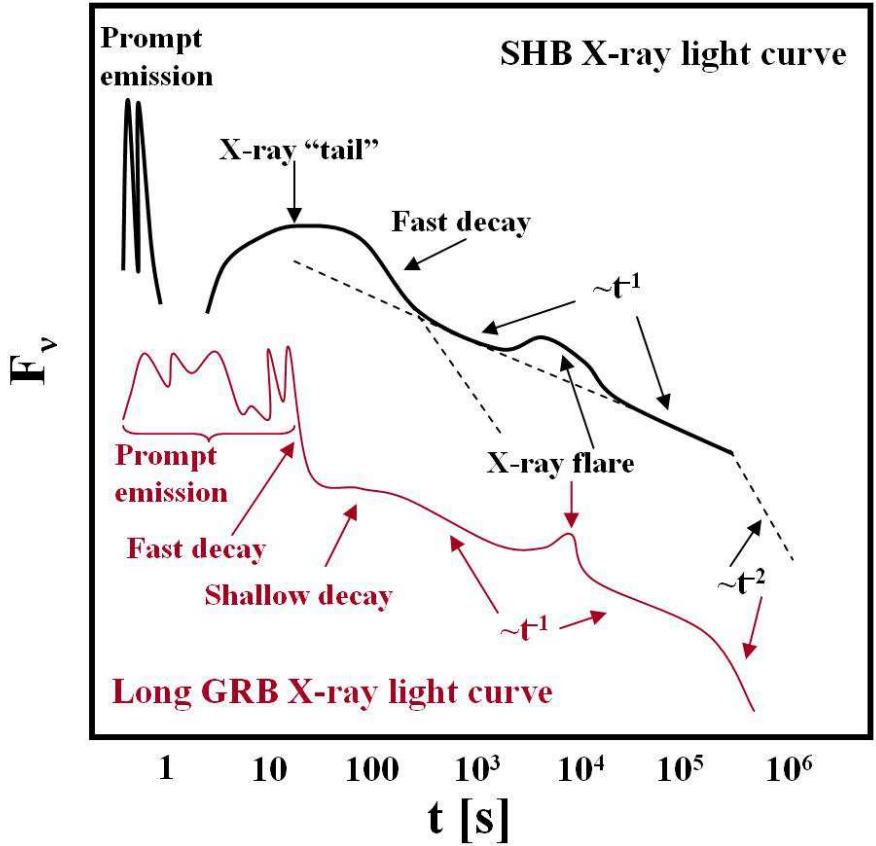}
\caption{\label{FIG: lc_cartoon} An illustration of the main
features observed in SHB X- and $\gamma$-ray light curves ({\it
thick black line}) and comparison to long GRBs ({\it thin red
line}). The dashed lines in the SHB light curve represent
alternative features that are observed in some bursts. The hard SHB
prompt emission is followed in some bursts by a soft X-ray tail. The
tail is well separated both in time and in spectrum from the prompt
emission. If a `tail' is observed, it is followed by a steep decay
that in most bursts turns into a shallower $\sim t^{-1}$ decay. Late
X-ray flares can be observed at any time. So far a single SHB
(051221A) has shown a late break in its light curve ($\sim t^{-2}$)
that is interpreted as a jet break. Prompt emission of long GRBs, in
contrast, is typically followed by a fast decay that flattens into a
very shallow decay, before it steepens again to a $\sim t^{-1}$
decay. X-ray flares and late light curve breaks are observed in long
GRBs as well. }
\end{figure}

In bursts that exhibit a bright X-ray tail, it is followed by a
steep temporal decay ($\alpha \gtrsim 2$; $\alpha \equiv -{\rm
d}log(F_\nu)/{\rm d}log(t)$) before the typical late afterglow
($\alpha \sim 1$) is observed. This fast decay is clearly seen in
SHBs 050709 and 050724. A fast decay phase is also observed in SHB
051210 \citep{Parola06}. Here the first X-ray observation is after
$80$ s and the flux is already falling rapidly, until a flare is
observed between $100-200$ s. The energy in this X-ray flare is
comparable to the prompt emission energy. After the flare the X-ray
flux decays as $t^{-2.6}$ and there is no evidence for a regular
late afterglow decay. In other bursts the regular X-ray decay
($\alpha \sim 1$) is observed starting at early times (e.g., SHB
050509B). Fig. \ref{FIG: lc_cartoon} illustrates the main features
observed in SHB X-ray afterglows.

The overall early X-ray emission of SHBs is different from the
complex early X-ray afterglows of long GRBs (e.g.,
\citealt{Nousek06}). Long GRBs do not show similar X-ray tails,
which are separated from the prompt emission\footnote{Long GRBs do
show X-ray flares but only rarely these have an energy that is
comparable to that of the prompt emission}, and while there is a
typical phase of fast X-ray decay in long GRBs, this decay marks the
end of the prompt emission and is not the end of a distinctive X-ray
component as it is in SHBs. The phase of shallow X-ray decay, which
is often observed in long GRBs, was not observed so far in SHBs.
Fig. \ref{FIG: lc_cartoon} presents a qualitative comparison between
the afterglows of long and short GRBs.

An early optical afterglow was detected so far only in one SHB
(060313; \citealt{Roming06}). In this burst the optical emission is
constant at first ($t =100-1000$ s) followed by a shallow decay,
superimposed with rapid variability. The X-ray afterglow of this
burst is uncorrelated with the optical emission.

\subsection{Observed redshift distribution}\label{SEC: z_obs}
It was suspected that SHBs  take place at cosmological distances as
soon as it was realized that they may be a distinct phenomenon from
long GRBs. Indirect evidence for their cosmological origin were the
nearly isotropic angular distribution in the sky
\citep{Briggs96,Balazs98} and the value of\footnote{ $V/V_{max}$ is
the ratio of the volume that is enclosed within the distance in
which the event is actually observed, and the one which is enclosed
within the maximal distance to which the same event would still be
detectable, assuming an Euclidian space. Under the assumption that
sources are distributed uniformly in distance, $\langle
V/V_{max}\rangle = 0.5$ in Euclidian space. A smaller value,
therefore, suggests sources at cosmological distances where the
smaller the value of $\langle V/V_{max}\rangle$, the larger is the
typical distance to the sources.} $\langle V/V_{max}\rangle ~<~0.5$
\citep{Katz96,Schmidt01}. These indicators also suggested that
observed SHBs are closer than observed long GRBs. \cite{Guetta05}
find that for SHBs $\langle V/V_{max}\rangle = 0.39 \pm 0.02$, which
is significantly larger than the value $\langle V/V_{max}\rangle =
0.29 \pm 0.01$ they find for long GRBs. \cite{Magliocchetti03}
calculate the two-point angular correlation function for \BATSE SHBs
and find a $\sim 2\sigma$ deviation from isotropy on angular scales
$\sim 2^\circ-4^\circ$ (a similar deviation is not found for long
GRBs). They suggest that this deviation is induced by nearby
large-scale structure through a low-redshift population of SHBs.

The first limits on the distance to individual SHBs was obtained by
examination of Interplanetary Network (IPN) localizations with small
error boxes\footnote{Earlier examination of IPN error boxes was
carried out by \cite{Schaefer98}, without deriving constraints on
the bursts' distances. See also \cite{Schaefer06}.}
\citep{Nakar06a}. The lack of bright galaxies inside the small IPN
error boxes of six bright SHBs implied that the distances to these
bursts exceed $100$ Mpc (this limit was derived under the assumption
that the SHB rate follows UV, blue or red light). The limits on the
distances translate to energy output lower limits of order $10^{49}$
erg, implying that these bursts can be detected by \BATSE at
distances greater than Gpc. This result provided another strong
indication that SHBs constitute a bone fide cosmological population.

The final confirmation of the cosmological origin of SHBs came with
the secure identification of their host galaxies, following
sub-arcsecond localizations of afterglows detected by \Swift and
\hete. The measured redshifts of the host galaxies are listed in
Table 2. 
All together there are 3 SHBs (050709, 050724 \& 051221) with
sub-arcsecond localizations that fall on top of galaxies which are
considered to be ``securely" identified as the hosts. The afterglow
of SHB 050509B  was detected only by the \Swift X-ray telescope
(XRT) and therefore was localized only up to within several
arcseconds. The error circle is located in the outskirts of a giant
elliptical. The confidence that this is indeed the host galaxy is
estimated to be $3-4\sigma$ \citep{Gehrels05,Bloom06}. The XRT error
circle of SHB 060502B is $\approx 20''$ from a bright early type
galaxy at z=0.287. \cite{Bloom07} suggest that this galaxy is the
host of SHB 060502B and they estimate the confidence of this
association to be $\approx 90\%$. Finally, the afterglow of SHB
050813 was also detected only by the XRT and its location falls on
the outskirts of an early type galaxy that is part of a
group/cluster at $z=0.72$ \citep{Prochaska05} and on top of another,
fainter, early type galaxy that is part of an apparent  cluster at
$z \approx 1.8$ \citep{Berger06a}. Therefore, this burst is most
likely at redshift 0.72 or 1.8 and its host is an early type galaxy
in a cluster or a group.

\begin{figure}[t]
\includegraphics[width=12cm]{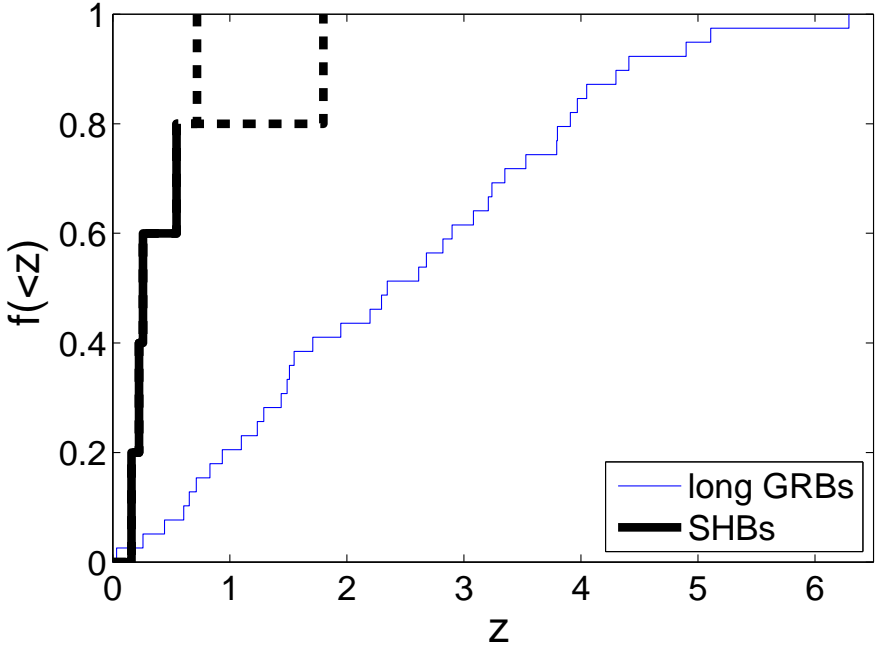}
\caption{\label{FIG: z_dist} The cumulative distribution of observed
SHB redshifts (including \hete SHB 050709 and not including the low
significance redshift association of SHB 060502B) [{\it thick line}]
compared to the of cumulative distribution of \Swift long GRBs with
known redshifts [{\it thin line}]. The dashed thick lines
corresponds to the two possible values of the redshift of SHB 050813
(0.72 or 1.8).}
\end{figure}

The cumulative distribution of the observed SHB redshifts (including
the one detected by \hete) is compared in figure \ref{FIG: z_dist}
to the distribution of \Swift long GRBs. The most striking result of
this comparison is that SHBs are detected at much lower redshifts
than typical long GRBs - while the median redshift of SHBs is $z
\approx 0.25$, the median redshift of  \Swift long GRBs is $z
\approx 2.5$. Note that some of the observed SHBs are bright enough
to be detected out to a much higher redshift. For example the prompt
emission as well as the afterglow of SHB 051221 would be detected by
\Swift also at a redshift of $z \approx 2$. Nevertheless, caution is
required when using the observed redshift distribution to draw
quantitative conclusions. The redshifts of more than half of the
\Swift SHBs are unknown and thus the observed distribution is likely
to be biased by selection effects that are needed to be quantified
before the entire sample can be used.

\cite{Gal-Yam05} study the four best-localized IPN SHBs ($<10 ~\rm
arcmin^2$), searching for significant luminosity overdensities
inside the error boxes. They identify two putative host/cluster
associations with SHBs. SHB 790613 is found to be significantly ($3
\sigma$ c.l.) associated with the rich galaxy cluster Abell 1892
($z=0.09$) while SHB 000607 is found to be associated (at the $2
\sigma$ c.l.) with a bright galaxy at $z=0.14$. They derive lower
limits on the redshifts of the two other bursts they examine of
$0.25[0.06]$ at $1[2]\sigma$ c.l.

Further clues about the redshift of SHBs are found by correlating
their locations with volume limited samples of galaxies and
clusters. \cite{Tanvir05} find a correlation between the locations
of \BATSE SHBs and the positions of galaxies in the local Universe,
suggesting that between 10 and 25 percent of \batse SHBs originate
at very low redshifts ($z < 0.025$). These SHBs
may be extra-galactic versions of the giant flares from  SGR 1806-20
(see \S\ref{SEC: SGR}) although they show a stronger correlation
with galaxies with Hubble type earlier than Sc.
So far, none of the well localized SHBs ($\approx
15$ bursts) are associated with such a nearby galaxy.
\cite{Ghirlanda06} find a positive $2\sigma$ angular
cross-correlation signal, on scales $<3^\circ$, between SHBs and
galaxy clusters at $z < 0.45$. The significance of the signal
increases when only clusters at $z<0.1$ are considered.

\cite{Berger06c} explore the XRT error circles of SHBs with detected
early X-ray afterglows but without identified hosts. They find that
the galaxies within these error circles are fainter than 23-26 mag,
and therefore are most likely at higher redshifts than the
identified SHB hosts. Based on comparison to the redshifts of
galaxies in the GOODS \citep{Cowie04,Wirth04} and the HUDF
\citep{Coe06} \cite{Berger06c} conclude that at least 1/4 of the
Swift SHBs are at $z>0.7$. This conclusion is valid if the
progenitors of these SHBs did not travel long distances from their
hosts so these are still within the XRT error circles.

We conclude that SHBs were shown to be genuine cosmological events
which are observed, on average, at significantly lower redshifts
than long GRBs. Indirect methods suggest that a non-negligible
fraction of SHBs originate even closer - in the local universe,
while non-detection of bright galaxies in some XRT error circles
suggest that a significant fraction of the SHBs originate at $z \sim
1$. These results still require validation by future observations.
The conclusions that can be drawn from the observed redshift
distribution on the {\it intrinsic} redshift distribution of SHBs
are discussed in \S\ref{SEC: redshift dist}.

\subsection{Host galaxies and cluster associations}\label{SEC: host obs}

Various properties of known SHB host galaxies are listed in Table 3
\citep[see][for recent SHB host galaxies brief reviews]{Berger06a,Bloom06c}. 
Host galaxies include both early- and late-type galaxies, as well as
field and cluster galaxies. This diversity is in striking contrast
to the properties of the host galaxies of long GRBs, which are
typically dwarf starburst galaxies and are always (so far) found in
the field. Moreover, the typical specific star-formation rate (SFR)
in a long GRB host is $\approx 10$ \Msunyr $(L/L_*)^{-1}$
\citep{Christensen04} while the specific SFR of SHB late-type hosts
is significantly smaller, $\lesssim 1$ \Msunyr $(L/L_*)^{-1}$. A
conservative comparison between the typical stellar ages in the
hosts of long and short GRBs rejects the null hypothesis that the
two populations are drawn from the same parent distribution at
$99.75\%$ confidence \citep{Gorosabel06}. These differences between
the host galaxies of short and long GRBs are among the main
foundations of the recent validation that long and short GRBs are
two different physical phenomena.

SHBs that take place in early-type galaxies are clearly associated
with an old stellar population. SHBs in late-type star-forming
galaxies, on the other hand, are not necessarily associated with
young stars. The hosts of both SHBs 050709 and 051221 show evidence
for a significant population of old ($\sim 1$ Gyr) stars
\citep{Covino06,Soderberg06a}. Moreover, an examination of the exact
location of SHB 050709 within its host shows that it is not
associated with the main star-forming regions in that galaxy
\citep{Fox05}.

\cite{Berger06b} explore the association of \swift SHBs with
clusters. They determine, using optical spectroscopy, that SHBs
050709, 050724 and 051221A are not associated with galaxy clusters.
Using X-ray observations they find that except for SHB 050509B there
is no evidence for other associations of  \swift SHBs with clusters
that are brighter than $3 \cdot 10^{-14}\rm~ erg/s/cm^2$, or with
mass $M
> 5 \cdot 10^{13} M_\odot$, assuming a typical redshift $z = 0.3$. These
results suggest that about $20\%$ of  SHBs are associated with
massive clusters, which is consistent, within the uncertainties,
with the fraction of stellar mass in such clusters
\citep{Fukugita98}.

Finally, it is important to remember that since host galaxies were
identified only for less than half of the \Swift bursts, this sample
likely suffers from selection effects. The distribution of the SHB
offsets from their host centers (Table 3) 
is also expected to be biased by similar selection effects. As we
discuss later (\S\ref{SEC: binary offset}), some progenitor models
predict brighter afterglows within late-type galaxies, making such
hosts easier to detect \citep{Belczynski06}. Additionally, the most
popular progenitor model (compact binary merger) predicts that some
of the events will take place far from their host galaxies ($\gtrsim
100$ kpc). This idea is also supported by the large possible offset
of SHB 050509B (see Table 3). 
In such cases the afterglow is expected to be very dim and even when
an afterglow is detected, the association with the parent host can
be unclear.

\begin{table}[t]

 {\bf \caption{Putative host galaxy properties}}\label{Table: host}
 \begin{tabular}{lc@{~~~~~~}lc@{~~~~~~}lc@{~~~~~~}lc@{~~~~~~}lc@{~~~~~~}lc@{~~~~~~}lc@{~~~~~~}lc@{~~~~~~}lc@{~~~~~~}l}

 \hline
 SHB & z & Type$^a$ & $L/L_* ~^b$ & SFR      & Offset & Offset  & Assoc. $^c$ &  SN$^d$ & Ref.\\
     &   &          &             & $\rm M_\odot/yr$& (kpc)  & ($r/r_e$)&        &   limit & \\
  \hline
  \hline
050509B& 0.225 & E(c)   &  3  & $<0.1$  & $44^{+12}_{-23}$ & $13^{+3}_{-7}$ & $3-4\sigma$ & -13 & 1 &\\
050709 & 0.16  & Sb/c   & 0.1 & 0.2  & 3.8              & 1.8          &  Secure     & -12 & 2 &\\
050724 & 0.258 & E/S0   &  1.5  & $<0.03$ & 2.6              & 0.4        &  Secure     & &3  &\\
$050813$ & 0.72 or & E/S0(c/g)&      &                  &             &             &  & & 4&\\
       & 1.8   & E/S0(c)&    &    &    &  &  &  &\\
051221A& 0.546 & Late   & 0.3  & 1.4   &  0.76           &  0.29        &  Secure     & -17.2 & 5 &\\
060502B& 0.287(?) &Early   &  1.6    & 0.6      &$73\pm 19$&        &  $\approx 90\%$ &          & 6&\\
\hline
IPN SHBs: &  &   &   &    &  &  &  &\\
\hline
$790613$& 0.09 &  E/S0(c)   &   &   &    &  &      $3\sigma$ & &7 &\\
000607& 0.14 &  Sb        &  1  & 0.3  &    &  & $2\sigma$ & &7 &\\

\end{tabular}
\newline
\newline
$^a$ Hubble type of the host galaxy. (c)[(g)]: The host belong to a
galaxy cluster [group].\\
$^b$ Early-type: the rest frame K-band host luminosity in units of
K-band $L_*$. Late-type: the rest frame B-band host luminosity in
units of B-band $L_*$.\\
$^c$ The confidence level of the association between the putative
host galaxy and the SHB. The association is considered to be secured
when a sub-arcsecond localization of the afterglow coincides with
the
putative host.\\
$^d$ The limit on the rest-frame absolute magnitude after $\approx
7-14$ days ($M_R$ for SHBs 050509B and 050709 and $M_V$ for SHB
051221) of a supernova associated with the SHB.\\
\newline References - 1. \cite{Gehrels05,Fox05,Bloom06} 2.
\cite{Fox05,Prochaska05,Covino06} 3. \cite{Berger05} 4.
\cite{Prochaska05,Berger06a} 5. \cite{Soderberg06a} 6.
\cite{Bloom07} 7. \cite{Gal-Yam05}

\end{table}

\subsection{Limits on a supernova component}\label{SEC: SN limit}

Despite the low redshifts of the first two SHB afterglows and the
extensive optical follow-up, no associated supernova was detected.
The limits are so deep (see Table 3) 
that an object that is $\sim 10$ times fainter than the faintest
observed supernova (of any type) would have been detected. These
observations are again in striking contrast to the observations of
long GRBs. So far a supernova was detected in almost any long GRB
with a redshift low enough to enable such a detection (see a
discussion of GRBs 060505 and 060614 in \S\ref{SEC: putative SHBs}).
The lack of associated supernovae is another line of evidence that
support the distinct nature of long and short GRBs.

\subsection{Observed rate}\label{SEC: obs rate}
The whole-sky detection rate of \BATSE was $\approx 170$ SHBs per
year \citep{Meegan97}. This rate can be translated to an observed
local rate ($z \ll 1$) using the observed \swift redshift
distribution. The first three \Swift SHBs with known redshift
constitutes a complete sample (see discussion in \S\ref{SEC:
sample}), unbiased by selection effects other than the \Swift
sensitivity, which is roughly comparable to that of \BATSE for SHBs
\citep{Band06}. Two out of these three bursts are within a distance
of $1$ Gpc, implying that at least $15\%$ of all \Swift SHBs are
within this distance (at the $2\sigma$ c.l.). Since the SHB redshift
distributions of \Swift and \BATSE bursts are not expected to be
very different (given the similar detector thresholds) the observed
local rate is ${\cal R}_{SHB,obs} \sim 10 ~\rm Gpc^3 ~yr^{-1}$
\citep{NakarGalYamFox06}. This rate is larger by about an order of
magnitude than estimated local rate of the long GRB
\citep{Guetta05b}. It is also higher than estimates of the SHB rate
that were done before the detection of SHB afterglows
\citep{Schmidt01,Ando04,Guetta05} by a factor of $10-100$.

\subsection{Identifying an SHB} \label{SEC: SHB identification}
An important observational challenge is to be able to distinguish
between SHBs and long GRBs on a burst-to-burst basis\footnote{Here,
the term SHB refers to any burst that is physically associated with
the population that constitutes the majority of \batse GRBs with
$T_{90} < 2$ s. Note that such a burst may not have $T_{90} < 2$
s.}. This was not so important prior to the detection of afterglows,
when SHBs were studied statistically, using mainly prompt emission
observations. However, now, when most recent scientific progress is
based on a small sample of bursts with afterglows, this topic
becomes crucial. It will probably remain an important issue even
when the sample size of SHBs with afterglows will increase
significantly if, as happened for long GRBs, much of the progress
will be based on a few bursts with unique properties (e.g., very low
redshifts).

Currently, no satisfactory tests can provide a definite burst
classification based on the prompt emission properties alone. The
test which is commonly used  ($T_{90} < 2$ s) is based on the \BATSE
duration distribution and, as evident from Fig. \ref{FIG: T90}, is
rough and somewhat arbitrary ($T_{90}<3$ s would serve just as
well). Clearly, there are long GRBs with $T_{90}<2$ s and SHBs that
last longer than $2$ s. Usage of this criterion is further
complicated by the possible contribution of the X-ray tail, which
follows some SHBs, to $T_{90}$ \citep{Norris06}. As demonstrated in
Fig. \ref{FIG: T90_HR}, considering the hardness ratio does not
significantly improve the ability to make a correct identification.
A third prompt emission property that can be used is the spectral
lag \citep[see \S\ref{SEC: prompt temporal}; ][]{Norris01}, which in
long GRBs is typically positive and, on average, longer than in
SHBs. Unfortunately, the lag distributions of the two populations
overlap as well. While a burst with a long ($\gtrsim 0.1$ s)
positive lag probably belongs to the long group, the opposite is not
true, since there are long GRBs with short lags. Here, the
lag-luminosity anti-correlation that was found for long GRBs may be
of help \citep{Norris00}. Long GRBs with short spectral lags ($\sim
10$ ms) tend to be the brightest ones ($\gtrsim 10^{52}$ erg/s).
Therefore, SHBs seem to occupy a separate region in lag-luminosity
space - short lags ($\lesssim 10$ ms) and lower luminosity
($\lesssim 10^{52}$ erg/s). This method, although promising, still
needs to be validated by increasing the sample size of SHBs with
known luminosity. Finally, since there are outliers to the
lag-luminosity relation of long GRBs, even this method cannot
provide absolute classification.

Tests that assign probability to the classification using the prompt
emission alone (e.g., as suggested by \citealt{Donaghy06}) are also
questionable. First, the sample that is currently used  to derive
such tests is the \BATSE sample, while the tests are applied to
bursts detected by other detectors (e.g., \Swift and \hete). Such an
application may be compromised since statistical properties of an
observed sample depend strongly on detector properties. Second, any
decomposition of the observed distribution to a sum of underlying
distributions \citep[e.g., ][]{Horvath02} depends on the assumed
shape of the underlying functions, which is not uniquely determined.
For example, the X-ray tails observed in some SHBs suggest that the
duration distribution of SHBs alone may be bimodal - one peak
corresponds to SHBs with faint X-ray tails and the other incudes
SHBs for which the X-ray tails are bright enough to affect $T_{90}$
(according to this speculation, the second longer peak is not
observed because long GRBs dominate the observed distribution above
$\sim 2$ s).

Based on the small sample of SHBs with afterglows it seems that a
much better classification can be done based on  environmental
properties \citep{Donaghy06}, i.e., the host galaxy type, its
star-formation rate, the location of the burst within the host
(\S\ref{SEC: host obs}), and on the existence of an accompanying
supernova (\S\ref{SEC: SN limit}). This classification is also based on
physical grounds $-$ progenitors of long GRBs are associated with
massive stars, while SHBs occur preferentially (according to the
current small sample) in places where the probability to find
massive stars is small. Put differently, if a  burst with a duration
of $100$ s will be detected in an early type galaxy and without an
accompanying supernova, it is most likely related physically to SHBs
and not to long GRBs (see discussion about GRBs 060505 and 060614 in
\S\ref{SEC: putative SHBs}).

A more quantitative test can be carried out when the afterglow is
imaged by the {\it Hubble Space Telescope (HST)}. \cite{Fruchter06}
explored 42 \hst images of long GRB host galaxies and found that
afterglow locations within the hosts are more concentrated in bright
blue pixels than those of core-collapse supernovae. This
concentration is most likely a result of the very short lifetime of
the progenitors of long GRBs. A quantitative test to determine if a
burst is long can be done by analyzing its \hst image in the same
way \cite{Fruchter06} did, and comparing the result with the known
distribution of long GRB locations.

GRB 050416A demonstrates  the difficulty to use the prompt emission
properties to classify an event. It is a soft burst with a
border-line duration ($T_{90} =2.4$ s) and a negative spectral lag
\citep{Sakamoto06}. As it turns out, this specific burst took place
in the brightest blue \hst pixel of a star-forming galaxy ($\approx
4 \rm ~M_\odot/yr/(L/L_*)$) and its afterglow shows the probable
signature of a supernova component \citep{Soderberg06b}, implying
that it is most likely a long burst.

In this review I used the ``classical'' criterion, namely $T_{90} <
2$ s, when compiling Table 2 
and as the sample of SHBs with afterglows or significant afterglow
constraints. Most of the bursts in this sample are short enough,
$T_{90} \ll 1$ s, to confidently assume that they are bone fide
SHBs. The only two SHBs that have border-line durations and that
have significant effect on theoretical interpretation are SHBs
050724 and 051221A. The former is observed within an early-type
galaxy and therefore is not a long GRB. SHB 051221A has a short
spectral lag compared to long GRBs, even when its high luminosity is
considered \citep{Gehrels06}, supporting its classification as a
SHB. Nevertheless, the possibility that GRB 051221A is not a genuine
SHB cannot be excluded.

\subsection{Additional putative SHBs}\label{SEC: putative SHBs}
Given the difficulty to identify SHBs, a growing number of GRBs that
do not pass the rough cut of $T_{90} < 2$ s were suggested, for
various reasons, to physically belong to the SHB population. Table 4 
lists all such \swift GRBs as well as \hete GRB 060121.

\begin{table}[t]
{\bf \caption{GRBs with $T_{90} \geqslant 2$ s that were proposed to be SHBs}}
 \begin{tabular}{lc@{~~~~~~}lc@{~~~~~~}lc@{~~~~~~}lc@{~~~~~~}lc@{~~~~~~}l}
 \hline
 SHB&$T_{90}^a$& z & $E_{\gamma,iso}$  &  Ref.\\
    & [s]    &   &  $10^{49}$ erg    &\\
  \hline
  \hline
050911$^\dagger$ & 16  &0.165(?)&2(?)&\cite{Tueller05_GCN3964,Berger06b}\\
051114A& 2.2 &        &    &\cite{Sakamoto05_GCN4275}\\
051211A& 4.2 &        &    &\cite{Kawai05_GCN4359}\\
051227 & 8   &        &    &\cite{Hullinger05_GCN4400}\\
060121$^{\dagger\dagger}$ & 2   &4.6[1.7]&$24[4]\times 10^3$&\cite{deUgartePostigo06,Levan06b}\\
060505 & 4   &0.089   & 1.2&\cite{Fynbo06,Ofek07}\\
060614 & 102 &0.125   &80  &\cite{Gal-Yam06,Fynbo06}\\
       &     &        &    &\cite{DellaValle06,Gehrels06}\\
061006& 130 &        &    &\cite{Krimm06_GCN5704}\\
061210& 85  &        &    &\cite{Palmer06_GCN5905}\\

\end{tabular}
\newline
\newline



A compilation of \swift GRBs that were appeared in the literature as
putative SHBs although their duration is $2$ s or longer. \hete GRB
060121 is included as well since its redshift is constrained.

$^a$ - $T_{90}$ as measured from \Swift observations in the $15-350$
keV energy band.\\
$^\dagger$ - The putative redshift is based on a possible
association
with the cluster EDCC 493 \citep{Berger06b}. \\
$^{\dagger\dagger}$ - A photometric redshift. Both $z\approx 4.6$
and $z \approx 1.7$ are consistent with the observations
\citep{deUgartePostigo06}.
\end{table}

The most interesting bursts in this sample are GRBs 060121, 060505
and 060614. Each one of these bursts affects the SHB theory if it is
related physically to this population. Unfortunately, it is unlikely
that an unambiguous classification will ever be achieved in any of the
cases. I mention some of the relevant theoretical implications of
these bursts being SHBs in the theoretical sections.

GRB 060121 is a border line burst \citep[$T_{90} = 1.97 \pm 0.06$ s;
][]{Donaghy06} with an optical afterglow \citep{Levan06b} and a
faint host with a photometric redshift probability with two peaks
$z=1.7 \pm 0.4$ and $z=4.6 \pm 0.5$ \citep{deUgartePostigo06}. The
source rest-frame duration is therefore less than $1$ s. This burst
also has short spectral lag, however, this cannot be used here to
support the suggestion that it is a SHB, since short lags are
expected in long bursts with similar peak flux ($\approx 3 \cdot
10^{53}$ erg/s at z=1.7).

GRB 060505 is another border-line GRB ($T_{90}=4\pm 1$s) which is
located on top of a highly star forming knot in the spiral arm of a
galaxy at $z=0.089$ \citep{Fynbo06,Ofek07}. This burst does not show
any evidence of supernova emission down to a very strict limit of
$M_B >-12.6$ and therefore was suggested as putative SHB.

The duration of GRB 060614 is $102$ s putting it, as long as
duration is concerned, safely in the long GRB group
\citep{Gehrels06}. However, \cite{Gal-Yam06} do not find any
supernova emission down to $M_B
>-12$ in \hst images of this z=0.125 burst \citep[see
also][]{Fynbo06,DellaValle06}. Moreover, the star-formation rate of
the host is low ($< 1 \rm ~M_\odot/yr/(L/L_*)$) and its
lag-luminosity puts it away from the long GRB population and
together with the other \swift SHBs. Additionally, its location in
the \hst image is on a faint pixel compare to the long GRB sample of
\cite{Fruchter06}. Finally, the light curve of this burst is
composed of an initial hard episode, which lasts $\sim 5$ s, that is
followed by a variable soft emission that lasts $\approx 100$ s.
Based on these observations \cite{Gal-Yam06} and \cite{Gehrels06}
suggest that this might be a long duration burst that is associated
physically with the SHB population.

\subsection{SHB interlopers}
It is possible that the SHB population defined according to its
\BATSE duration, contains sub-classes which are not physically
associated with the dominant SHB population, or with long GRBs.
Actually, the recent $\gamma$-ray giant flare from SGR $1806$-$20$
implies that this is indeed the case. Such possible  sub-classes are
discuss below.

\subsubsection{Extra-galactic SGR giant flares} \label{SEC: SGR}
Soft gamma-ray repeaters (SGRs) are compact sources of persistent
X-ray emission and repeating bursts of soft gamma-rays
\citep[see][for a review]{Woods06}, which are believed to be
highly-magnetized young neutron stars (known as magnetars; e.g.,
Duncan \& Thompson 1992; Paczy\'nski 1992). At infrequent intervals,
these sources emit extreme flares of high energy radiation,
releasing more than $10^{44}$ ergs in the form of gamma-rays alone.
These ``giant flares'' are characterized by a short ($\sim 0.1$
sec), hard ($\sim 300$ keV) and very intense spike that is followed
by a long ($\sim 300$ sec) pulsating soft ``tail'' (not to be
confused with the X-ray tail observed in some SHBs). The maximal
observed isotropic equivalent luminosity of such giant flares is
$\sim 10^{47}$ erg/s following the eruption of SGR~1806$-$20 on
December 2004 \citep[][for as short review of this event see
\citealt{Taylor06}]{Hurley05,Palmer05}.

Only the initial short and hard spike of an extragalactic giant
flare would have been detected by {\it BATSE} because its luminosity
is much higher than that of the pulsating tail, in which case it
would be classified as a SHB \citep{Duncan01,Eichler02}. With a
luminosity of $\sim 10^{47}$ erg/s such flares can be detected by
\BATSE up to $\approx 50$ Mpc, suggesting that a significant
fraction of the \BATSE SHB sample is comprised of similar
extra-galactic flares
\citep{Dar05b,Dar05a,Palmer05,Hurley05,Nakar06a}. This fraction,
however, cannot be very large given that none of the current \Swift
SHBs took place in a nearby galaxy, and therefore they are all much
more energetic than the giant flare from SGR 1806-20. Additional
constraints on the rate of SGR giant flares was obtained by
exploration of IPN error boxes \citep{Nakar06a, Ofek07b}, the lack
of a \BATSE SHB overdensity toward the virgo cluster
\citep{Palmer05,Popov06} and the small fraction of SHBs with a
blackbody spectrum \citep[][; the spectrum of the SGR giant flare is
consistent with  a blackcody]{Lazzati05}. All these methods yield an
upper limit on the fraction of giant flares out of the \BATSE SHB
sample, $f_{SGR}$, while the single observed event puts a lower
limit on this fraction. Having the most stringent and least
assumption dependent result, \cite{Ofek07b} finds that
$0.01<f_{SGR}<0.14$ at $95\%$ confidence.

Interestingly, \cite{Tanvir05} find evidence that a local population
(within $\approx 100$ Mpc) of SHBs may constitute $\approx 10\%$ of
the total SHB population. At first glance it looks as if these are
the expected extra-galactic SGR Giant flares. However,
\cite{Tanvir05} find an even stronger correlation when galaxies with
a Hubble type later than Sc are excluded from the sample.
SGRs are thought to be produced by core collapse of young stars and
to be associated with star forming regions, as supported by the
Galactic SGRs \citep[e.g.,][]{Gaensler01}. Therefore, the stronger
correlation in the case that the latest type galaxies are excluded
is surprising.
Following these observations \cite{Levan06}
discussed an alternative channel to produce SGRs, a merger of two
white dwarfs, which should take place also in early-type galaxies
\citep[see also ][]{Thompson95,King01}.

Distinguishing between an SGR giant flare and  a SHB is not trivial,
even if the location of the burst coincides with a nearby galaxy. An
example is GRB 051103, localized by the IPN
\citep{Golenetskii05_GCN4197} in the vicinity of M81 and M82
($\approx 3.6$ Mpc from Earth). The error box includes one of the
star forming arms of M81 \citep{Ofek06,Frederiks07}. Lacking
additional observations, it is currently unknown whether this was an
SGR giant flare in M81/82 or a background SHB.

\subsubsection{Very short GRBs}
\cite{Cline99,Cline01} and \citet{Cline05} suggest that  SHBs with
$T_{90} < 0.1$ s compose a sub-class that is distinct from the rest
of the SHB population. The main evidence that they present
are a value of $\langle V/V_{max}\rangle \approx 0.5$ and an
anisotropic distribution on the sky of the very short GRBs. Both
properties suggest a nearby population and are different than those
of longer SHBs. They also find that \konus very short GRBs show on
average harder spectrum above $3$ MeV than longer SHBs.
\cite{Cline01} and \citet{Cline05} suggest that these events may be
produced by primordial black hole evaporations or by nearby
extra-galactic SGR giant flares. Note that these observations
suggest that if indeed this is a subclass, very short GRBs take
place in the local universe. So far there are $5$ \swift SHBs with
$T_{90} < 0.1$ s, none of which are associated with a nearby galaxy.

\section{Relativistic outflows and the prompt emission}\label{SEC: Rel+propmpt}
The theory of long GRB prompt and afterglow emission was explored
extensively. The generally accepted picture is that a stellar object
undergoes a catastrophic event leading to the formation of a central
engine. The engine rapidly releases energy in a compact region
($\sim 10^6-10^7$ cm). This energy is deposited in the form of
radiation, heat and/or electromagnetic field and leads to the
acceleration of an ultra-relativistic outflow. At first, this flow
is optically thick and it ``carries'' the energy out, while cooling
adiabatically, to large distances ($\gtrsim 10^{13} ~\rm cm$), where
it becomes optically thin. From this point and on, dissipation of
the flow's energy, most likely by internal processes (i.e., without
interaction with the ambient medium), results in observed radiation.
The dissipation heats electrons to highly relativistic velocities
and these radiate in the presence of magnetic field and/or radiation
field (e.g., synchrotron and inverse compton radiation) that is
either advected by the flow from the source or generated by the
dissipation process. This emission is observed as the prompt
gamma-rays. At larger radii ($10^{16}-10^{18}$ cm) the energy that
remains in the relativistic flow is transferred into the
circum-burst medium, generating a decelerating blast wave. This
blast wave, which propagates into the external medium, is the source
of the afterglow emission.

This picture is based on the observed luminosity, time scales and
spectral and temporal evolution of long GRB prompt and afterglow
emission. Since most of these properties are common to long and
short GRBs, the same model is applied  also to SHBs and appears to
give an adequate explanation to their observations\footnote{This is
true only for the physical processes that produce the prompt and
afterglow emission. There are two separate branches of theoretical
research of short and long GRBs that explore the progenitors and the
processes the lead to the formation of the central engine (see
\S\ref{SEC: progenitors}).}.

Numerous review articles describe GRB theory, focusing on the
general model outlines described above
\citep[e.g.,][]{Piran99,Piran05,Meszaros06}. I will review here only
the main theoretical aspects of this model while quantitatively
confronting it (in several cases for the first time) with SHB
observations.

\subsection{Relativistic effects}\label{SEC:Rel affects}
The ultra-relativistic velocity of the source of the prompt and
afterglow emission directly affects their observational properties.
I briefly describe here some of these effects, serving as a basis
for further discussion.

\subsubsection{Time scales}\label{SEC: Rel time scales}

\begin{figure}[!t]
\includegraphics[width=12cm]{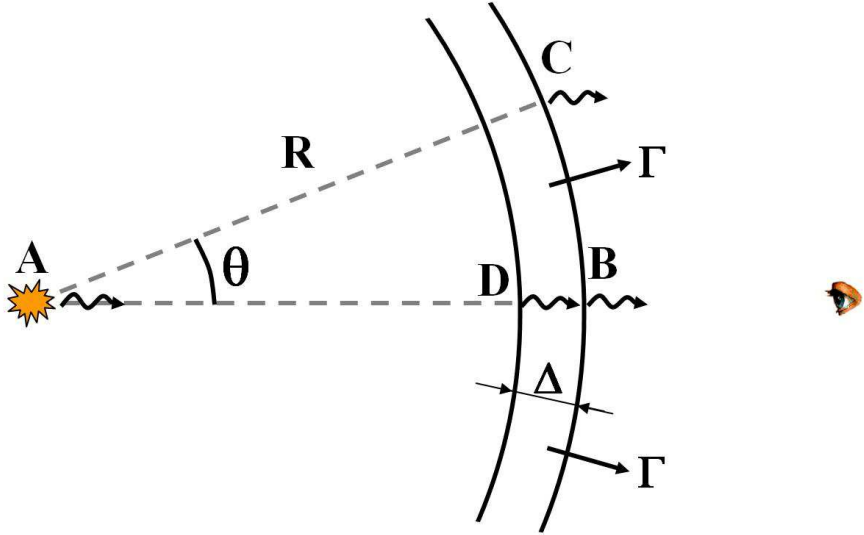}
\caption{\label{FIG: Rel Time Scales} Illustration of the emission
sites of four photons from a relativistic spherical shell with a
width $\Delta$ that expands with a Lorentz factor $\Gamma$. These
photons set three typical time scales (see the text). Photon `A' is
emitted from the origin at the time of the explosion. Photons
`B',`C' and `D' are all emitted when the shell front is at radius
$R$. `B' is emitted from the front of the relativistic shell on the
line-of-sight. 'C' is emitted from the front of the shell at an
angle $\theta$. `D' is emitted from the rear of the shell on the
line-of-sight.}
\end{figure}

Consider a spherical shell with a width $\Delta$ that is expanding
with a relativistic Lorentz factor $\Gamma \gg 1$ and emits photons
at radius $r=R$ from points `B', `C' \& `D' (Fig. \ref{FIG: Rel Time
Scales}). Photon `A' is emitted from the origin at the time of the
explosion (when the shell is ejected; Fig. \ref{FIG: Rel Time
Scales}). The different arrival times of these four photons
to the observer set three different time scales: \\
{\bf Line-of-sight time}: $t_{AB}$ is the time interval between two
photons emitted by the shell front along the line-of-sight one when
the shell is ejected, $r=0$, and the second when the shell is at
$r=R$:
\begin{equation}\label{EQ: t_los}
    t_{los} = \frac{R}{2c\Gamma^2}.
\end{equation}
This time scale is also comparable to the time between any two
photons that are emitted over the expansion time, i.e., the time
over which the radius doubles (which is also the adiabatic cooling
time). If the Lorentz factor of the emitting shell is not constant,
but decays as $\Gamma \propto R^{-g}$ (e.g., during SHB external
shock
$g=3/2$) then $t_{los} = R/(2[1+2g]c\Gamma^2)$.\\
{\bf Angular time}: $t_{BC}=R\theta^2/2c$ is the time interval
between two photons emitted at radius $R$, one along the line of
sight and the other at an angle $\theta \ll 1$. Light abberation
from relativistic sources implies that most of the photons that
arrive to the observer are from angles $\lesssim 1/\Gamma$. This
sets the angular time over which photons that are emitted from the
shell at radius $R$ arrive to the observer:
\begin{equation}\label{EQ: t_ang}
    t_{ang}=\frac{R}{2c\Gamma^2}.
\end{equation}
{\bf Shell width time}: The time scale that is set by the shell
width is not affected by the relativistic motion:
\begin{equation}\label{EQ: t_delta}
    t_{\Delta}=\Delta/c.
\end{equation}

Note that if we set $t=0$ to the arrival time of photon `A' then
photons from radius $R$ arrive to the observer between $t_{los}$ and
$\sim 2 t_{los}$  ($t_{los} \approx t_{ang}$). As I will discuss
later, this property makes it difficult for a single spherical shell
to produce high amplitude rapid variability in the light curve
(\S\ref{SEC: prompt internal VS. external}).

In the scenario discussed above the emitting material and the
emission front are propagating at the same Lorentz factor $\Gamma$.
This is not always the case. For example, according to the afterglow
theory the emission is generated by a blast wave (i.e., emission
front) that propagates at a lorentz factor that is larger by a
factor of $\sqrt{2}$ than the fluid that is crossing the shock
(i.e., emitting material). In such case the line-of-sight time (Eq.
\ref{EQ: t_los}) depends on the emission front Lorentz factor while
the angular time (Eq. \ref{EQ: t_ang}) depends on the Lorentz factor
of the emitting material.

\subsubsection{Causal connection and quasi-sphericity}\label{SEC: Rel qausi-spher}
Consider a spherical shell that is expanding at a Lorentz factor
$\Gamma \gg 1$, and a signal (e.g., a sound wave) that is
propagating on its surface at velocity $c\beta_s$ (as measured in
the local rest frame). Between the radius $R$ and $R+dR$ this signal
propagates an angular distance $d\theta_s = \beta_s dR/(\Gamma R)$.
Therefore, during the expansion time (i.e., the time over which the
radius doubles) the signal propagates $\theta_s \approx
\beta_s/\Gamma $ (logarithmic terms are neglected). Since $\beta_s
\leq 1$ the angular size of a causally connected patch of the shell
is $\sim 1/\Gamma$. Therefore, a relativistic shell is `frozen' over
angular scales that are larger than $1/\Gamma$. Relativistic light
abberation implies that most of the photons that are emitted from an
angle $\gtrsim 1/\Gamma$ with respect to the line-of-sight as
observed from the origin, do not reach the observer. The combination
of the two effects implies that any event occurring at angles that
are larger than $1/\Gamma$ cannot be observed directly and are also
causally disconnected from the region that can be seen. Thus, the
observer is completely ignorant of anything outside of the
$1/\Gamma$ angular region.

While this property makes it difficult to study the angular
structure of relativistic outflows, it makes the early stages of
GRBs considerably simpler to model. As long as the angular structure
of the flow  varies over angles that are larger than $1/\Gamma$,
each part of the outflow behaves as if it was a part of a spherical
flow with the local properties. Thus we can apply spherically
symmetric models to the prompt emission and to the initial phases of
the afterglow. Throughout the paper I will refer to this property as
quasi-sphericity.

\subsubsection{High latitude emission}\label{SEC: Rel high latitude}
Consider an infinitesimally thin spherical shell that expands
relativistically with a Lorentz factor $\Gamma=(1-\beta^2)^{-1/2}$
and radiates an arbitrarily short pulse at radius $R$. What would be
the observed shape of this pulse? The first photons that arrive to
the observer are emitted on the line-of-sight and therefore arrive
at $t_{los}$. At later times photons arrive from larger and larger
angles with respect to the line-of-sight, $\theta$, such that any
given observer time is mapped onto a given angle (`latitude'). The
relativistic blueshift decreases with increasing latitude (and time)
resulting in a decay of the energy flux. Additionally, due to the
decreasing blueshift, a given observer frequency corresponds to an
emitted rest frame  frequency that increases with latitude.
Therefore, the observed pulse shape at a given frequency window
depends on the emitted spectrum. In the case of a power-law spectrum
($F_\nu \propto \nu^{-\beta}$) the observed flux depends on the
Lorentz boost as $F_\nu \propto {\cal D}^{-(2+\beta)}$ where ${\cal
D} =\Gamma(1-\beta \cos \theta)$ is the inverse of the Doppler
factor. Using the relation between the observer time and the
emission angle, $t-t_{los} = 1 - R \cos (\theta) /c$, one obtains
${\cal D} \propto 1+(t-t_{los})/t_{ang}$ in the limit of $\Gamma \gg
1$, implying \citep{Kumar00,NakarPiran03}:
\begin{equation}\label{EQ: high latitude}
    F_{\nu} \propto
\left(1+\frac{t-t_{los}}{t_{ang}}\right)^{-(2+\beta)}.
\end{equation}
At late times ($t \gg t_{los}$ and $t \gg t_{ang}$) the light curve
decays as $t^{-(2+\beta)}$. This relativistic high latitude emission
is important since it predicts a prolonged observed flux, with
decreasing peak frequency, even in the case of an arbitrary short
burst of emission \citep[e.g., a `naked' afterglow, see \S\ref{SEC:
naked aft}; ][]{Kumar00}. It also constrains the amount of possible
variability from a spherically symmetric source \citep[][see
\S\ref{SEC: aft variability}]{NakarPiran03}.

\subsection{The Lorentz factor of the outflow}\label{SEC: LF}
Perhaps the most prominent feature of GRBs is that they are
(special) ultra-relativistic sources. This is a well-established
result, relying on several independent indications, some of which
are (almost) model independent. Here I discuss two methods, starting
with the opacity constraint, the most robust and model independent
method\footnote{In the case of long GRBs the most direct evidence of
a mildly relativistic motion during the afterglow phase are the
resolved radio images of the afterglow of GRB 030329
\citep{Taylor04,Taylor05}.}, where a lower limit on the Lorentz
factor is set by requiring that the source of the prompt emission is
optically thin. The second method provides a measurement of the
Lorentz factor $\Gamma$ during the early afterglow and can be
applied when the onset of the afterglow is observed. It depends on
the afterglow model and is valid only when the observed afterglow
emission is produced by an external shock (see \S\ref{SEC: afterglow
theory}).

These two methods (as well as several others) were used to carefully
analyze only long GRB observations and therefore the Lorentz factor
of SHBs was never properly constrained. Here I briefly present the
derivation of the theoretical constraints and then apply them to the
observations of SHBs. The main result is that, like long GRBs, SHBs
are ultra-relativistic. However, while observations of long GRBs
require $\Gamma \gtrsim 100-300$, for most of the SHBs $\Gamma
\gtrsim 10-50$ is consistent with the observations.

\subsubsection{Opacity constraints}
The prompt emission of GRBs (long and short) is non-thermal
(\S\ref{SEC: spectrum prompt}), implying that the source is
optically thin to the observed photons. On the other hand, if a
non-relativistic source is assumed, a calculation of the optical
depth to Thomson scattering, $\tau_T$, based on the enormous
observed luminosity of MeV $\gamma$-rays, results in $\tau_T \sim
10^{13}$ \citep{Schmidt78}. This discrepancy was known as the
compactness problem and at first was used to argue that GRBs are
Galactic. This conflict  was alleviated when it was realized that
the source of the emission may be moving at relativistic velocities
towards the observer \citep[e.g.,][]{Guilbert83,PiranShemi93}. The
most comprehensive calculation of the opacity limit on the Lorentz
factors of long bursts appears in \cite{Lithwick01}. Here I carry
out similar analysis, adapting it to the possibly different prompt
emission spectra of SHBs.

The non-thermal spectrum of GRBs implies that the source is
optically thin to Thomson scattering on $e^-e^+$
pairs\footnote{Opacity to $\gamma\gamma$ pair production provides
less stringent constraints on SHB Lorentz factors.}. An inevitable
source for such pairs is the annihilation of photons with rest frame
energy $\epsilon^,_{ph}
> m_ec^2$, where  $m_e$ is the electron mass. Therefore, the
Thomson optical depth for a given pulse during the prompt emission
phase is\footnote{Here I assume that he source is moving directly
toward the observer. If the source is moving at some angle with
respect to the line of sight then the optical depth increases and so
does the lower limit on $\Gamma$.}:
\begin{equation}\label{EQ: tau_T1}
    \tau_T \approx \frac{\sigma_T
    N_{ph} f(\epsilon^,_{ph}>m_ec^2)}{4\pi R^2},
\end{equation}
where $\sigma_T$ is the Thomson cross-section, $N_{ph}$ is the total
number of emitted photons within the pulse,
$f(\epsilon^,_{ph}>m_ec^2)$ is the fraction of photons that create
pairs and $R$ is the radius of the source. Relativistic motion of
the source has two effects. First, it reduces the energy of the
photons in the rest frame, thereby reducing $f$. Second, for a given
observed time of the pulse, $\delta t$, it increases the emission
radius as $R \sim c\delta t\Gamma^2$. The time scales and the
luminosities of individual pulses in long and short GRBs are similar
(see \S\ref{SEC: prompt temporal}), but the spectrum may be
different. While the spectrum of most long GRBs is best described by
a Band function \citep[smoothly broken
power-law;][]{Preece00,Ghirlanda02}, the best fits spectrum of most
SHBs is a low energy power-law and an exponential cut-off
\citep[][PLE; see \S\ref{SEC: spectrum prompt}]{Ghirlanda04}.  While
this spectral difference may be a result of observational selection
effects, PLE spectrum should be used when conservatively deriving
the lowest Lorentz factor that is consistent with all current
observations. The reason is that PLE spectrum is consistent with all
available data of most bursts and it is less constraining than
spectra with high energy power-law. Note also that Gev photons,
which support spectral high-energy power-law, were observed in
several long GRBs \citep[e.g.,][]{Schneid92,Sommer94,Hurley94} while
there is no report in the literature of a photon harder than $10$
MeV that was observed from a SHB. Hopefully, \glast observations
will unambiguously determine the high energy spectra of SHBs,
enabling more stringent limits on SHB Lorentz factors.

\begin{figure}[!t]
\includegraphics[width=13cm]{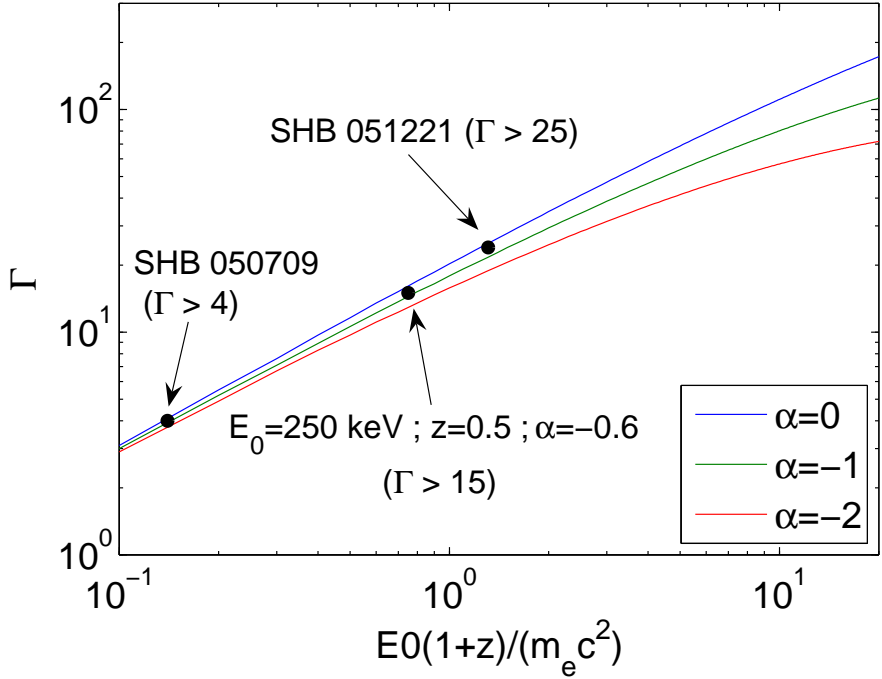}
\caption{\label{FIG: LF} The lower limit on the Lorentz factor of
the prompt emission source as a function of the rest frame spectral
typical energy $E_0(1+z)$ in units of $m_ec^2$. The limit is derived
by the opacity constraint (Eq. \ref{EQ: gamma constraint1}) using
three different low energy power-law slopes $\alpha$. The dots mark
three bursts, SHB 050709, SHB 051221A, and a typical SHB
\citep[according to][]{Ghirlanda04} at z=0.5. \citep[prompt emission
properties are taken from ][]{Villasenor05,Golenetskii05_GCN4394}
 }
\end{figure}

Using a power-law spectrum with index $\alpha$ and an exponential
cut-off at $E_0$ (Eq. \ref{EQ: prompt spectrum}), Eq. \ref{EQ:
tau_T1} becomes:
\begin{equation}\label{EQ: tau_T2}
    \tau_T \approx 10^{14} S_{\gamma,-7} d_{L,28}^2 \delta t_{-2}^{-2}
    \frac{m_ec^2}{E_0} \Gamma^{-(4-\alpha)} {\rm exp}\left[-\frac{\Gamma m_e c^2}{E_0 (1+z)}\right],
\end{equation}
where $S_\gamma$ is the observed gamma-ray fluence of the pulse,
$d_L$ is the luminosity distance to the burst (at redshift $z$) and
throughout the paper $N_x$ denotes $N/10^{x}$ in c.g.s units (except
for $E_0$ which is traditionally used in this context as the
spectral cutoff energy). Requiring $\tau_T<1$ results in the
following constraint on the Lorentz factor:
\begin{equation}\label{EQ: gamma constraint1}
 \frac{\Gamma m_e c^2}{E_0 (1+z)} + (4-\alpha) {\rm ln}(\Gamma)
 +{\rm ln}\left[\frac{E_0}{m_ec^2}\right] \gtrsim 30.
\end{equation}
The logarithmic dependence on $S_\gamma$, $\delta_t$ and $d_L$ is
neglected in Eq. \ref{EQ: gamma constraint1} (the range of the
observed values of SHB pulses may affect the value of Eq. \ref{EQ:
gamma constraint1} by less than $50\%$). Figure \ref{FIG: LF}
presents the lower limit on $\Gamma$ as a function of $E_0$ for
three values of $\alpha$. This lower limit is $\Gamma \gtrsim 15$
for the majority of the bursts analyzed by \citet{Ghirlanda04},
assuming that they are cosmological, while for SHBs 051221 and
050709 the opacity lower limits are $\Gamma > 25$ and $\Gamma > 4$
respectively. These lower limits are significantly lower than those
obtained for long GRBs \citep[$\Gamma \gtrsim 100$;
e.g.,][]{Lithwick01}. Note however that for both populations only
lower limits on the Lorentz factor are available and, while the
typical Lorentz factor of SHBs could be lower than that of long
GRBs, a precise comparison between the real values of $\Gamma$ is
impossible. Additionally, the smaller lower limits on SHB Lorentz
factors depend on  the best-fit function of their spectra which
might be affected by observational selection effects. Hopefully the
high energy spectra of SHBs will be securely determined by the
upcoming \glast mission.


\subsubsection{Constraints from the onset of the afterglow}
Within the framework of all the models discussed in this review, the
afterglow is produced by a blast wave that propagates into the
circum-burst medium, which is generated by the interaction of the
relativistic outflow with the ambient medium. Once most of the
energy is deposited into the external medium the blast wave assumes
a self-similar profile \citep{Blandford76} and the X-ray and optical
emissions are expected to decay as power-laws (roughly as $t^{-1}$,
where $t$ is the observer time since the trigger of the prompt
emission; see \S\ref{SEC: afterglow theory}). During the self
similar phase:
\begin{equation}\label{EQ: gamma constraint2}
\Gamma (t) \approx 40 \left(\frac{E_{k,iso,50}}{n_0} \right)^{1/8}
\left(\frac{t}{100~{\rm s} }\right)^{-3/8} (1+z)^{3/8},
\end{equation}
Where $E_{k,iso}$ is the isotropic equivalent kinetic energy that
remains in the relativistic outflow after the prompt emission phase
and $n$ is the density of the external medium, which is taken to be
constant (as expected in the environment of SHBs; see \S\ref{SEC:
progenitors}). Note that the dependence on both $E_{k,iso}$ and $n$
is weak. Plugging the earliest observed time at which the afterglow
shows a regular decay that fits the self-similar model into Eq.
\ref{EQ: gamma constraint2} provides a lower limit on the Lorentz
factor of the outflow.

Figure \ref{FIG: Aft_lc} shows three SHBs for which a regular
afterglow, decaying as $\sim t^{-1}$, is observed at $t \approx 100$
s. Therefore, assuming that this emission results from an external
shock, the Lorentz factor of the ejecta in these bursts is
constrained to be $\Gamma \gtrsim 40$ \citep[e.g.,][use this method
to constrain the Lorentz factor of SHB 050509B]{Lee05}. This lower
limit is similar to the one obtained by the prompt emission opacity.
Note that the Lorentz factor of SHB 050709 is not well constrained
by this method, since the first observations of ``regular''
afterglow decay in this burst is more than a day after the burst.


\subsection{The composition of the relativistic outflow}\label{SEC: composition}

One of the main open questions about the physics of GRBs (long and
short) as well as that of other relativistic astrophysical phenomena
(e.g., micro-quasars and active galactic nuclei) is how to launch
(and often collimate) a relativistic outflow. Most models of the
``engine'' that produces the flow are composed of different
combinations of rotation, accretion and magnetic fields, where the
energy source of the flow can be any one of the above. Whatever the
energy source is, the outflow starts its way when a vast amount of
energy is deposited in a relatively baryonic free compact
environment. The evolution of the flow depends on the composition of
the deposited energy (heat, magnetic field, etc.) and the
environment in which the outflow propagates. Almost all suggested
models are applicable for both long and short GRBs, where the main
difference is the duration over which the engine is active, which is
roughly comparable to the observed duration of the burst
(\S\ref{SEC: prompt internal VS. external}). Another difference is
that the engine of long GRBs is believed to operate at the center of
a collapsing star, while in all current SHB progenitor models the
engine is ``exposed'' (\S\ref{SEC: progenitors}). Thus, relativistic
ejecta from SHBs do not have to penetrate through several solar
masses of surrounding material, and can be launched directly.

The relativistic outflow composition is currently unknown. The main
candidates are baryonic plasma (protons, electrons and likely
neutrons) and magnetized plasma (at various ratios of magnetic to
particle energy). Currently no observations of either long or short
bursts  conclusively point towards one of these cases. Since there
are no studies of this topic that are specific for SHBs, and given
that this topic is well covered elsewhere
\citep{ZhangMeszaros04,Piran05,Piran05b,Meszaros06,Lyutikov06}, I
provide here only a brief overview.

\subsubsection{Baryonic flow}\label{SEC: baryonic acceleration}
In this case the engine energy is deposited at the end of
acceleration into an extremely small load of baryons, thereby
accelerating the baryons to ultra-relativistic velocities. The most
basic model of pure baryonic outflow suggests that the outflow is
accelerated by radiation pressure. It was realized early on that
releasing the observed luminosity in the form of pairs and/or
radiation in a compact region ($<0.001$ light sec) results in the
relativistic expansion of a pairs-radiation fireball
\citep{Goodman86,Paczynski86}. If there is a negligible load of
baryons in such a fireball, most of the energy escapes as
quasi-thermal radiation. However, if the proton load is high enough,
then the accompanying electrons increase the optical depth and the
fireball remains optically thick until almost all the the energy is
converted into bulk motion kinetic energy
\citep{Shemi90,Piran93,Meszaros93,Grimsrud98,Daigne02b,NakarPiranSari05,LiSari06}.
The minimal (isotropic equivalent) proton load that results in full
conversion of radiation to bulk motion energy corresponds to a
maximal Lorentz factor:
\begin{equation}\label{EQ: Gamma_max}
\Gamma_{max} \approx 1000 E_{50}^{1/4} R_{0,6}^{-1/4} T_{-1}^{-1/4},
\end{equation}
Where $R_0$ is the radius in which the energy is released
(comparable to the engine size) and $T$ is the duration over which
the energy, $E$, is isotropically deposited. Together with the lower
limit on the Lorentz factor ($\sim 30$; see \S\ref{SEC: LF}) the
allowed range of baryonic mass is ($E=\Gamma M_{ej} c^2$):
\begin{equation}\label{EQ: M_ej}
5\cdot 10^{-8} M_\odot E_{50}^{3/4} R_{0,6}^{1/4} T_{-1}^{1/4}
\lesssim M_{ej} \lesssim 2\cdot 10^{-6} M_\odot E_{50}.
\end{equation}
Note that while the mass lower limit (and the Lorentz factor upper
limit) is model dependent and valid only in this scenario, the upper
limit on the baryonic mass is generic since it is derived by the
Lorentz factor opacity constraint (\S\ref{SEC: LF}). This implies
that in any model, the region in which the ejecta is accelerated
must be almost completely clean of baryons.

This simple model does not discuss collimation of the flow, which
may be done by a hydrodynamic interaction with the environment or by
magnetic forces. In both cases the interaction may affect the final
Lorentz factor (as discussed below). Additionally, any variability
of the outflow in this quasi-spherical model is generated by varying
the deposition rate of the energy  and/or the baryonic loading. If
neutrons are present in the fireball as well, as expected if the
engine is driven by accretion onto a compact object
\citep[e.g.,][]{Beloborodov03b}, these will be dragged by nucleon
collisions along with the protons up to high Lorentz factor and may
affect the evolution of the fireball at late times
\citep[e.g.,][]{Derishev99,Derishev99b,Pruet02,Beloborodov03,Beloborodov03b,VlahakisPengKonigl03,Peng05,Rossi06}.
In this case the mass limit is roughly applicable to the total
baryonic mass.

An additional, purely hydrodynamic acceleration mechanism of a
non-magnetized plasma flow was discussed recently by \cite{Aloy06}
(see also \citealt{Aloy05}). The acceleration is powered by
rarefaction waves that propagate into a relativistic flow in the
presence of large velocities tangential to a discontinuity. Such
hydrodynamical conditions are expected to be generated by the
interaction of the flow with a thick accretion torus that feeds the
compact engine. Thus, once the flow achieves a moderate Lorentz
factor (e.g., by radiation pressure as discussed above) additional
acceleration is provided by this process. As a result, the maximal
Lorentz factor in this model is not set by transparency
considerations and can significantly exceed the upper limit in Eq.
\ref{EQ: M_ej}. Rarefaction waves also provide natural collimation
to the flow and \cite{Aloy05} find that for the observed SHB
parameters, the resulting flow is a narrow jet ($\sim 0.1 ~\rm rad$)
with a core Lorentz factor of $\sim 1000$ and rather sharp edges.

\subsubsection{Magnetized flow}
Many authors suggested models in which magnetic fields  play an
important role in the dynamics (acceleration and possible
collimation) as well as in the final energetic content of the
outflow \citep[e.g.,
][]{Usov92,Levinson93,Usov94,Thompson94,ThompsonChris06,Katz97,Meszaros97b,Kluzniak98b,Drenkhahn02,Vlahakis03,Vlahakis03b,Lyutikov02,Lyutikov03,Lyutikov04,Lyutikov04b,Lyutikov06}.
These models differ in the initial ratio of magnetic to particle
energies and in the large scale magnetic field configuration. During
the evolution of the flow the energy can be transferred between the
different components of the flow (bulk, heat and magnetic) and
different models predict different final flow compositions (in the
sense that there are different final energy ratios between these
components). Some models predict that by the time that the prompt
emission is generated, most, or comparable amount, of the energy is
in a baryonic component (e.g.,
\citealt{Katz97,Vlahakis03,Vlahakis03b}) while in other models the
magnetic field is the dominant component at all times (e.g.,
\citealt{Meszaros97b, Lyutikov03}).

Introducing magnetic fields enables models to achieve final Lorentz
factors that are significantly higher than the one obtained in pure
baryonic models (e.g., $\sim 10^6-10^7$ in \citealt{Meszaros97b})
and it also relaxes the constraints on the minimal baryonic load,
allowing for a pure magneto-leptonic flow
\citep[e.g.,][]{Thompson94}. Another example is
\cite{VlahakisPengKonigl03} that suggest a scenario in which the
interplay between the magnetic and baryonic energy making it
possible to increase the baryonic load in the form of neutrons that
decouple from the flow while its Lorentz factor is still moderate.

Regardless of the acceleration mechanism, the final composition of
the flow is the relevant initial condition for the prompt emission
phase. As we see below it mainly affects the dissipation process
that gives rise to the prompt emission.

\subsection{The prompt emission}\label{SEC: prompt theory}
The prompt emission is generated (or at least re-processed) at a
radius in which the optical depth of the relativistic flow is low.
Any heat that is generated in the flow while it is optically thick
is quickly lost to adiabatic expansion, and therefore the outflow
energy must be dissipated into internal heat near the radius in
which the emission takes place. Dissipation processes vary according
to the composition of the flow. In a baryonic or weakly-magnetized
plasma the dissipation may result from strong shocks, while in
Poynting-flux-dominated flows the heat is produced by dissipation of
the magnetic field (e.g., via reconnection).

\subsubsection{Internal or external dissipation?}\label{SEC: prompt internal VS. external}

The dissipation can be done either by processes that are internal to
the flow (e.g., internal shocks) or by interaction with an external
medium (e.g., external shock). In the case of a baryonic flow an
important difference between the two dissipation modes is their
implications for the central engine. In internal dissipation models
of baryonic outflow the duration of the burst, $T$, is related to
the duration of the engine activity. Since the light-crossing time
of the engine is shorter by many orders of magnitude than the burst
duration, a long-lived engine is required. In these models the
observed variability time scale, $\delta t$, reflects the
variability of the flow, induced either by the engine itself or by
interactions at the base of the flow during the acceleration. On the
other hand, the external shock model
\citep[e.g.,][]{Rees92,Katz94,Dermer99}, which is the most popular
of the external dissipation models of baryonic outflow, allows for
an explosive source, i.e., the energy can be released on an engine
dynamical time scale. If the source is explosive, the duration of
the burst in this model is determined by the radius in which the
interaction takes place ($T \sim R/c\Gamma^2$), while the
variability time scale may be determined by the size of density
fluctuations in the external medium.

In long GRBs the consensus is that if the outflow is baryonic the
prompt emission results from internal dissipation (see however
\citealt{DermerMitman04}; for a comprehensive discussion of the
topic see \citealt{Piran05}). The main argument in favor of internal
dissipation is the rapid, high amplitude variability observed during
the prompt emission of long GRBs. Namely, the prompt emission flux
varies by $\sim 100\%$ over time scales, $\delta t \ll T$.  The main
obstacle to highly variable emission from external shocks is the
angular smoothing \citep{Fenimore96} and detailed hydrodynamical
analysis shows that quasi-spherical external shocks cannot produce
the observed variability \citep{Sari97}. In  quasi-spherical
symmetry the angular time ($t_{ang}$) is always the longest observed
time scale and therefore it is comparable to the total duration of
the burst, thereby washing out any variability that arises during
the emission (for a definition of quasi-sphericity and relevant time
scales see \S\ref{SEC:Rel affects}). Thus, viable external shock
models must invoke spatial variability on small angular scales ($\ll
1/\Gamma$). The main attempt to do so suggests a clumpy external
medium \citep{Dermer99,DermerMitman04}. In the context of long GRBs
this model encounters difficulties with achieving the required
variability together with a reasonable efficiency
\citep{Sari97,Nakar02b}. furthermore, in this model the width of the
relativistic shell, $\Delta$, sets a lower limit on the variability
time scale ($\delta t \gtrsim \Delta/c$) implying $\Delta \ll
R/\Gamma^2$. Any variability during the acceleration of the flow
will result in $\Delta \sim R/\Gamma^2$ and therefore this model
requires the source of the flow to be extremely, some may say
unphysically, steady.

Many SHBs show variability that is similar to the one observed in
long bursts \citep[\S\ref{SEC: prompt temporal};][]{Nakar02} and
therefore the arguments discussed above in favor of internal
dissipation if the outflow is baryonic are applicable in the case of
variable SHBs as well. In the case of SHBs there are additional
arguments against external shock as the prompt emission dissipation
process. First, variable external shocks require the ambient medium
to have numerous dense clumps. Such an environment may be formed by
winds from massive stars, the progenitors of long GRBs. However, the
proposed progenitor models for SHBs predict a rather smooth
circum-burst medium. Second, the afterglow is most likely produced
by external shock. Therefore, if the prompt emission arises from an
external shock as well, then the afterglow is expected to join
smoothly with the prompt emission. However, the observed afterglows
show a clear distinction between the prompt emission, the following
X-ray tail and the late X-ray afterglow (Fig. \ref{FIG: Aft_lc}). To
produce all these phases by external shocks, even if the prompt
emission is not variable, requires a contrived external density
profile. Finally, as discussed in \S\ref{SEC: afterglow
observations}, there are SHBs with very faint X-ray afterglows.
These bursts lack bright external shock emission following the
prompt emission and therefore, for any reasonable external density
profile, the prompt emission should arise from a different process.

If the outflow is Poynting-flux-dominated then the prompt emission
energy is generated by dissipation of the outflow magnetic energy.
This dissipation may be affected by interaction with the external
medium \citep[e.g.,][]{ThompsonChris06}. The observed prompt
emission variability can be generated in a Poynting-flux-dominated
flow by emission process that is not isotropic in the rest frame of
the expanding outflow \citep[e.g.,][]{Lyutikov03,ThompsonChris06}.
In such case not all the $1/\Gamma$ angular region is observed but
only small regions within it where the the emission happens to point
towards the observer. This way the angular smoothing is avoided and
high variability may be archived also in external dissipation
process (see more details in \S\ref{SEC: prompt_magnetized}). Note
that in the two specific models of the prompt emission in a
Poynting-flux-dominated flow that are currently available
\citep{Lyutikov03,ThompsonChris06}, the duration of the burst
reflects that activity time of the engine.

To conclude, if the outflow is baryonic then variable SHBs (which
are the majority) are most likely produced by internal dissipation.
In a Poynting-flux-dominated flow the dissipation process can be
either internal of external. All viable models in both flow types
relate the observed duration of the burst to the engine activity
duration
\citep[e.g.,][]{Kobayashi97,Nakar02c,Janka06,Lyutikov06,ThompsonChris06}.
SHB that shows a smooth prompt emission light curve may arise from
an external shock in case that the afterglow joins smoothly with the
prompt emission (no such cases were observed so far).

\subsubsection{Internal shocks}

In a baryonic or weakly magnetized flow the suggested dissipation
process is internal shocks
\citep[e.g.,][]{Rees94,Narayan92,Paczynski94}, namely, shocks that
arise in a variable flow when plasma portions with different
velocities collide. In this model each pulse in the light curve
corresponds to a single episode of collision that takes place at a
radius $R \sim c\delta t \Gamma^2 = 3 \cdot 10^{12} {\rm ~cm}
~\delta t_{-2} \Gamma_2^{2}$. The total duration of the burst
corresponds to the total width of the outflow, and is therefore not
related to $R$. At these radii the plasma density is low enough so
the flow is optically thin to Thomson scattering, as implied by the
observations. The low plasma density also implies that the mean free
path for a binary collision between plasma particles (e.g., protons)
is much larger than $R$ and therefore these shocks are collisionless
and are mediated by collective electromagnetic plasma instabilities
(\S\ref{SEC: collisionless_shocks}). Collisionless shocks are
believed to efficiently accelerate particles to relativistic
velocities (\S\ref{SEC: collisionless_shocks}) and the propagation
of the accelerated relativistic electrons in the plasma magnetic
field is thought to be the source (vie e.g., synchrotron process) of
the observed prompt emission.

The main advantage of internal shocks is that they are expected to
arise naturally in the baryonic outflow and that they can easily
reproduce the observed variable light curve
\citep[e.g.,][]{Kobayashi97}. In fact the observed variability
reflects the variable activity of the engine\footnote{If the flow is
accelerated by radiation pressure alone than the prompt emission
light curve directly reflects the energy deposition rate and/or
baryonic load modulation at the base of the flow
\citep{Kobayashi97,Nakar02c}. If the flow is accelerated
continuously also at large radii (e.g., by rarefaction waves as
discussed in \S\ref{SEC: baryonic acceleration}) then it reflects
the structure of the flow at the end of the acceleration phase
\citep{Janka06,Aloy06}.} and the duration of the burst is comparable
to, or at most larger by an order of magnitude than \citep{Janka06},
the duration of engine activity. Additionally, the simplest
interpretation of the emission as synchrotron radiation roughly
predicts the correct observed frequency. The relative velocities
between different plasma portions are mildly relativistic and
therefore internal shocks heat the protons to mildly relativistic
temperatures. Assuming that the shocks couple the electron and
magnetic energies to the proton energy, the electrons have typical
Lorentz factor $10^2-10^3$, and gyrate in a $\sim 10^6$ G magnetic
field (in the plasma rest frame) producing $\sim 100$ keV
synchrotron emission (in the observer frame)\footnote{Note however
that this model predicts larger variance in the observed typical
frequency than suggested by \BATSE observations.}.

The main disadvantage of the internal shocks model is its
efficiency. Internal shocks are expected to radiate only a small
fraction of the total flow energy, $\sim 1\%$
\citep[e.g.,][]{Kobayashi97,Daigne98,Kumar99}. Being mildly
relativistic, internal shocks convert only a small fraction of the
bulk kinetic energy into internal energy, and only a fraction of
this internal energy is carried by radiating electrons. As a result,
most of the flow energy is not radiated and remains to be dissipated
by an external shock, resulting in a bright afterglow. In long GRBs
the afterglow is not as bright as expected, suggesting that
typically as much as $10-50\%$ of the relativistic flow energy is
emitted in the form of gamma-rays and in some cases the gamma-ray
efficiency may even be higher
\citep{Freedman01,Lloyd04,Granot06,Fan06a}. As I show in \S\ref{SEC:
gamma eff} the efficiency of the prompt emission in SHBs is most
likely high as well \citep[see also][]{Bloom06} . Therefore,
efficiency is potentially a problem of this model in SHBs as well.
The same extensions of the internal shocks model that were suggested
in order to resolve this issue in the context of long GRBs are
applicable to SHBs as well, e.g., very large amplitude Lorentz
factor variability \citep{Beloborodov00,Guetta01} and repeated shock
crossings \citep{Kobayashi01}.

The interpretation of the prompt emission as optically thin
synchrotron radiation predicts a spectrum that is softer than
$dN/d\nu \propto \nu^{-2/3}$ \citep{Rybicki79}. This prediction is
in conflict with the observed spectrum  of most SHBs
\citep[\S\ref{SEC: spectrum prompt};][]{Ghirlanda04}. This problem,
raised by \cite{Preece98} in the context of long GRBs, is more
severe in short GRBs where the low energy spectrum is harder. Again,
different models that were put forward to explain the hard spectrum
of long GRBs are also applicable here. Some of these models are:
random magnetic fields on very short length-scales, leading to
jitter radiation instead of synchrotron \citep{Medvedev00}, effects
of synchrotron self absorption or anisotropic electron pitching
angles \citep{Lloyd00b,Granot00}, Upscatter of optical self-absorbed
photons by inverse Compton \citep{PanaitescuMeszaros00} and
Photospheric effects \citep{Meszaros00}.

On the high energy end, most SHB spectra are consistent with having
an exponential cut-off, while long GRBs show a power-law
\citep[\S\ref{SEC: spectrum prompt};][]{Ghirlanda04}. If this
cut-off is intrinsic, and not an observational selection effect,
then it most likely reflects the energy distribution of radiating
electrons, suggesting that in SHBs, unlike in long GRBs, the
electrons are not accelerated to very high energies during the
prompt emission phase. In fact, the observed spectrum in most SHBs
can be produced by a mildly relativistic shocks that only couple the
electrons and the protons, without further accelerating electrons to
higher (power-law) energies as required in long GRBs. In this case
there are no electrons with Lorentz factors above $10^2-10^3$ and an
exponential cut-off in the spectrum is expected at $\sim 100$ keV. I
speculate here that such difference may arise from different
magnetization level of the outflow, where collisionless shocks
during the prompt emission of SHBs are magnetized while those in
long GRBs are not. Numerical simulations \citep{Spitkovsky05}
suggest that first order Fermi acceleration of particles may not
take place in magnetized shocks while it has been recently suggested
\citep{Lyubarsky06} that in such magnetized shocks the electrons can
reach equipartition with the protons (with a quasi-thermal energy
distribution) through the the synchrotron maser instability. The
reason that strong magnetic fields at the shock front may suppress
electron acceleration is that they prevent downstream thermal
electrons from crossing the shock back into the upstream, thereby
suppressing the Fermi process. On the other hand, theoretical
considerations \citep{Milosavljevic06} and numerical simulations
\citep{Spitkovsky05}  suggest that in unmagnetized shocks some of
the thermalized downstream electrons can cross the shock back into
the upstream and thus start the Fermi acceleration cycles (see
\S\ref{SEC: collisionless_shocks}).

\subsubsection{Prompt emission from a magnetized flow}\label{SEC: prompt_magnetized}
If the outflow energy is carried mostly by Poynting flux up to the
point where the interaction with the external medium starts, the
dissipation of the bulk energy into internal energy cannot occur
through internal shocks\footnote{Note that outflow models that are
dominated by magnetic energy at small radii but convert their energy
into particle kinetic energy during the acceleration, can use
internal shocks to produce the prompt emission.}. In this case the
energy source of the  radiation is most likely magnetic dissipation
processes (e.g., reconnection). Currently there are two models of
magnetized flows treating the prompt emission in some detail.

\cite{Lyutikov03} suggest that electromagnetic current-driven
instabilities dissipate magnetic energy into heat and high energy
particles \citep[see also ][]{Lyutikov06}. The propagation of these
high energy particles within a strong magnetic field, naturally
present in this model, is the source of the observed prompt
$\gamma$-ray emission. In this model the dissipation takes place
close to the deceleration radius ($\sim 10^{15}-10^{16}$ cm) while
the outflow Lorentz factor is $\Gamma \sim 100-1000$. Therefore, the
observed variability time scale implies that the emission is not
spherical (the observed variability, $\delta t$, is much shorter
than $t_{ang}$ in this case). \cite{Lyutikov03} suggest that the
reconnection of the magnetic field produces, additionally to high
energy particles, also relativistic turbulence with a Lorentz factor
$\gamma_t \gg 1$, as measured in the rest frame of the expanding
shell. This bulk relativistic motion of the emitting matter causes
the radiation to be beamed into an angel of $\approx 1/(\Gamma
\gamma_t) \ll 1/\Gamma$. This way, each observed pulse is coming
from a small angular region ($\ll 1/\Gamma$) on the expanding shell,
allowing $\delta t \ll t_{ang} = R/(2c\Gamma^2)$. In this model the
burst duration is greater or equal to $t_{ang}$, so that $\delta t
\ll T$.

\cite{ThompsonChris06} suggests a flow which is a combination of a
strong magnetic field and pair-radiation plasma. In this model the
radius in which the outflow becomes optically thin is determined by
the pair enrichment of the ambient medium through the interaction
with the flow radiation
\citep{Thompson00,Meszaros01,Beloborodov02,Kumar04}. The magnetic
dissipation is triggered by the interaction with the pair loaded
external medium and is regulated to take place around the radius
where the outflow becomes optically thin. The prompt emission in
this model is generated by inverse Compton scattering of the seed
photons that are carried in the flow by pairs which are accelerated
by the reconnecting magnetic field. Highly variable light curve is
obtained  by emission that is directed in the rest frame of the
relativistic outflow. The peak spectral energy of the prompt
emission (observed at $\sim 500$ keV) reflects the temperature of
these seed photons (Lorentz boosted to the observer frame).
Therefore, it is determined by the radius at which the flow starts
expanding freely, and by the Lorentz factor at this radius. In long
GRBs the observed spectrum corresponds to a Lorentz factor of a few
at a radius of $\sim 10^{10}$ cm which may be naturally explained as
the radius of the progenitor star. It is unclear at this point if
this model can be applied to SHBs as well, since the recent
observations indicate that in SHBs the free expansion should also
start at a large radius. For example, \cite{Thompson06} find that in
the case of SHB 050709 the peak spectral energy corresponds to a
flow that is still mildly relativistic and thermalized at $\sim
10^{10}$ cm. Currently it is unclear how the thermalization radius
can be so high in SHBs.

\subsection{High energy Cosmic-rays and neutrinos}
\subsubsection{High energy Cosmic-rays}
Cosmic-ray acceleration is typically discussed in the context of
long GRBs \citep[for a recent review see][]{Waxman04}.  Long GRBs
were suggested as the source of extra-galactic ultra-high energy
cosmic rays, with observed energies between $10^{18}-10^{20}$ eV
\citep[UHECRs;][]{Waxman95,Milgrom95,Vietri95,Wick04,DermerAtoyan06b}.
This idea is supported by the similar total flux of long GRB
gamma-rays and UHECR and by the small number of other candidate
UHECR sources. Cosmic rays are suggested to be accelerated in long
GRB internal and reverse shocks, where diffusive shock acceleration
(a.k.s first order Fermi acceleration; \S\ref{SEC:
collisionless_shocks}) may be able to accelerate particles to
$10^{20}$ eV. External shocks are not a promising UHECR acceleration
site via diffusive shock acceleration
\citep{Gallant99,Milosavljevic06c} and therefore long GRBs are
expected to be UHECR sources only if the outflow is baryonic
(\S\ref{SEC: composition}), unless particles are accelerated in the
external shock by an alternative mechanism \citep[e.g., second order
Fermi acceleration;][]{Dermer99,Dermer01,Dermer06}.

The three major factors that limit shock acceleration are time,
confinement and cooling. Assuming that the coherence length of the
magnetic field in the upstream (unshocked wind) is larger than
$R/\Gamma$, the criterion that the acceleration is faster than
adiabatic cooling is $R_L<R/\Gamma$ where $R_L$ is the Larmor radius
of accelerated particles in the upstream field and R [$\Gamma$] is
the radius [Lorentz factor] of the relativistic wind (note that the
shock itself is mildly relativistic in the wind rest frame). This
condition also guarantees that the accelerated particle is confined
and that it has sufficient time to be accelerated. Writing this
condition using the parameters of relativistic winds one obtains
\citep{Waxman95}:
\begin{equation}\label{EQ confinment}
    E_{p,max} \approx 10^{20} {\rm ~eV} \sqrt{\varepsilon_{B,-1}^{us}} \Gamma_{2}^{-1} L_{w,51}^{1/2},
\end{equation}
where $E_{p,max}$ is the maximal energy to which protons can be
accelerated and $L_w$ is the wind luminosity. $\varepsilon_B^{us}$
is the fraction of magnetic field energy density in the unshocked
wind\footnote{$\varepsilon_B^{us}$  is not to be confused with the
magnetic field density in the shocked downstream, $\varepsilon_B$,
which is the site of synchrotron emission.} out of the total wind
energy density (including rest mass energy) as measured in the wind
rest frame. The requirement that the acceleration is faster than
synchrotron cooling is satisfied for $R>10^{13} {\rm ~cm}~
E_{p,20}^3 \Gamma_{2}^{-2}$ \citep{Waxman95}. If SHB winds are
baryonic then both conditions can be satisfied during the internal
shocks and the reverse shock, given that $100 \lesssim \Gamma
\lesssim 1000$ and that the wind is mildly magnetized. Therefore,
SHBs may be able to accelerate $10^{20}$ eV cosmic rays just like
long GRBs. However, The total SHB gamma-ray energy flux is lower by
two orders of
magnitude than that of long GRBs and the observed UHECRs (Table 1). 
Therefore, while SHBs may be sources of high energy cosmic rays,
they are unlikely to be the dominant source of the observed UHECRs.

\subsubsection{High energy neutrinos}
High energy neutrinos ($\gtrsim$ TeV) are also usually discussed in
the context of long GRBs \citep[for reviews
see][]{Waxman00,MeszarosRazzaque06}. In long GRBs, four sites were
suggested as primary sources of high energy neutrinos: (i)
Acceleration of a neutron rich outflow \citep{Bahcall00,Derishev99}.
(ii) Internal shocks \citep{Waxman97,waxman99,Rachen98,Dermer03}
(iii) Reverse shocks \citep{WaxmanBahcall00}. (iv) In the course of
a long GRB jet drilling through the envelope of its massive-star
progenitor \citep{MeszarosWaxman01}.

The last site is  irrelevant for SHBs, as the progenitors are not
massive stars. All other processes may take place in SHBs if their
outflow is baryonic. If the outflow is Poyniting-flux-dominated then
large neutrino flux is unexpected unless neutrinos are efficiently
produced in the forward shock \citep{Dermer02,Li02}. This may be
possible in the less likely case that $\sim 10^{19}$ eV protons are
accelerated in the forward shock.

The energy of neutrinos produced during the acceleration of a
neutron rich outflow is $\sim 10$ GeV and therefore these are hard
to detect (due to neutrino detector sensitivity at this energy
range). Reverse shock neutrinos are produced by interaction of $\sim
10^{20}$ eV cosmic rays with UV photons. The UV flux in SHB reverse
shocks is significantly fainter compare to long GRBs (\S\ref{SEC:
early aft theory}) and therefore this process is inefficient in
SHBs. Internal shock neutrinos are produced by interaction of
$\gtrsim 10^{16}$ eV cosmic rays with the prompt gamma-rays. The
neutrinos are produced by the decay of pions that result from this
interaction. The efficiency of this process depends on the
efficiency of the cosmic ray acceleration, the observed gamma-ray
flux and the variability time scale, all of whom may be similar in
both phenomena. In long GRBs \cite{Waxman97} estimate that in
optimal conditions about 10\% of the burst energy is in the form of
$\sim 10^{15}$ eV neutrinos. Assuming a similar efficiencies for
long and short GRBs out of the total gamma-ray emission (Table 1)
implies that the neutrino flux from short GRBs is:
\begin{equation}\label{EQ Enu}
    \epsilon_\nu^2\Phi_\nu (\epsilon_\nu \approx 10^{15}\rm~eV) \sim 5 \cdot 10^{-11}
{\rm~GeV/cm^2/s/sr}.
\end{equation}


\section{The afterglow - Theory}\label{SEC: afterglow theory}
X-ray and optical emission observed  hours, days and weeks after
some SHBs resemble the afterglow observed in long GRBs. Any GRB
model that is not $100\%$ efficient in converting energy into
gamma-rays predicts an interaction of the remaining relativistic
ejecta with the external medium, which is expected to produce
radiation at longer wavelengths. Therefore, various models of the
prompt emission, that were developed for long GRBs and are
applicable to SHBs as well, also predict similar afterglow emission.
Indeed, the existence of SHB afterglows was predicted long before
these were observed \citep{Panaitescu01b,Nakar02}. Differences
between long and short GRB afterglows are expected mostly due to
different total energies, different properties of the ambient medium
and, possibly, different outflow geometries. I give here a brief
description of the standard external shock afterglow model and
emphasize the interpretation of observed SHB afterglows in this
context. I extend and adjust the basic model in places where the
specific SHB parameter space requires it. The reader is referred to
\cite{Piran99,Piran05} and \cite{Meszaros06} for further details on
the model\footnote{SHB afterglow models that I do not cover here are
the ``cannonball'' model \citep[for review see][]{Dar04} and a
cylindrical external shock model \citep{Wang05}}.

\subsection{Standard model - synchrotron radiation from a spherically symmetric  adiabatic blast wave}\label{SEC: afte standard theory}
The remaining relativistic flow energy, after the end of the prompt
emission phase, is transferred to the external medium by driving a
shock-wave into it - the external shock. At early times the energy
content in the ejecta and the freshly-shocked external medium is
comparable. At this time the ejecta composition has a strong effect
on the dynamics and the emission. I discuss this early, but more
complicated, phase later (\S\ref{SEC: early aft theory}). After all
the energy is dissipated into the external medium the only memory of
the initial conditions is the total amount of energy and its angular
distribution (i.e. the initial geometry of the flow). This phase,
generally referred to as the late afterglow, is discussed here.

Consider an adiabatic and spherical\footnote{The spherical
approximation can be used here as long as the blast wave properties
do not vary over angles that are smaller than $1/\Gamma$
(quasi-sphericity of relativistic flows). $E$ in this model
corresponds to the isotropic equivalent kinetic energy of the
outflow.} blast wave with energy $E$ that propagates with Lorentz
factor $\Gamma$ into an external medium composed of proton-electron
plasma with a constant density $n$. The random motion Lorentz factor
of the shocked plasma in the blast wave rest frame is $\approx
\Gamma$ and therefore the total blast wave energy, as measured in
the observer frame, is $\approx M(R) \Gamma^2 c^2$, where $M(R)$ is
the mass accumulated by the blast wave up to a radius $R$. Energy
conservation dictates \citep[e.g.,][]{Sari98}:
\begin{equation} \label{EQ: Gamma,R,t}
  \gamma  \approx  50 \left(\frac{E_{50}}{n_{-2}}\right)^{1/2} R_{17}^{-3/2}
    \approx 6 \left(\frac{E_{50}}{n_{-2}}\right)^{1/8}\left(\frac{t_d}{1+z}
\right)^{-3/8},
\end{equation}
\begin{equation}
   R  \approx 4 \cdot 10^{17} cm \left(\frac{E_{50}}{n_{-2}}\right)^{1/4}\left(\frac{t_d}{1+z}
\right)^{1/4},
\end{equation}
where $\gamma$ is the Lorentz factor of the freshly-shocked external
medium (the Lorentz factor of  the shock itself is $\Gamma =
\sqrt{2}\gamma$) and $t_d$ is the observer time in days \citep[$t
\approx R/4\gamma^2$; ][]{Waxman97c,Sari97b}. Conservation of mass,
momentum and energy fluxes across the shock imply that the proper
number and energy density of the freshly shocked plasma (in the
plasma rest frame) are $4\gamma n$ and $4\gamma^2 n m_p c^2$
respectively. The profiles of the hydrodynamic quantities (pressure,
density and velocity) behind the shock are self similar
\citep{Blandford76} and the entire shock energy is concentrated in a
very thin shell with a width $\Delta \approx R/10\Gamma^2$. The
simplest model assumes that the shock couples the electrons to the
proton energy and accelerates all the electrons to a power-law
$dN/dE \propto E^{-p}$ starting at a minimal Lorentz factor
$\gamma_{e,m} =
 \gamma \varepsilon_e (m_p/m_e) (p-2)/(p-1)$, where $p>2$ and $\varepsilon_e$
is the fraction of the internal energy that is carried by the
electrons. After crossing the shock the electron and the proton
temperatures are not coupled any more, and the electrons cool down
by radiation and $pdV$ work. This model assumes further that the
magnetic field in the shocked plasma carries a constant fraction
$\varepsilon_B$ of the internal energy, where the magnetic field of
freshly shocked plasma is $B=(32 n m_p  c^2 \varepsilon_B)^{1/2}
\gamma$. Note that by definition $\varepsilon_e+\varepsilon_B < 1$.

This model has five free parameters: $E$, $n$, $\varepsilon_e$,
$\varepsilon_B$ and $p$. These parameters can, in principle, take
almost any value, but observation of long GRBs together with the
properties of the host galaxies of SHBs suggests a ``natural" range
for these parameters. Assuming that the processes that produce the
prompt emission in long and short GRBs are similar, we expect
similar efficiency, implying that the $\gamma$-ray isotropic
equivalent energy output is a reasonable estimator of $E$
\citep{Freedman01,Granot06,Fan06a}, and therefore $E \sim 10^{50}$
erg (see Table 2). 
Observed SHB host galaxies, and the location of the bursts within
them, indicate that the progenitors are associated with an old
stellar population. Moreover, various progenitor models suggest that
the peculiar velocity of the progenitors can be high enough to
escape from the gravitational potentials of their hosts. Therefore,
possible environments of SHBs are the inter-stellar medium (ISM),
intra-cluster medium (ICM) or the inter-galactic medium (IGM). In
all of these cases the gas density would be  rather constant on the
scales relevant for SHBs, and the corresponding density values are
$10^{-2}-1 ~\rm cm^{-3}$ (ISM), $10^{-3}-10^{-2} ~\rm cm^{-3}$ (ICM)
and $\sim 10^{-6} ~\rm cm^{-3}$ (IGM). Finally, the values of
$\varepsilon_e$, $\varepsilon_B$ and $p$ are determined by
collisionless shock physics, which take place on microphysical
scales and are therefore unlikely to depend on global properties
such as the total energy, the density profile, etc. It is therefore
reasonable to use the values observed in long GRBs (see also
\S\ref{SEC: collisionless_shocks}): $\varepsilon_{e} = 0.05-0.5$,
$\varepsilon_{B} = 10^{-3}-0.1$ and $2\lesssim p \lesssim 3$
\citep[e.g.,][]{Panaitescu01,Yost03}.

Within this parameter range the radio to X-ray afterglow is
dominated by synchrotron radiation. The synchrotron spectrum has
three characteristic frequencies
\citep[e.g.,][]{Meszaros97a,Waxman97a,Sari98,Granot99b}: $\nu_m$ -
the typical synchrotron frequency ($\propto \gamma B \gamma_e^2$)
that corresponds to the synchrotron frequency of an electron with
Lorentz factor $\gamma _{e,m}$; $\nu_c$ - the synchrotron frequency
that corresponds to $\gamma _{e,c}$, which is the Lorentz factor
above which radiative cooling is significant; and $\nu_a$ - the
synchrotron self-absorption frequency. The observed spectrum is
usually four-segment broken power-law with breaks at the
characteristic frequencies, where the power-law indices depend on
the order of these frequencies. The value of $\nu_a$ and of the peak
value of $F_\nu$ depend on the order of the characteristic
frequencies as well. In the parameter range relevant for SHBs the
typical frequency order would be $\nu_a<\nu_m<\nu_c$ during the
relativistic phase (i.e., cooling is slow). In this case the
characteristic frequencies are\footnote{Eq. \ref{EQ: aft synch spec}
is generalized to include self-synchrotron Compton cooling (see
\S\ref{SEC: SSC}) by taking $\nu_c \rightarrow \nu_c (1+Y)^{-2}$
where $Y$ is defined in Eq. \ref{EQ: Y1}. As I show, in SHBs $Y \ll
1$.}:
\begin{equation}\label{EQ: aft synch spec}
\begin{array}{c}
 \nu_a \approx  0.2 {\rm ~GHz}~  (1+z)^{-1} \varepsilon_{e,-1}^{-1}  \varepsilon_{B,-2}^{1/5} E_{50}^{1/5}
    n_{-2}^{3/5},\\
  \nu_m \approx  5 \cdot 10^{10} {\rm ~Hz}~  (1+z)^{1/2} \varepsilon_{e,-1}^2  \varepsilon_{B,-2}^{1/2} E_{50}^{1/2}  t_d^{-3/2},\\
  \nu_c \approx  8 \cdot 10^{18} {\rm ~Hz}~
(1+z)^{-1/2}\varepsilon_{B,-2}^{-3/2} E_{50}^{-1/2} n_{-2}^{-1}
 t_d^{-1/2},\\
 F_{\nu,m} \approx  5 {\rm ~\mu Jy}~  (1+z) \varepsilon_{B,-2}^{1/2} E_{50}
    n_{-2}^{1/2} d_{L,28}^{-2},
    \end{array}
\end{equation}
where $F_{\nu,m}$ is the flux at $\nu_m$. The observed flux is a
broken power-law with the following spectral and temporal indices:
\begin{equation}\label{EQ: aft spherical power-laws}
    F_\nu \propto \left\{
    \begin{array}{cl}
      \nu^2t^{\frac{1}{2}}&\nu<\nu_a \\
      \nu^{\frac{1}{3}}t^{\frac{1}{2}}&\nu_a<\nu<\nu_m\\
      \nu^{-\frac{p-1}{2}}t^{-\frac{3}{4}(p-1)}&\nu_m<\nu<\nu_c \\
      \nu^{-\frac{p}{2}}t^{-\frac{3p-2}{4}}&\nu_c<\nu
    \end{array}\right. .
\end{equation}
A detailed discussion of the light curve in cases of other orders of
the break frequencies can be found in \cite{Granot02} while the case
of $1<p<2$ is discussed in \cite{Dai01}.

Once the blast wave accumulates enough mass it decelerates to
non-relativistic velocities and makes a transition from the
relativistic self-similar Blandford-Mackee solution
\citep{Blandford76} to the Newtonian self-similar  Sedov-Von
Neumann-Taylor solution \citep{Sedov46,Neumann47,Taylor50}. The
transition between the two solutions is gradual and takes about one
decade in time around:
\begin{equation}\label{EQ: t_NR}
    t_{NR} \sim 0.5 ~{\rm yr}~ \left(\frac{E_{50}}{n_{-2}}\right)^{1/3}
    (1+z)
\end{equation}
At $t \gg t_{NR}$ the blast wave velocity is $v \propto t^{-3/5}$
and the flux at the radio frequencies, $\nu_R$, which is typically
the only observed wavelengths at this time, behaves as:
\citep{Waxman98,Frail00}:
\begin{equation}
    F_{\nu,R} \propto \nu^{-\frac{p-1}{2}}t^{-\frac{3(p-1)}{2}+\frac{3}{5}};~~~~\nu_a,\nu_m<\nu_R<\nu_c
~~\&~~ t \gg t_{NR}
\end{equation}

\subsubsection{Comparison to the observations}
The handful of observed SHB afterglows were analyzed within the
framework  of the standard afterglow model in several papers
\citep{Hjorth05b,Fox05,Berger05,Lee05,Bloom06,Panaitesc06,Campana06,Burrows06,Grupe06,Soderberg06a}.
The simple model gives a reasonable explanation of the general
properties of most of the observed afterglows. Moreover, most of the
observed properties were predicted by the standard model
\citep[e.g.,][]{Panaitescu01b} and while the current observations
are too sparse to confirm its validity, they are certainly
encouraging. For example, this model predicts that the emitting
electrons are accelerated in unmagnetized relativistic collisionless
shocks to a power-law distribution with index $p$. Long GRB
afterglow observations as well as theory of particle acceleration in
relativistic shocks suggests  $2 \lesssim  p \lesssim 3$
(\S\ref{SEC: collisionless_shocks}). This prediction determines,
with little freedom, the optical and X-ray spectral and temporal
evolution (Eq. \ref{EQ: aft spherical power-laws}). Most SHB
afterglows conform with the evolution of the afterglow as predicted
by this range of $p$ values. A few examples are SHBs 050709
\citep[$p \approx 2.5$;][]{Fox05,Panaitesc06}), 050724
\citep[p=2-3;][]{Berger05,Panaitesc06} and 051221 \citep[$p \approx
2.2$;][]{Soderberg06a,Burrows06} where broad-band modeling (radio to
X-ray) gives a reasonable description of the observed afterglows.

Unfortunately, the current quality of the data is not sufficient to
break all the degeneracies between the other four parameters of the
model ($E$, $n$, $\varepsilon_e$ and $\varepsilon_B$), and to
determine their values. Nevertheless, in most cases where the simple
standard model fits the data, the default parameters that were
predicted based on the observations of long GRBs (i.e., $E \sim
E_{\gamma,iso}$, $n<1 \rm~cm^{-3}$ and $\varepsilon_e$ and
$\varepsilon_B$ not much smaller than equipartition), give a
satisfactory description of the data \citep[e.g.,][]{Panaitesc06}.

On the other hand, some of the observed details cannot be explained
by the simple, spherically symmetric, self-similar model. The most
prominent features that require deviation from the basic model are
the early X-ray tail (see \S\ref{SEC: X-ray tail}),  the late flares
observed in the X-ray afterglows of several bursts (Fig. \ref{FIG:
Aft_lc}) and the late break in the light curve of SHB 051221.
Extensions of the basic model that can explain at least some of
these features are discussed below.

\subsection{Synchrotron self-Compton}\label{SEC: SSC}

The same electrons that produce synchrotron radiation also upscatter
synchrotron photons by the inverse Compton process (synchrotron
self-Compton - SSC; e.g.,
\citealt{Meszaros94,Waxman97b,Wei98,Wei00,PanaitescuKumar00,Dermer00,Sari01,Zhang01b}).
SSC photons themselves are upscattered again by the same
relativistic electrons to even higher energies producing a second
generation of inverse Compton photons. The process continues to
higher and higher generations until the photon energies, as seen in
the electrons' rest frame, are above the Klein-Nishina energy
($m_ec^2$), where the scattering cross-section drops rapidly
(roughly linearly) with photon energy. For typical long GRB
parameters the first generation of the SSC is below the
Klein-Nishina limit and is usually calculated without considering
the effect of the Klein-Nishina cross-section. This first SSC
generation is predicted to produce bright GeV emission that would be
detected by sensitive GeV future observatories (e.g., {\it GLAST}).
The second SSC generation for long GRB afterglows is well above the
Klein-Nishina limit and is completely suppressed. As I show below
for SHBs the situation is different as the first generation SSC
component is already affected by the Klein-Nishina cross-section,
though it is not entirely suppressed. I therefore repeat the
analysis presented in \cite{Sari01} augmented by a simple estimate
of the effect of the Klein-Nishina cross-section
\citep[c.f.,][]{LiWaxman06}. The result is that in all but the
brightest SHBs the SSC emission is expected to be significantly less
energetic than the synchrotron emission.

Photons that are scattered by an electron with a Lorentz factor
$\gamma_e$ are below the Klein-Nishina limit as long as the observed
frequency satisfies:
\begin{equation}\label{EQ: nu_KN}
   h \nu \lesssim h \nu_{KN}(\gamma_e) = m_e c^2 \frac{\Gamma}{\gamma_e}
\end{equation}
where $h$ is the Planck constant and $\Gamma$ is the plasma Lorentz
factor ($\gamma_e$ is measured in the plasma rest-frame). In the
slow cooling regime (i.e., $\nu_m<\nu_c$, as expected for SHBs) most
of the synchrotron energy is emitted around $\nu_c$ and most of the
SSC energy is emitted by $\sim \gamma_c$ electrons that upscatter
$\sim \nu_c$ photons. Therefore, the Klein-Nishina limit can be
neglected only if $\nu_c/\nu_{KN}(\gamma_c)\lesssim 1$:
\begin{equation}\label{EQ: nuc/nuKN}
    \frac{\nu_c}{\nu_{KN}(\gamma_c)} \approx 10^4 E_{50}^{-1} n_{-2}^{-3/2}
    \varepsilon_{B,-2}^{-5/2} (1+Y)^{-2}.
\end{equation}
Here I included the effect of SSC cooling on $\nu_c$ using the $Y$
parameter (defined in Eq. \ref{EQ: Y1} and found to be small). Eq.
\ref{EQ: nuc/nuKN} shows clearly that in SHBs the Klein-Nishina
limit cannot be neglected and that the first SSC branch in this case
is strongly affected by the Klein-Nishina cross-section\footnote{ In
long GRBs the isotropic equivalent blast wave energy is larger by
$2-3$ orders of magnitude and the external medium density is usually
higher. Therefore, the Klein-Nishina cross-section can usually be
neglected for the first SSC generation.}.

The total energy in the first SSC generation can be estimated as
follows. In the relevant parameter space $\nu_m<\nu_{KN}(\gamma_c)
<\nu_c$ where $F_\nu \propto \nu^{(1-p)/2}$, implying that for
$2<p<3$ most of the upscattered energy is emitted by $\sim \gamma_c$
electrons that upscatter $\sim \nu_{KN}$ photons.
Therefore, the ratio of the SSC luminosity to the synchrotron
luminosity is (c.f. Eq. 3.1 of \citealt{Sari01}):
\begin{equation}\label{EQ: Y1}
    Y\equiv \frac{L_{IC}}{L_{syn}}=\frac{U_{syn}(\nu<\nu_{KN}[\gamma_c])}{U_B}=
    \frac{\eta_{rad} \eta_{KN} U_e}{(1+Y)U_B}=\frac{\eta_{rad} \eta_{KN} \varepsilon_e}{(1+Y)\varepsilon_B},
\end{equation}
where $U_{syn}$, $U_B$ and $U_e$ are the energy density of the
synchrotron emitted photons,  the magnetic field, and the
relativistic electrons respectively (measured in the plasma rest
frame). $\eta_{rad}$ is the fraction of the electrons' energy that
is radiated:
\begin{equation}\label{EQ: eta_rad}
    \eta_{rad}=\left(\frac{\nu_m}{\nu_c}\right)^\frac{p-2}{2}
    ~~~~;~~~~\nu_m<\nu_c.
\end{equation}
$\eta_{KN}$ is the fraction of the radiated energy which is in
frequencies smaller than $\nu_{KN}(\gamma_c)$:

\begin{equation}\label{EQ: eta_KN}
    \eta_{KN}=\left(\frac{\nu_{KN}(\gamma_c)}{\nu_c}\right)^\frac{3-p}{2}
    ~~~~;~~~~\nu_m<\nu_{KN}(\gamma_c)<\nu_c.
\end{equation}
Expressing Eqs. \ref{EQ: eta_rad} \& \ref{EQ: eta_KN} using the five
free parameters of the model and plugging them into Eq. \ref{EQ: Y1}
yields:
\begin{equation}\label{EQ: Y2}
    Y=0.01e^{4.3(2.5-p)}\varepsilon_{e,-1}^{p-1}  \varepsilon_{B,-2}^\frac{3-p}{4} E_{50}^{1/2}n_{-2}^\frac{5-p}{4} t_d^{-\frac{p-2}{2}}
\end{equation}
This equation is valid for $2<p<3$ and as long as
$\nu_m<\nu_{KN}(\gamma_c) <\nu_c$, as expected in SHBs. Note that in
this range of $p$ values the dependence on $\varepsilon_B$ and $t$
is very weak while by definition $\varepsilon_e < 1$. Therefore, $Y
\ll 1$ in all SHBs but the most energetic ones that happens to
explode in relatively dense environments. The implications of such
low $Y$ values is that SSC cooling is inefficient and can be ignored
in Eq. \ref{EQ: aft synch spec} and that in general SSC GeV emission
from most SHB late afterglows will be dim and is unlikely to be
detected by planed GeV observatories.

\subsection{Early afterglow and the reverse shock}\label{SEC: early aft theory}

According to the internal-external dissipation model, between the
prompt emission and the late afterglow there is an intermediate
phase, during which the energy in the ejecta is dissipated into the
external medium. The composition of the ejecta plays an important
role in the physical evolution during this phase. In long GRBs a
baryonic flow may show a clear observable signature (the reverse
shock optical flash and radio flare) which is not expected in a
Poynting flux dominated flow. As I show below for SHBs, even if the
flow is baryonic and there is a strong reverse shock, bright early
optical emission is not expected.

Interaction of baryonic ejecta with the external medium generates
two shocks. One is propagating into the external medium (forward
shock) and the other is propagating back into the ejecta (reverse
shock). Between these two shocks there are two shocked regions that
are separated by a contact discontinuity. Similar shock structure
can be found in many astrophysical environments such as supernova
remnants and stellar winds. As I show below, in SHBs the reverse
shock is mildly relativistic. This implies that during a single
crossing the shock dissipates most of the bulk motion energy of the
outflow into internal energy. After the shock crosses the outflow
once and a rarefaction wave is reflected, only the forward shock
remains, and is the source of the late afterglow. The reverse shock
is short-lived and might produce an observable signature which may
be distinguishable from the late afterglow emission. If, however,
the outflow is dominated by Poynting flux, then the large magnetic
pressure prevents the formation of a strong reverse shock. The exact
prediction for magnetized flow depends on the magnetization level of
the plasma \citep[e.g.,][]{Lyutikov06,Zhang05}.

Observed signatures of reverse shocks were explored extensively in
the context of long GRBs
\citep[e.g.,][]{Meszaros97a,Sari99,Kobayashi00,
 Sari00,Fan02,Soderberg02,Zhang03,Nakar04,Beloborodov05,McMahon06}
and the same results are applicable, with some minor changes, to
SHBs \citep{Fan05a}. I summarize the predictions for reverse shock
emission, making the required adjustments to SHBs.

Consider a homogenous cold baryonic shell expanding relativistically
into homogenous cold ambient medium. The problem is well defined by
the shell (isotropic equivalent) energy $E$, width $\Delta$ in the
lab frame, initial Lorentz factor $\Gamma_0$ and the ambient density
$n$. As the ejecta plow through the external medium, a forward shock
and a reverse shock are produced. The strength of the reverse shock
is determined by the dimensionless parameter \citep{Sari95}:
\begin{equation}\label{EQ: xi}
    \xi \equiv (l/\Delta)^{1/2}\Gamma_0^{-4/3}
\end{equation}
where $l \equiv (3E/(4\pi nm_pc^2))^{1/3}$ is the Sedov length. If
$\xi \lesssim 1$ the reverse shock is relativistic and the shell
decelerates, while dissipating most of its bulk motion energy,
within a single shock crossing of the shell at $ R_\Delta \approx
l^{3/4}\Delta^{1/4}$. In this case $R_\Delta = R_{dec}$, where
$R_{dec}$ is the deceleration radius, namely the radius in which all
the ejecta decelerates by a factor of two or more. For $\xi \gg 1$
the RS is Newtonian and many crossings are required to decelerate
and shock dissipate a significant fraction of the ejecta energy. In
this case $R_{dec}$ is the radius in which the forward shock
collects an amount of mass that is equivalent to $\sim
E/(c^2\Gamma_0^2)$.

The Lorentz factor of the outflow is expected to vary by at least a
factor of order unity, causing the shell to expand once it gets out
to a radius $R_{spread} \approx \Delta_0 \Gamma_0^2$ where $\Delta_0
\approx cT$ is initial width of the shell ($T$ is the burst
duration):
\begin{equation}\label{EQ: delta}
    \Delta \approx max (\Delta_0,R/\Gamma_0^2),
\end{equation}
Correspondingly, $\xi$ is constant up to $R_{spread}$, and it
decreases at larger radii. The value of $\xi_0 \equiv \xi(\Delta_0)$
determines the evolution of the reverse shock. If $\xi_0 < 1$ then
$R_{dec}<R_{spread}$ and the relativistic reverse shock crosses the
shell before it starts spreading. If $\xi_0 \gg 1$ the reverse shock
is initially Newtonian and the shell begins to spread during the
first crossing of the reverse shock. As the shell spreads the
reverse shock grows stronger and at the radius where the reverse
shock finally crosses the shell, $\xi \approx 1$ and the reverse
shock is mildly relativistic (its Lorentz factor is $\approx 1.25$).

For typical SHB parameters,
\begin{equation}\label{EQ: xi0}
  \xi_0 =  43 \left(\frac{E_{50}}{n_{-2}}\right)^{1/6}
  \Gamma_{0,2}^{-4/3} T_{-1}^{-1/2} \gg 1.
\end{equation}
Therefore, in most (and maybe all) SHBs the reverse shock is mildly
relativistic ($\xi \approx 1$) and does not depend on the initial
Lorentz factor, the total energy or the burst duration. The
deceleration radius in this case is:
\begin{equation}\label{EQ: Rdec}
    R_{dec} \approx 5 \cdot 10^{16} {\rm ~cm}~ \left(\frac{E_{50}}{n_{-2}}\right)^{1/3}
  \Gamma_{0,2}^{-2/3}
\end{equation}
Corresponding to an observer time:
\begin{equation}\label{EQ: tdec}
    t_{dec} \approx  \frac{R_{dec}}{2c\Gamma_0^2}(1+z) \approx 90 {\rm ~s}~ (1+z) \left(\frac{E_{50}}{n_{-2}}\right)^{1/3}
  \Gamma_{0,2}^{-8/3}.
\end{equation}
At $t_{dec}$ the reverse shock dies away and its optically thin
synchrotron emission (at $\nu>\nu_a$) peaks and then starts fading
rapidly \cite[e.g.,][]{Sari97b}. $t_{dec}$ also marks the onset of
the ``late'' afterglow - a self similar adiabatic solution - as the
forward shock turns into a single blast wave (and indeed Eqs.
\ref{EQ: tdec} and \ref{EQ: gamma constraint2} are similar). Note
that in SHBs $t_{dec}$ does not depend on $T$ and that $t_{dec} \gg
T$. This is different than long GRBs where $\Delta_0$ is
significantly larger and $\xi_{0} \lesssim 1$, in which case
$t_{dec} \approx T$. Thus, in SHBs a gap is expected between the
prompt emission and the afterglow while in long GRBs the two are
expected to overlap.

The emission from the reverse shock is calculated as in the standard
afterglow model, taking into account that the reverse shock is
mildly relativistic while the forward shock is ultra-relativistic.
The strength of the shock defines the typical energy per particle in
the shocked region. This in turn sets the typical frequency of the
emitted synchrotron radiation. Thus, while the forward shock is
radiating in the X-ray band, the reverse shock is radiating at low
frequencies. The reverse shock signature in long GRBs is an optical
flash that peaks soon after, or during, the prompt GRB emission and
then decays as $\sim t^{-2}$ \citep[e.g.,][]{Sari99}. $R \sim 18$
mag is expected from the reverse shock of a typical long GRB and in
extreme cases it can be brighter by several orders of magnitude
\citep{Nakar04}. An accompanying radio flare is expected to peak
hours to days later \citep[e.g.,][]{Kulkarni99,Nakar04}. In SHBs,
where the energy and typical density are lower and the reverse shock
is not expected to be very strong, the predictions are different.
For SHB ``canonical'' parameters ($E_{50}=1$ , $n=0.01 {\rm
~cm^{-3}}$, $\varepsilon_e=0.1$, $\varepsilon_B=0.01$) most of the
reverse shock emission is radiated around  $\sim 100$ GHz and the
early optical emission is faint, $\sim 2 \rm~\mu Jy$ ($R\sim 23$
mag) at $z=0.2$\footnote{If the ejecta is mildly magnetized so the
reverse shock is still mildly relativistic but $\epsilon_B \sim
1/3$, then the optical emission from the reverse shock is brighter
($F_{\nu,opt}$ is roughly linear with $\epsilon_B$ at this parameter
range). Additional increase in the reverse shock optical luminosity
is expected if only a small fraction of the electrons is
accelerated, thereby significantly increasing $\nu_m$.}. In this
case the optical emission from the forward shock is expected to be
brighter than the reverse shock emission. Actually, the reverse
shock emission from SHBs is more likely to be detected by early
radio observations. In the above case a peak flux of  $\sim 0.1$ mJy
is expected in  $8$ GHz  at $t \approx 20$ min.

To conclude, early reverse shock optical flash is not expected from
most SHBs (assuming canonical parameters), even if the ejecta is
baryonic. A reverse shock signature might however be detected by
very early deep radio observations.

\subsection{``Naked'' afterglow}\label{SEC: naked aft}
When the relativistic outflow that produces the prompt emission is
quasi-spherical, an afterglow (late soft emission) is expected even
if no emission from the external shock is observable (e.g., due to
low external density). This is the high latitude emission of the
prompt gamma-rays source (\S\ref{SEC: Rel high latitude}), that
decays at late times ($t \gg t_{los}$ and $t \gg t_{ang}$;
\S\ref{SEC: Rel time scales}) as $F_\nu \propto
\nu^{-\beta}t^{-(2+\beta)}$. This type of afterglow
 \citep[``naked'';][]{Kumar00,Dermer04,Page06} is suggested to be
responsible for the first phase of long GRB X-ray
afterglows\footnote{Note that identification of high latitude
emission can be tricky. At early times it requires an identification
of $t_{los}$ of the specific shell that dominates the late naked
afterglow emission. As evident from Eq. \ref{EQ: high latitude}, a
significant underestimate of $t_{los}$ leads to an overestimate of
the decay index at early time.} \citep{Nousek06,Zhang06}.


SHBs 051210 \citep{Parola06} and 050813 \citep{Roming06} show X-ray
afterglows that may be high latitude emission. The afterglow of
051210  also  shows a flare at $t \sim 100$ s. Assuming that no
external shock emission exists, this flare is probably a result of
the engine activity, in which case $t_{los} \sim 100$ s.
\cite{Parola06} find that assuming this value of $t_{los}$, the late
time decay is consistent with high latitude emission. In order for
external shock emission to be fainter than the high latitude
emission, they estimate that the external density is lower than $n <
4 \cdot 10^{-3} ~\rm cm^{-3}$.

\subsection{X-ray dark afterglows and $\gamma$-ray efficiency}\label{SEC: gamma eff}
Figure \ref{FIG: ratio_gx} presents the distribution of the
dimensionless ratio $f_{x\gamma} \equiv F_x t/S_\gamma$, where $F_x$
is the X-ray ($0.2-10$ keV) energy flux at time $t$ and $S_\gamma$
is the prompt emission gamma-ray fluence ($15-150$ keV), of \Swift
long and short GRBs at one day (\S\ref{SEC: obs late aft}). Within
the framework of the standard afterglow model (\S\ref{SEC: afte
standard theory}) and as long as the blast wave is quasi-spherical:
\begin{equation}\label{EQ: LX_Eg_ratio}
    f_{x\gamma} \equiv \frac{F_x t}{S_\gamma} \approx \left\{
    \begin{array}{cr}
    10^{-2} \kappa^{-1} \varepsilon_{e,-1}^{3/2}  \varepsilon_{B,-2} E_{k,50}^{1/3}n^{1/2} ~~~&~~~\nu_x<\nu_c \\
    2 \cdot 10^{-3} \kappa^{-1} \varepsilon_{e,-1}^{3/2}  t_d^{-1/3} ~~~&~~~\nu_x>\nu_c
                                          \end{array}\right.
\end{equation}
where,
\begin{equation}\label{EQ: kappa}
    \kappa \equiv \frac{E_\gamma}{E_k}
\end{equation}
represents the $\gamma$-ray efficiency of the prompt emission. $E_k$
(equivalent to $E$ in \ref{SEC: afte standard theory}) is the
kinetic energy of the blast wave and $E_\gamma$ is the energy
emitted in $\gamma$-rays (all energies in this section are isotropic
equivalent). The exact power of the parameters in Eq. \ref{EQ:
LX_Eg_ratio} (e.g., $\varepsilon_e$) depends weakly on $p$ (assuming
here $2.1<p<3$). For simplicity I use approximate power values. I
also neglect weak dependence (power-law indices below $1/4$) on
parameters, since these cannot affect the result significantly (the
lack of dependence on $n$ is exact). Following \S\ref{SEC: SSC} the
SSC cooling is neglected as well.

Assuming that microphysics of collisionless shocks does not vary
significantly between bursts (either long or short) $\varepsilon_e
\approx 0.1$ and $\varepsilon_B \approx 0.01$ are adopted. Under
this assumption, for long GRBs at late time ($\sim 1$ d; but before
the jet-break), one expects $\nu_c<\nu_x$ in which case
$f_{x\gamma}$ is almost a direct measure of the $\gamma$-ray
efficiency, $\kappa$. As evident from Fig. \ref{FIG: ratio_gx}, for
long GRBs $f_{x\gamma}(1 d)=0.1-10^{-3}$ implying $\kappa_{LGRB}
\approx 0.01-1$. This result is well known
\citep{Freedman01,Lloyd04,Granot06,Fan06a}. Figure \ref{FIG:
ratio_gx} also shows that the values of $f_{x\gamma}$ for SHBs with
observed X-ray afterglows are comparable to those of long GRBs,
$\sim 0.01$.

For some SHBs the circum-burst density can be low, in which case it
is not clear wether $\nu_c$ is above or below the X-ray band at 1
day. If $n \gtrsim 0.01~\rm cm^{-3}$ then $\nu_x \lesssim \nu_c$ and
$\kappa_{SHB} \approx 0.1$. If the density is significantly smaller
then $\nu_x > \nu_c$ and $\kappa$ decreases as $n^{1/2}$. Since SHBs
with observed afterglows are typically located within their host
galaxy light, most likely $n \gg 10^{-4}~\rm cm^{-3}$, so $\kappa
\approx 0.01-0.1$. We can conclude that at least some SHBs (those
with observed X-ray afterglow) the gamma-ray efficiency is most
likely similar to that of long GRBs \citep[see][for a specific
exploration of the efficincy of SHB 050509B]{Bloom06,Lee05}.

Early X-ray afterglow from long GRBs is always detected. In
contrast, there are several SHBs with tight upper limits on any
early ($< 100 s$) X-ray emission. The values of $f_{x\gamma}
\lesssim 5 \cdot 10^{-5}$ for these bursts are exceptionally low
(Fig. \ref{FIG: ratio_gx}). Making the plausible assumptions that
the gamma-ray efficiency of these bursts is typical ($\kappa \sim
0.1$) as are the initial Lorentz factor and the microphysical
parameters, these values of $f_{x\gamma}$ indicate that these events
occurred in extremely low density environments, $n \lesssim 10^{-5}
\rm ~cm^{-3}$, typical for the inter galactic medium. As I will
discuss below this result suggests a long-lived progenitor with high
natal velocity (\S\ref{SEC: binary offset}). An alternative would be
to relax any of the above assumptions. For example, assuming an
inter-stellar density ($n \gtrsim 0.01 ~\rm cm^{-3}$), the low
$f_{x\gamma}$ value can be explained by ultra-efficient gamma-rays
production ($\kappa \gtrsim 100$), by unusually low electron and
magnetic field energies ($\epsilon_e^{3/2}\epsilon_B \lesssim
10^{-6}$) or by low initial Lorentz factor $\Gamma_0 \lesssim 20$.
The latter case can explain the faint early afterglow because if
$\Gamma_0$ is low the deceleration time (and therefore the afterglow
onset) is at $t \gg 100$ s.

\subsection{Angular structure of the outflow}\label{SEC: beaming}
The angular structure of the relativistic outflow is of great
physical interest. As discussed in \S\ref{SEC: Rel qausi-spher},
during the prompt emission phase the observer is sensitive only to a
small patch of the flow (a solid angle of $\sim 1/\gamma^2$ sr), and
is ignorant to any of the flux emitted outside of this patch. This
flux determines the total energy emitted by the burst as well as the
range of the viewing angles over which a given burst can be
detected. Therefore the angular structure is a crucial ingredient in
determining the total burst energy and in extracting the total SHB
rate out of the observations. Furthermore, the structure of the jet
is most likely determined by the engine, and different progenitor
and central engine models predict different angular structures.

Currently, the small number of SHB afterglows prevents the
determination of the exact angular structure of the outflow (even in
long GRBs it is still an open question). Therefore, I discuss here
only the simplest jet structure: a uniform energy over a half
opening angle $\theta_j \ll 1$ and no energy outside of this angle
(a.k.a 'top hat')\footnote{Observational signatures of the `top hat'
jet are very similar to those of a structured jet where the energy
per solid angle is $ \propto \theta^{-2}$, if the jet opening angle
is replaced by the viewing angle \citep{Lipunov01,Rossi02,Zhang02}.
Therefore, all the discussion here is valid for this case as well
(replacing $\theta_j$ with the viewing angle). Note that such a
structured jet results in a larger beaming-corrected energy and
lower beaming-corrected rate than top hat jet.}. This simple model
turns out to be consistent with the few available observations
(excluding the observed flares discussed in \S\ref{SEC: aft
variability}) and it is also supported by numerical simulations
\citep{Janka06}. A comprehensive review of jets in GRBs can be found
in \cite{Granot_jets06}, where the dynamics and observational
signatures of relativistic blast waves with different angular
structures are discussed.

Two key processes govern the observational signatures of a
decelerating collimated relativistic blast wave. The first is
relativistic beaming of the radiation from an emitting element with
a Lorentz factor $\Gamma$ to an angle of $1/\Gamma$. This implies
that the radiation from a jet with half opening angle $\theta_j$ is
confined to an angle $\theta_j+1/\Gamma$. The second process is
hydrodynamic lateral spreading of the jet, which is limited by the
angular size of causally connected regions (see \S\ref{SEC: Rel
qausi-spher}),  $\approx  1/\Gamma$. Therefore, the half opening
angle of a jet with an initial half opening angle $\theta_{j,0}$ is:
$\theta_j \lesssim \theta_{j,0} + 1/\Gamma$. If the observer viewing
angle (with respect to the axis of the jet), $\theta_{obs}$, is
smaller than $\theta_{j,0}$, the two effects combine to produce a
light curve that is initially similar to that of the spherical blast
wave discussed in \S\ref{SEC: afte standard theory}. Only when the
blast wave decelerates  and $\Gamma$ becomes comparable to
$1/\theta_{j,0}$ the jetted nature of the blast wave ``reveals''
itself to the observer in the form of a light curve break
\citep{Rhoads97,Rhoads99,SariPiranHalpern99}.

This jet-break in the light curve results from a combination of the
two processes discussed above. When the radiation beaming angle
becomes wider than the jet size, the ``missing'' emission manifests
itself as a faster decay of the light curve. Simultaneously, when
the jet center become causally connected with the jet edges, the jet
spreads sideways and the energy per solid angle decreases resulting
in  a faster deceleration with the radius. Therefore, at:
\begin{equation}\label{EQ: t_b }
  t_j \approx 2 ~{\rm d}~ (1+z)
  \left(\frac{E_{k,iso,50}}{n_{-2}}\right)^{1/3}
  \left(\frac{\theta_{j,0}}{0.2}\right)^{8/3}
\end{equation}
the light curve `breaks' roughly simultaneously in all observed
bands. The asymptotic decay at $t_j \ll t \ll t_{NR}$ (defined in
Eq. \ref{EQ: t_NR}) is\footnote{The hydrodynamics of lateral
spreading is a major open question. Numerical simulations
\citep{Granot01a,Cannizzo04} and approximate analytical treatment
\citep{Kumar03} suggest that the spreading velocity in the co-moving
fluid frame is significantly smaller than $c$. The details of the
lateral spreading determine the relative importance of each one of
the two processes that cause the observed break. In any case,
observational signatures do not depend strongly on the spreading
details. Eq. \ref{EQ: jet } is derived assuming maximal lateral
spreading (comparable to the light speed), but is a fairly good
approximation for spreading at any other velocity (including no
spreading at all).} \citep{SariPiranHalpern99}:
\begin{equation}\label{EQ: jet }
    F_\nu \propto \left\{
    \begin{array}{lr}
      \nu^2\,t^0 &\nu<\nu_a \\
      \nu^{1/3}\,t^{-\frac{1}{3}}&\nu_a<\nu<\nu_m\\
      \nu^{-(p-1)/2}\,t^{-p}~~~&\nu_m<\nu<\nu_c \\
      \nu^{-p/2}\,t^{-p}&\nu_c<\nu
    \end{array}\right. .
\end{equation}
Comparison with the temporal decay at $t \ll t_b$ (Eq. \ref{EQ: aft
spherical power-laws}) shows that in the optical and the X-ray
(where $\nu > \nu_m$) a roughly similar break is expected from
$\alpha \approx 1$ to $\alpha \approx 2$. Therefore, when an
achromatic break, with roughly the predicted power-law indices, is
observed, it is usually interpreted as an indication that the
outflow was initially collimated to a half opening angle:
\begin{equation}\label{EQ: theta_j}
  \theta_{j,0} \approx 0.15 ~{\rm rad}~
  \left(\frac{E_{k,iso,50}}{n_{-2}}\right)^{-1/8}
  \left(\frac{t_j/(1+z)}{1 ~{\rm d}}\right)^{3/8}.
\end{equation}

To date, three SHBs have known redshift and late afterglow light
curves (Fig. \ref{FIG: Aft_lc}). The limits on the opening angle of
these bursts are:\\
{\it SHB 050709}: In this burst there are optical and X-ray
detections at several different epoches \citep{Fox05,Hjorth05b}.
\cite{Fox05} suggest an optical break based on three \hst
observations, but \cite{Watson06} have shown that when additional
optical data are considered, the optical data set cannot be fitted
by a single or a broken power-law. Additionally, the last X-ray
observation ($t = 16$ d) shows evidence of an intense flare. Given
the small number of observations and the apparent variability, it is
impossible to reliably determine an underlying power-law behavior
that is consistent in all wavelengths. Therefore, unfortunately, the
opening angle of this burst cannot be constrained.

{\it SHB 050724}: Observations include  a detailed X-ray light curve
\citep{Campana06,Grupe06} but only sparse detections in the IR,
optical and radio \citep{Berger05}. At early times the X-rays are
dominated by the X-ray tail (\S\ref{SEC: X-ray tail}), and following
its decay two flares are observed, one at $\sim 1000$ s and the
other around $\approx 10^4-10^5$ s. Between the two flares the
power-law decays as $ \sim t^{-1}$ and late \chandra observations at
$t \approx 3$ weeks show that after the second flare the afterglow
resumes its shallow $ \sim t^{-1}$ decay. These observations suggest
that $t_j > 20$ d as there is no evidence for a jet-break in the
data, implying \citep{Grupe06}: $\theta_{j,0} \gtrsim 0.4 ~{\rm
rad}~ (E_{k,iso,50}/n_{-2})^{-1/8}$. Possible caveats are that a jet
signature may have been obscured by the flaring activity or, if the
density is indeed as high as suggested by \citet{Panaitesc06},
$\gtrsim 0.1 ~\rm cm^{-3}$, the Newtonian transition may already
affect the light curve after 3 weeks, preventing the detection of a
jet. If instead one adopts the beginning of the second flare
($\approx 10^4$ s) as a lower limit on the jet-break time, the
opening angle is $\theta_{j,0} \gtrsim 0.06 ~{\rm rad}~
(E_{k,iso,50}/n_{-2})^{-1/8}$. Assuming  `top hat' jet structure
these limits imply a total (beaming corrected) energy of
$E_{\gamma,tot}=E_{\gamma,iso}\theta_j^2/2 \gtrsim 0.7 [0.02]
10^{49} ~{\rm erg}~ (E_{k,iso,50}/n_{-2})^{-1/4}$ (the bracketed
value is for the more conservative limit on $\theta_{j,0}$).

{\it SHB 051221}: Observations include a detailed X-ray light curve
\citep{Burrows06} and continuous optical monitoring
\citep{Soderberg06a}. This afterglow shows a continuous X-ray decay
as $t^{-1.2}$ between $100$ s and $4$ d with a short  episode of
flattening around $t =1$ hr. At $t \approx 5$ d the X-ray decay
becomes much steeper (roughly as $t^{-2}$). The optical light curve
decays as $\approx t^{-0.9}$ until $4$ d where it falls below the
detection limit. The $3 \sigma$ optical upper limits after four days
require a simultaneous optical break. Here the interpretation of the
break at $4$ d as a jet signature seems quite suggestive given its
achromatical nature and the regular decay before and after the break
at the predicted decay rates. The corresponding jet opening angle is
$\theta_{j,0} \approx 0.16 ~{\rm rad}~ (E_{k,iso,51}/n_{-2})^{-1/8}$
implying a total energy $E_{\gamma,tot}\approx 3 \cdot 10^{49} ~{\rm
erg}~ (E_{k,iso,51}/n_{-2})^{-1/4}$.

Since there is only one ``clean'' case (SHB 051221) for which  a
reasonable constraint on the beaming can be obtained, the opening
angle distribution of SHB jets is not well constrained. Current
observations suggest that the average beaming factor ($f_b \equiv
1-\cos\theta_j$) satisfies $1 \ll \left\langle f_b^{-1}
\right\rangle < 100$, but at this point no secure conclusion can be
drawn. The fine details of the outflow angular structure remain
obviously unconstrained.

\subsection{Afterglow variability}\label{SEC: aft variability}
Afterglow model ingredients discussed so far are regular and as such
they predict smooth and regular light curves (smoothly broken
power-law). These models are simplified and the true physical
environment is expected to be more complicated with irregularities
giving rise to light curve variability. Thus, afterglow variability,
or the lack of it, can probe irregularities in the external shock
and late engine activity.

The external shock cannot produce large amplitude, rapid flux
variability ($\delta F/F \gtrsim 1$ over $\delta t/t \ll 1$ where
$F$ is the afterglow flux and $t$ is the time since the burst). The
reason is that during the external shock $t_{ang} \sim 10 t_{los}
\sim 10 t_{\Delta}$ (the width of the emitting shell following the
external shock is $\Delta \approx R/(10\Gamma^2)$; \S\ref{SEC: Rel
time scales}). Emission from a radius $R$ is observed between $t =
t_{los}$ and $t \approx t_{ang} + t_{los} \approx t_{ang}$ and
therefore spherical variations are smoothed over $t_{ang}$, implying
$\delta t \sim t_{ang} \sim t$. Even variations in a well localized
angular region will be smoothed over $t_\Delta$ implying $\delta t
\gtrsim t/10$. For an external shock to produce high amplitude
variability on shorter time scales a very small region (relative to
$\Delta$ and $R/\Gamma$) should increase its luminosity so that it
becomes much brighter than the remainder of the observed shell, and
then quickly decay without affecting the emissivity of its vicinity.
Such behavior is not expected from hydrodynamical processes.
Producing significant variability even over time scales as short as
$t_\Delta$ requires a unique configuration. On the other hand, if
the engine is active at late times, it can easily produce
variability with arbitrary amplitude on time scales as short as its
dynamical time ($< 1$ ms), just as we observe during the prompt
emission. Therefore, whenever a high amplitude, rapid variability is
observed, it is likely a result of late engine activity.

High latitude emission (\S\ref{SEC: Rel high latitude}) from a
quasi-spherical shell sets another  interesting constraint on the
variability induced by an external shock \citep{NakarPiran03}. A
spherical external shock cannot produce light curves that decay
faster than $F_{\nu} \propto \nu^{-\beta} t^{-(2+\beta)}$. Rapidly
decaying light curves must result either from late engine activity
or from an aspherical blast wave on angular scales $\lesssim
1/\Gamma$.

Alternative explanations for afterglow temporal variability are:\\
(i) {\it Variable external density}
\citep[e.g.,][]{Wang00,Lazzati02,Ruiz01,Dai02,NakarPG03,NakarPiran03,Ruiz05,Ioka05,Eldridge06,Peer06,Nakar06b}:
This variability source is discussed mostly in the context of
turbulent ISM or massive stellar wind environment. In the case of
SHBs massive stellar winds are most likely irrelevant (\S\ref{SEC:
progenitors}), while density inhomogeneities resulting from ISM
turbulence are expected to have a low amplitude (of order unity).
The main distinction of density induced variability is that it is
chromatic. The light curve varies both above and below $\nu_c$, but
these variations are uncorrelated \citep{Nakar06b}. The resulting
fluctuations can be either 'bumps' or `dips' (relative to  the
unperturbed light curve). A turbulent ISM can produce only rapid,
low amplitude, fluctuations on top of a smooth power law decay
\citep{Wang00}. Significant density fluctuations, which are not
expected in the case of SHBs, can produce high amplitude light curve
variability but only over a long duration ($\delta t \sim t$).
Detailed limits on the light curve fluctuations in the case of
increasing density are discussed in \citet{Nakar06b}.

(ii){\it angular variability of the ejected energy} \citep[``patchy
shell'' e.g.,][]{KumarPiran00b,NakarPG03,NakarOren04,Ioka05}: If the
outflow energy has an intrinsic angular structure it will produce a
variable afterglow light curve. As the blast wave decelerates, the
angular size of the observed region ($\sim 1/\gamma$) increases. As
a result, the effective (average) energy of the observed region, and
hence the observed flux, vary. If the angular structure has a
typical scale then the fluctuation amplitude decreases with time as
 more patches enter the observed region. Averaging over
larger random structures leads to a decay of the fluctuation
envelope as $t^{-3/8}$ \citep{NakarPG03,NakarOren04}. A surprising
result is that although the variability is not limited by the
angular time, the gradual change in the visibility of the patches
dictates that the variability time scale is $\Delta t \sim t$
\citep{NakarOren04}. An important feature in this scenario is the
break of the axial symmetry and therefore the possible production of
linear polarization. Both the amplitude and the angle of
polarization are correlated with the light curve variations
\citep{Granot03,NakarOren04}.

(iii){\it Energy injection} \citep[e.g., ``refreshed
shocks''][]{Rees98,KumarPiran00a,Sari00,Bjornsson02,GranotNakar03,Fox03,Nousek06,Zhang06,GranotKumar06}:
Energy can be injected into the external blast wave by late-arriving
shells. Such shells can be produced by late engine activity, or a
slow shell, ejected during the initial burst, can overtake the
decelerating material behind the (initially faster) afterglow shock.
In both of these cases the blast wave energy increases monotonically
and, therefore, the observed flux is expected to rise above the
unperturbed light curve, in all wavelengths,  during energy
injection. When the injection stops the original decay is resumed.
The resulting light curve in the case of discrete episodes of energy
injection has a distinctive step-like structure. The time scale of
the steps ($\delta t$) depends on their timing relative to the
jet-break. $\delta t \sim t$ before the brake and $\delta t < t$ is
possible after the jet-break \citep{GranotNakar03}. In any case
$\delta t$ is limited by the time associated with the width of the
shell behind the afterglow shock, $t_\Delta$, implying $\delta t
\gtrsim t/10$. A variant of this model is the two-component jet
model \citep[e.g.,][]{Berger03,Konigl04,Peng05,Granot05,Wu05,Jin07}
where the opening angel of the slow ejecta is larger than the
opening angel of the fast ejecta.

(iv) {\it Microlensing event}
\citep{LoebPerna98,Garnavich00,Granot01,Mao01,Gaudi01,Koopmans01,Ioka01,Baltz05}:
The apparent radius of the afterglow ($\sim R/\Gamma$) increases
with time (as $t^{5/8}$ in the quasi-spherical phase) and at
cosmological distances its angular size after a day is $\sim \rm
~\mu as$, comparable to the Einstein radius of a solar mass
gravitational lens at these distances \citep{LoebPerna98}.
Therefore, a microlensing event may induce afterglow brightening,
which is expected to be seen simultaneously in all wavelengths. Some
color dependence is expected as a result of the color dependent
afterglow surface brightness \citep{Granot99a,Granot01}. Note that
the probability for an afterglow microlensing is small
\citep{Koopmans01,Baltz05} and therefore only a small fraction of
the observed afterglow variability may be attributed to
microlensing.

All well-sampled SHB X-ray afterglows show strong variability. In
almost all cases $\delta t \sim t$, implying that the source of the
variability is not necessarily late engine activity (although it may
be). An exception is SHB 050709 where \cite{Fox05} find that during
the last \chandra observation (exposure time of 5 hr, taken at $t =
16$ d) most of the flux arrived during the first $6$ ks (a constant
emission rate is rejected at 99.9\% confidence), suggesting a high
amplitude flare, $\Delta F/ F \approx 10$, with very rapid
variability $\delta t / t \approx 0.005$. Such a flare, if real,
requires a source that is still active 16 days after the burst. Note
that the energy emitted in this X-ray flare was $\sim 10^{45}$ erg,
about four orders of magnitude below that of the GRB itself.

Large flares observed in other bursts (e.g., SHBs 050724, 051221 and
060313) are too bright and rapid to be explained by external density
enhancement \citep{Nakar06b} and  most likely result from energy
fluctuations of the external shock (either a patchy shell or
refreshed shocks) or by internal dissipation of flow ejected during
 late engine activity. Note that SHB 051221 shows the very
indicative step-like light curve expected by refreshed shocks
\citep{Burrows06,Soderberg06a} while flux in SHB 050724 afterglow
returns to its original level after it flares \citep[assuming a
constant underlying decay;][]{Campana06,Grupe06} suggesting that the
flares result from hot spots on the blast wave, or that they are
unrelated to the external shock (i.e., late engine activity).
Interestingly, similar variable afterglows are observed in long
GRBs, such as the step-like light curve of GRB 030329
\citep[e.g.,][]{lipkin04} or the X-ray flares after which the
original underlying flux level and decay rate are resumed in many
\swift X-ray afterglows \citep[e.g.,][]{Nousek06}.

\subsection{The early X-ray ``tail''}\label{SEC: Xtail theory}
Currently, two \swift and \hete SHBs (050709 \& 050724) exhibit soft
X-ray tails, which last for $\sim 100$ s and are separated by $\sim
10$ s from the short and hard bursts of prompt gamma-rays
(\S\ref{SEC: X-ray tail}). The variability of such a tail is an
important clue to its origin. Similarly to the prompt emission,
rapid high amplitude variability implies continuous engine activity,
while low variability emission can be explained by external
interaction. In these two bursts no apparent rapid variability is
seen, although the low signal to noise ratio in these two bursts may
hide such variability, if it exists. Among the several \BATSE bursts
identified by \cite{Norris06} to have a similar temporal structure,
there are some in which the soft tail is highly variable. Similarly,
the X-ray tail of the nearby GRB 060614, which may be an extreme
member of the SHB family \citep{Gal-Yam06,Gehrels06}, is also highly
variable. Therefore, while at this point continuous engine activity
is not absolutely necessary to explain the observed X-ray tails,
there are suggestive hints in this direction. Note that X-ray flares
observed hours and days after the bursts, might require late engine
activity as well.

If the X-ray tail can be explained by external dissipation then it
is intriguing that its duration and timing are similar to the
expectation from reverse shock emission (Eq. \ref{EQ: tdec}).
According to the canonical microphysical assumptions, reverse shock
emission is not expected to contribute to the X-ray emission
(\S\ref{SEC: early aft theory}). However, it is possible that the
composition of the outflow (e.g., its magnetization level) can
affect particle acceleration or the magnetic field of the shocked
plasma in a way that results in bright X-ray emission. For example,
a large $\varepsilon_B$ together with effective acceleration of only
a small fraction of the electrons (resulting in a high value of
$\gamma_m$) can produce soft X-ray emission that begins  $\sim 10$ s
after the prompt emission, lasts for $\sim 100$ s and has a
comparable energy to the one observed in the prompt emission.

\cite{MacFadyen05} suggest a binary SHB progenitor that is composed
of a neutron star and a non-compact companion (\S\ref{SEC: AIC}). In
this model the SHB takes place once the neutron star collapses to a
black hole (driven by accretion). The X-ray tail is produced by
interaction of relativistic ejecta with the non-compact companion.
The time scale as well as the typical emission frequency and the
total radiated energy from this interaction can fit  observed X-ray
tails. This model, however, can explain only a single smooth
episode.

If however X-ray tails are produced by extended engine activity then
there are several different schemes that were suggested in order to
extend the duration of hyperaccretion of $\sim 0.1 ~\rm M_\odot$
onto a black hole (the most popular central engine model) beyond the
natural timescale of $\lesssim 1$ s (\S\ref{SEC: central engine}).
Examples include magnetic barrier that chokes the accretion
\citep{vanPuttenOstriker01,Proga06}, disk fragmentation
\citep{Perna06} and a complex evolution of the neutron star
disruption in a NS-BH binary \citep{Faber06b}. Alternatively,
viscosity that is lower then expected can prolong the accretion time
(\S\ref{SEC: central engine}). \cite{Fan05b} argue that if the X-ray
tail results from a low accretion rate, the energy of the accretion
cannot be extracted by neutrino-anti neutrino annihilation, since
the efficiency of this process falls rapidly when the accretion rate
is lower than $0.01 ~\rm  M_\odot/s$. For a detailed discussion of
accretion disk lifetime and the efficiency of different energy
extraction mechanisms see \S\ref{SEC: central engine}. A different
solution may be that the central engine is a hyper-magnetized
neutron star \citep[e.g.,][;\S\ref{SEC: Magnetar} and \S\ref{SEC:
Binary merger}.II]{Usov92}. The activity duration of this type of
engine may be significantly longer than several seconds
\citep[e.g.][]{Gao06,Fan06b,Dai06}.

\subsection{Macronova}
An interesting possible source of late time emission is radioactive
decay of matter that was ejected sub-relativistically during the
burst. There are several possible scenarios in which such ejecta are
launched along with the relativistic outflow that produce the SHB.
For example, during the coalescence of a neutron star with another
compact object (neutron star or a black hole) considerable amount of
mass, of the order of $0.01 \rm M_\odot$, may be ejected from the
system \citep[e.g.,][]{Rosswog99,Ruffert01}. During the merger the
density and temperature of the ejected mass can be high enough to
produce heavy neutron rich elements, most of them radioactive
\citep[e.g.,][]{Lattimer74,Lattimer76,Symbalisty82,Eichler89,Rosswog99,Ruffert01}.
Another example of a source of radioactive sub-relativistic ejecta
may be nucleosynthesis in the wind that is driven by a
neutrino-dominated hyper-accreting disk, which is the leading
central engine model \citep[e.g.,][]{MacFadyen03}.

Sub-relativistic radioactive ejecta also power the emission from
supernova explosions. The higher expected velocities in the SHB
case, together with the lower amount of mass, lead to a fainter and
faster evolving, yet supernova reminiscent, emission. Radioactive
energy dominates the observed emission of sub-relativistic ejecta
since initial thermal energy is quickly lost to expansion. The
properties of the observed emission are determined by the interplay
between the typical radioactivity lifetime (which depends on the
nuclear composition of the outflow), and the maximal time (and
radius) at which the ejecta are still dense enough to reprocess this
energy into a blackbody radiation. If a significant amount of
radioactive heat is generated while the ejecta are dense enough we
may detect the emitted blackbody emission. If most of the energy is
released at later times the radioactively generated photons may be
observed directly.

\cite{Li98} were the first to discuss this possibility. Having the
unknown composition ejected matter during a compact binary merger in
mind, they postulate a radioactive source which decays as a
power-law in time. Assuming that $0.01 ~M_\odot$ is ejected at a
velocity of $c/3$, and that 0.1\% of the rest mass energy is
converted to heat, they find that after one day the bolometric
luminosity can be as high as $10^{44}$ erg/s ($M_{bol} \sim -21$).
Such a bright emission is already ruled out by tight constraints on
the optical emission from SHB 050509B \citep{Hjorth05}.

\cite{Kulkarni05}, introducing the term ``Macronova'', presents
detailed light curve calculations for two specific nuclear heat
sources, free neutron decay and radioactive Ni 56. He finds that
neutron-heated ejecta may generate an optical emission that peaks
hours after the SHB with a luminosity of $\sim 10^{41}$ erg/s ($M
\sim -14$). Such emission can be detected from $z=0.2$ by a
10-meter-class telescope. Ni 56 heated ejecta, on the other hand,
are very hard to detect in optical observations, unless the velocity
of the ejecta is unreasonably slow (much below the escape velocity
from a neutron star surface), otherwise the ejecta becomes optically
thin before most of the Ni 56 decay. Most of the radioactive energy
in this case would escape as hard X-ray photons.

To conclude, macronova emission may be detected if SHBs eject a
significant amount of radioactive mass with a favorable decay time,
and even then only by rapid (less than $1$ d post burst) and deep
optical observations. The characterizing signatures of macronova
optical emission are blackbody colors and non-correlated optical and
X-ray emission.

\subsection{Relativistic collisionless shocks}\label{SEC: collisionless_shocks}
The microphysics of collisionless shocks is a major open question in
GRB physics \citep[for a recent review see][]{Waxman06}. The mean
free path of a proton to undergo a binary collision of any type
during GRB external or internal shocks is larger by many orders of
magnitude than the size of the system. Therefore, any interaction
between the particles must be due to collective processes. Such
electromagnetic interactions are known to take place in anisotropic
plasma and they give rise to collisionless shocks in GRBs. The
external shock is ultra-relativistic and the ``upstream'' (e.g., the
unshocked ISM) is unmagnetized. Internal shocks, if they take place,
are mildly relativistic and the upstream may be magnetized by
magnetic field that is advected all the way from the central engine.

Afterglow observations require collisionless shocks to generate the
two major ingredients of the observed emission:  (i)  The shock
should generate a near equipartition magnetic field while a $\mu$G
upstream field is some eight orders of magnitude below equipartition
(the fraction of the generated magnetic energy out of the total
energy is  parameterized by $\varepsilon_b$). The generated field
must be sustained along the downstream. (ii) The shock should
accelerate electrons to Lorentz factors that are much higher than
the thermal temperature of the ``downstream'' (the shocked ambient
medium). The outcome of the acceleration process is parameterized by
$\varepsilon_e$ and $p$ (see \S\ref{SEC: afte standard theory}).

There is currently no complete ``first principles" theory that
explains how collisionless shocks  accelerate particles or generate
and sustain the magnetic field. $\varepsilon_b$, $\varepsilon_e$ and
$p$ are simply phenomenological (almost certainly oversimplified)
model parameterizations, which are constrained by the observations.
In long GRBs these parameters are reasonably constrained
\citep[e.g.,][]{Panaitescu01,Yost03} and so is the requirement that
the magnetic field is sustained long after the shock \citep[$\sim
10^8 \lambda_s$, where $\lambda_s$ is the plasma skin
depth;][]{Rossi03}. In SHBs the current observations are not
detailed enough to constrain these parameters, but the similarities
to long GRB afterglows suggests that the microphysics is similar as
well. This is not very surprising given that the microphysical
processes are most likely affected only by local properties such as
the shock velocity and the upstream magnetization and not by global
properties such as the total outflow energy or  large scale
gradients in the external density.

Unmagnetized  collisionless shocks are not well understood.
\cite{Moiseev63} suggested that unmagnetized Newtonian shocks are
generated by the filamentation mode of the  Weibel instability
\citep{Weibel59}, a kinetic instability that arises when the
velocity distribution of plasma particles is anisotropic. The
typical scale of this instability is $\lambda_s$. \cite{Medvedev99}
and \cite{Gruzinov99} suggested that the same process may take place
in relativistic GRB shocks \citep[see
however][]{LyubarskyEichler06}. The fillamentation modes of this
instability \citep[e.g.,][]{Fried59,Silva02,Bret04,Bret05} result in
the breakdown of the plasma into current filaments and in the
generation of a magnetic field in the plane of the shock. In this
picture this magnetic field, which grows to equipartition levels,
provides the effective collisionality that produces the shock.
\cite{Medvedev99} suggested that this magnetic field is also
responsible for the observed synchrotron emission. The generated
field coherence scale is $\lambda_s$ and basic considerations
suggest that such a field should decay over a similar scale after
the shock \citep{Gruzinov01,Milosavljevic06b}. The field coherence
scale must grow quickly enough if it is to survive along the
downstream, as suggested by \cite{Silva03} and \cite{Medvedev05}. It
is currently unclear whether the coherence length of the field is
growing fast enough and therefore the fate of the generated field is
unknown.

An alternative idea is that the field is generated already in the
upstream by interaction between  shock accelerated protons and
unperturbed upstream \citep{Gruzinov01,Milosavljevic06c}.
Interestingly, \cite{LiWaxman06} used the requirement that the
acceleration time of the radiating electrons (assuming Fermi
acceleration) is shorter than the inverse Compton cooling time of
these electrons in the upstream, in order to put a lower limit on
the upstream field in front of the shock (a larger field reduces the
acceleration time). They find that in several long GRBs the field
must be larger than $0.2 n_0^{5/8}$ mG, where $n$ is the external
density, suggesting that the upstream field is amplified. Repeating
the same analysis for SHBs shows that no meaningful constraints can
be derived, and a $\mu$G field is consistent with the observations.
The reason is that for typical SHB parameters inverse Compton
cooling in the upstream is suppressed by the Klein-Nishina
cross-section (\S\ref{SEC: SSC}).

Acceleration of non-thermal particles in collisionless shocks is
believed to be done through diffusive shock acceleration
\citep[a.k.a. first order Fermi accleration ; for reviews
see][]{Drury83,Blandford87}. Acceleration in this process takes
place when a particle is scattered back and forth across the shock
by interactions with plasma waves on each side. With each shock
crossing the particle gains energy. GRB observations require that
the acceleration process will produce electrons with a high energy
power-law distribution with an index  $2\lesssim p \lesssim 3$.
Theoretical predictions of relativistic diffusive shock acceleration
typically suggest $p \approx 2.2$
\citep[e.g.,][]{Achterberg01,Ellison02,Lemoine03,Keshet05}, which
fits the observations nicely\footnote{The exact value of $p$ depend
on the details  of the interaction between the accelerated particles
and the magnetic turbulence
\citep[e.g.,][]{Ellison04,Lemoine06,Katz07,Keshet06}.}. However,
there are some simplifying assumptions that go into these
calculations that are not necessarily valid, and the full
acceleration process is far from being understood.

During the last several years there has been a large effort to
simulate relativistic collisionless shocks in three dimensions,
using the particle-in-cell (PIC) method
\citep{Silva03,Fonseca03,Frederiksen04,Jaroschek04,Jaroschek05,Nishikawa05,Spitkovsky05}.
These simulations show, as expected, that when two unmagnetized
plasma shells collide relativistically, shocks are formed by the
Weibel instability, building skin depth scale equipartition magnetic
field in the plane of the shock. None of the current simulations is
large enough (in time and space) to achieve a steady state shock or
to explore the decay of the generated magnetic field in the
downstream, due to computational resource limitations. None of the
simulations show a clear indication of diffusive shock acceleration.
This is not very surprising, given the simulation size and the fact
that the upstream field is practically zero. The reason is that in
these conditions there is no upstream scattering agent that can
reflect high energy particles back into the shock and start the
acceleration process. In reality the pre-existing upstream (e.g.,
ISM) magnetic field is necessary to enable reflection of particles
back into the shock over scales that are much larger than the skin
depth.

\section{Progenitors and the central engine}\label{SEC: progenitors}
In my mind, the most interesting open SHB question today is the
nature of the progenitor and the processes leading to the formation
of the central engine \citep[for a recent review see][]{LeeRuiz07}.
This branch of research has been active for many years and some of
the models were explored in great detail. Interest in this field was
boosted following the recent detection of SHB afterglows, as we now
have an idea about the typical environment and redshift in which SHB
progenitors dwell. An immediate implication of this information is
the understanding that SHB progenitors are different systems than
the progenitors of long GRBs, providing the proof that long and
short GRBs are two distinct physical phenomena. Nevertheless, the
identity of the progenitors of SHBs is still unknown.

Coalescence of a compact binary, either a double neutron star or a
neutron star (NS) and a black hole (BH) binary, is currently the
leading progenitor candidate. It is also the most extensively
explored model and at least NS-NS systems are observationally known
to exist. Therefore, this model will be the focus of this chapter
(\S\ref{SEC: Binary merger}). This progenitor candidate is of great
interest since NS-NS and NS-BH coalescence events are the most
promising sources of gravitational-wave (GW) signals accessible to
ground-based GW observatories. The evidence supporting the merger
model are indirect, and some aspects of this model seem to be
`uncomfortable' with the observations. Therefore, it is important at
this point to keep an open mind to the possibility that SHBs may
have a different origin than compact binary mergers. Other models
are discussed in \S\ref{SEC: Other models}.

The debate about the origin of long GRBs ended only with the
detection of SN spectral signatures in optical long GRB afterglow
\citep{Stanek03,Hjorth03}. Lacking similar direct evidence for SHBs,
the next best thing is an examination of indirect environmental
properties. The hope is that this information, together with the
requirement that the progenitor will be, in principle, able to
produce an SHB, will single out viable progenitor systems.
Therefore, before discussing specific models and their predictions,
I review the new constraints that the detection of SHB afterglows
pose on progenitor properties.

\subsection{Progenitors lifetime and intrinsic rate}\label{SEC: lifetime}
SHB progenitor lifetime is constrained using two independent methods
- (i) based on the observed redshift-luminositiy distribution, and
(ii) based on the host galaxy type distribution and the fractional
association of host galaxies with galaxy clusters. The results from
both methods are similar - SHB progenitor population is dominated by
long-lived ($\sim$ several Gyr) systems. Unfortunately, the current
sample available for this kind of analysis is small, and it is
unlikely that its size will significantly grow in the near future
(see below). Therefore, the conclusions that are based on this
sample should be considered with care.

\subsubsection{The sample}\label{SEC: sample}
The constraints on SHB progenitor lifetime  and on the intrinsic SHB
rate are statistical by nature. Therefore, the sample that is used
should be chosen carefully, and selection effects should be well
understood. When the first analysis exploring this topic was carried
out \citep[the end of the summer of 2005;
][]{NakarGalYamFox06,Guetta06} all four SHBs localized with
arcsecond precision were associated with host galaxies (SHBs 050509,
050709, 050724) or with a galaxy structure (SHB 050813). Therefore,
these four bursts constitute a complete sample in the sense that no
events were left out due to possible selection effects.
\cite{NakarGalYamFox06} also carried out an analysis based on an
extended sample of SHBs, including four IPN localized bursts
\citep[\S\ref{SEC: z_obs};][]{Gal-Yam05}. These constitute a
complete sample of IPN bursts selected based on the size of their
error boxes and locations far from the Galactic plane.

Unfortunately, the hosts and the redshifts of SHBs discovered after
these first well-localized four bursts, were not identified (in some
cases even an early X-ray afterglow was not detected), in part as a
a result of diminished observational effort. As a result, any
additional SHB for which the redshift and/or host is determined
(e.g., SHB 051221) cannot be currently included in an unbiased
sample. The reason is that selection effects that prevent easy
identification of the hosts of these bursts may strongly bias any
sample that does not compensate for the undetected bursts.
Hopefully, it will be possible in the future to assemble a larger
unbiased sample in order to obtain more definite conclusions.

\subsubsection{Constraints from SHB redshift distribution}\label{SEC: redshift dist}
The {\it observed} redshift-luminosity  two-dimensional (2D)
distribution is determined by the {\it intrinsic} redshift and
luminosity distributions, modified usually by the detector
sensitivity. Thus, the goal of the analysis is to use the observed
distribution to constrain the intrinsic redshift distribution
\citep{Piran92,Gal-Yam04b,Ando04,Guetta05,Guetta06,NakarGalYamFox06}.
Then, the lifetime of the progenitor can be constrained by making
the most probable assumption that the progenitor birth rate follows
the cosmic star-formation history, and that the intrinsic redshift
distribution is a convolution of the cosmic birth rate with the
lifetime distributions. The main disadvantage of this method is that
for small observed samples the intrinsic distributions are
under-constrained. It is only possible to assume several physically
motivated parametric functional forms of the intrinsic distributions
and constrain their parameter values.

The two-dimensional observed redshift, $z$, and luminosity, $L$,
distribution is derived from the intrinsic distributions via:
\begin{equation}\label{EQ: NLz}
\frac{d\dot{N}_{obs}}{dLdz} = \phi(L)
\frac{R_{SHB}}{1+z}\frac{dV}{dz} S(P),
\end{equation}
where $\dot{N}_{obs}$ is the {\it observed} SHB rate and $\phi(L)$
is the {\it intrinsic} peak luminosity function, which is assumed to
be independent of z. $0 \leq S(P) \leq 1$ is the probability for
detection, including redshift determination when redshift
information is needed, of a burst with a peak photon flux $P$, which
in turn depends on $L$ and $z$ as well as on the spectrum of the
bursts. $R_{SHB}(z)$ is the intrinsic SHB rate per unit comoving
volume and comoving time. Since SHB progenitors are most likely of a
stellar origin it is expected that:
\begin{equation}\label{EQ: R_SHB}
    R_{SHB}(z) \propto \int_z^\infty SFR(z')f\left(t(z)-t(z')\right)
    \frac{dt}{dz'} dz',
\end{equation}
where $SFR(z)$ is the star formation rate at redshift $z$ (per unit
comoving volume and comoving time), $t(z)$ is the age of the
universe at redshift $z$, and $f(\tau)d\tau$ is the fraction of SHB
progenitors that are born with a lifetimes between $\tau$ and
$\tau+d\tau$.

$S(P)$ can describe a single detector or a combination of several
detectors, each weighted by its field of view and operational time.
In principle if $S(P)$ is well known and if the observed sample is
large enough then the  intrinsic distributions can be extracted from
Eq. \ref{EQ: NLz}. In reality  we have to work with a limited sample
as well as poorly understood $S(P)$. In the case of GRBs (long and
short) there is a large sample of bursts, observed by \BATSE, for
which only the peak flux distribution is available while the
redshift (and thus luminosity) is unknown. The \BATSE sample can
constrain the intrinsic distributions by considering the observed
flux distribution which is an integration of Eq. \ref{EQ: NLz}:
\begin{equation}\label{EQ: logNlogS}
    \frac {d\dot{N}_{obs}}{dP}=\frac{d}{dP}\int_0^\infty dz \int_{\tilde{L}(z,P)}^\infty dL
    \frac{d\dot{N}_{obs}}{dLdz},
\end{equation}
where
\begin{equation}\label{EQ: Lmin}
    \tilde{L}(z,P)=4\pi d_L^2 k(z) P
\end{equation}
is the luminosity for which a burst at a redshift z has a peak flux
$P$. $d_L(z)$ is the luminosity distance and $k(z)$ depends on the
spectrum of the bursts and includes the k-correction as well as the
conversion from energy flux to photon flux.  $k(z)$ is assumed to be
a function of the redshift only.

If $\phi(L)$ is a single power-law, $\phi(L) = \phi_0 L^{-\beta}$,
with no upper or lower cutoff (within a luminosity range that is
discussed below) then the integral over $z$ in Eq. \ref{EQ:
logNlogS} does not depend on $P$ and thus the observed peak flux
distribution does not depend on $R_{SHB}$ and simply satisfies:
\begin{equation}\label{EQ: SPL_LogNLogS}
   \frac {d\dot{N}_{obs}(P)}{dP} \propto P^{-\beta} S(P).
\end{equation}
Note that since Eq. \ref{EQ: SPL_LogNLogS} is independent of the
burst redshifts $S(P)$ here is the probability for detection alone.
Similarly, the integral over $L$ in eq. \ref{EQ: NLz} results in:
\begin{equation}\label{EQ: Nz}
  \frac {d\dot{N}_{obs}}{dz} = (4\pi d_L^2 k(z))^{1-\beta} \phi_0 \frac{R_{SHB}}{1+z}\frac{dV}{dz}
\int P^{-\beta} S(P) dP,
\end{equation}
thereby eliminating the dependence on $S(P)$ up to a constant
normalization factor. Naturally, for $\beta < 2$ an upper cutoff
must exist while for $\beta>1$ a lower limit is necessary. However,
if the lower cutoff is low enough so that it affects only a
negligible volume, and if the upper cutoff is high enough so it
affects only the detection at high redshift, then Eqs. (\ref{EQ:
SPL_LogNLogS}) and (\ref{EQ: Nz}) are applicable (these cut-offs
also prevent the integral over $P$ in Eq. \ref{EQ: Nz} from
diverging). Therefore, if the observed peak flux distribution can be
fitted by Eq. \ref{EQ: SPL_LogNLogS} then the luminosity function
can be a single power-law. In this case data sets for which $S(P)$
is not well known can be readily used and Eq. \ref{EQ: Nz} enables a
comparison of the one-dimensional observed redshift distribution
with model predictions. Unfortunately the {\it observed} luminosity
distribution depends on $S(P)$ even when the luminosity function is
a single power-law. If $S(P)$ is well known, a better constraint on
the intrinsic distributions can be obtained by a comparison with the
two-dimensional luminosity-redshift distribution (Eq. \ref{EQ:
NLz}).

\begin{figure}[!t]
\includegraphics[width=13cm]{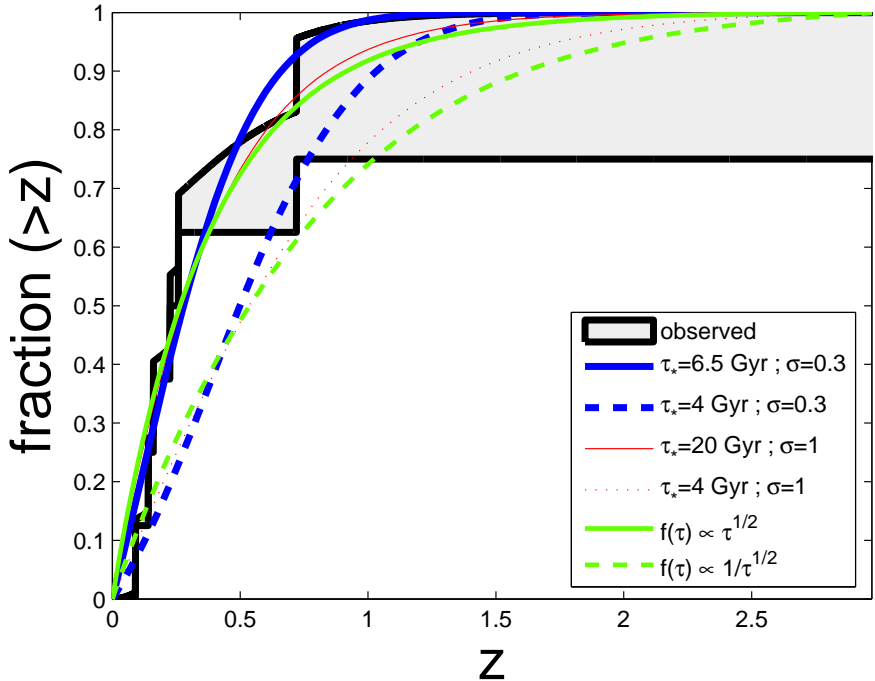}
\caption{ The cumulative {\it observed} redshift distribution as
predicted by various lifetime distributions when the luminosity
function is $\phi(L) \propto L^{-2}$ and the star formation history
is given by SF2 formula in \cite{Porciani01}. The cumulative
redshift distribution of the observed data (shaded area) is
bracketed between the lower solid line, which is the cumulative
redshift distribution of the six bursts with known redshifts, and
the upper solid line, which includes also the contribution of the
two bursts with upper limits (see \cite{NakarGalYamFox06} for
details). The figure demonstrates that only models which are
dominated by long-lived progenitors provide a good fit to the data
while models with short-lived progenitors under-predict the fraction
of bursts at low redshift $z\lesssim 0.3$. [taken from
\cite{NakarGalYamFox06}]}\label{FIG: z_dist_theory}
\end{figure}

The observed \BATSE peak flux distribution,
$\frac{d\dot{N}_{obs}(P)}{dP}$, can be fitted successfully by a
single power-law with an index $\beta = 2 \pm 0.1$
\citep{NakarGalYamFox06}. Therefore, it is consistent to assume a
single power-law luminosity function and to use Eq. \ref{EQ: Nz} on
a sample of bursts from different detectors with unknown $S(P)$ -
altogether 8 bursts (4 \swift/\hete bursts and 4 IPN bursts). Figure
\ref{FIG: z_dist_theory} \citep[from][]{NakarGalYamFox06} presents
the observed cumulative redshift distribution comparing to the
predictions of several models (Eqs. \ref{EQ: R_SHB} \& \ref{EQ: Nz})
taking the star formation history formula SF2 from \cite{Porciani01}
and using different functional forms of life time distribution: (i)
lognormal, $f(\tau)d\tau= (\tau\sigma\sqrt{2\pi})^{-1}
exp[-(ln(\tau)-ln(\tau_*))^2/2\sigma^2] d\tau$ with various values
of $\tau_*$ and narrow ($\sigma=0.3$) or wide ($\sigma=1$)
dispersions; (ii) power-law distributions $f(\tau) \propto
\tau^{\eta}$ with a lower cutoff at $20$ Myr and an upper cutoff
that is larger than the Hubble time. This figure vividly illustrates
that models with a typical delay are consistent with the data only
if this delay is long ($\gtrsim 4$ ~\rm Gyr). Models with no typical
lifetime (power-law distributions) must have a birthrate of
progenitors per unit logarithmic lifetime that increases
significantly as a function of lifetime (i.e., $\eta>-0.5$). The
reason that models dominated by short-lived systems do not fit the
data is that they under-predict the fraction of bursts at low
redshift ($z\lesssim 0.3$). Since the star formation rate is higher
at large redshifts ($\gtrsim 1$), while SHBs are found
preferentially at low redshift ($\approx 0.3$), although they are
bright enough to be detected at high redshfit, implies that a long
delay is required for the progenitors so they are born when the
universe is young and produce SHBs only when it is much older.
Quantitatively, \cite{NakarGalYamFox06} find that for lognormal
lifetime distributions (narrow and wide) $\tau_*>4[1]$ Gyr in
$95[>99.9]\%$ confidence and for a power-law lifetime distribution
the most probable power-law index is $\eta=0.6$ and $\eta>-0.5[-1]$
at $95[99.5]\%$ confidence\footnote{These values are calculated
assuming that SHB 050813 is at z=0.72
\citep{Gladders05,berger05b,Prochaska05}. If a redshift of 1.8 is
taken instead \citep{Berger06a} then a narrow lognormal lifetime
distribution is ruled out (cannot explain both the low and high
redshift bursts). For a wide lognormal lifetime distribution
$\tau_*>2[1]$ Gyr in $95[>99]\%$ confidence and for a power-law
lifetime distribution the most probable power-law index is $\eta=0$
and $\eta>-0.8[-1]$ at $95[98]\%$ confidence.}. These results are
similar to those obtained by \cite{Guetta06}.

The fact that the observed luminosity function fits a single
power-law does not necessarily imply that the intrinsic luminosity
function is a single power law as well. When an intrinsic luminosity
function that is not a single power-law is considered, the threshold
of the detector must be known (at least partially) and therefore the
sample that can be used is reduced only to the first three \swift
SHBs. However, in such case the full two-dimensional $L-z$
distribution can be considered. \cite{NakarGalYamFox06} find that
when a `knee shaped' broken power law (i.e., $\phi(L)$ is steeper
above the break) lifetime distribution is assumed the results are
very similar to the case of a single power-law. These results
suggest that the observations prefer long-lived progenitors (Gyrs)
for unimodal lifetime distributions. If a bimodal luminosity
function is considered then there are solutions in which the
progenitor lifetime is dominant by short-lived systems. However,
such luminosity functions are not expected unless SHBs are composed
of two separate populations.

This analysis does not include the recent indication found by
\cite{Berger06c} that at least 1/4 of the SHBs are at $z>0.7$
(assuming that we are not fulled by SHBs that take place at large
distances, $\sim 100$ kpc, from there hosts). If confirmed, this
result indicates that the lifetime of a non-negligible fraction of
the SHB progenitors is rather short (Gyr or less), implying that the
lifetime distribution is wide (includes both old and young
progenitors). It rules out a narrow lognormal ($\sigma=0.3$)
lifetime distribution or a power-law distribution with $\eta \gtrsim
0$. A wide lognormal ($\sigma=1$) lifetime distribution with
$4<\tau_*<8$ Gyr and a power-law distribution with $-0.5 \lesssim
\eta \lesssim 0$ are consistent with all current observations.

\subsubsection{Based on host galaxy types}
\begin{figure}[!t]
\includegraphics[width=12cm]{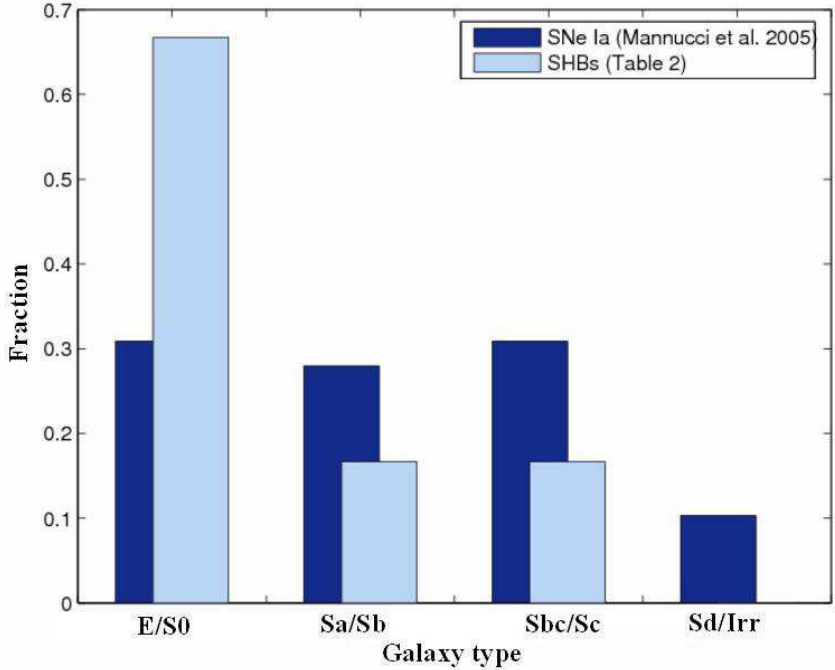}
\caption{ A comparison between the host galaxy types (E/S0, Sa/Sb,
Sbc/Sc and Sd/I) of SHBs (from Table 2) and SNe Ia
\citep{Mannucci05}. The fraction of SHBs in early type galaxies is
significantly larger than the fraction of SNe Ia observed in such
galaxies in the nearby Universe, indicating that the progenitor
systems of SHBs are probably longer-lived than those of SNe Ia.
[form \cite{Gal-Yam05}]}\label{FIG: HOst_SNeIa}
\end{figure}

An alternative method to estimate the progenitor lifetime is based
on the spectral types of the host galaxies \citep{Gal-Yam05}. The
stellar populations in early type galaxies (E \& S0) is entirely
dominated by old systems (several Gyr) while the population in
galaxies of later types is composed of a mix of old and young
systems. Therefore, the fraction of SHBs in different types of
galaxies is an independent measure of the progenitor lifetime.
\cite{Gal-Yam05} compared the distribution of SHB host types to that
of type Ia supernovae (SNe) from \cite{Mannucci05}, building on the
extensive effort done in undertaken to constrain SNe Ia progenitor
lifetime (Fig. \ref{FIG: HOst_SNeIa}). They find that a larger
fraction of SHBs, compared to SNe Ia, take place in early type
galaxies (at $93\%$ confidence), implying that it is most likely
that SHBs are older, on average, than SNe Ia. The typical time delay
of SNe Ia, $\tau_{Ia}$ is still under debate, but all different
models that assume a unimodal lifetime distribution agree that
$\tau_{Ia} \gtrsim 1$ Gyr \citep{Tonry03,Barris05,
Strolger04,Gal-Yam04b,Maoz04,Mannucci05} and therefore SHBs, which
are most likely older, have a lifetime of several Gyr. It was
recently suggested that the population of SNe Ia is composed of two
sub-groups - a long-lived component which is responsible for the
entire SN Ia population in early type galaxies, and a short-lived
component, which follows the star formation rate
\citep{Mannucci05,Mannucci06,Scannapieco05}. If this is the case,
the distribution of SHB host types is consistent with that of the
long-lived SN Ia component which is several Gyr old.

\cite{Zheng06} constrain the lifetime of SHB progenitors directly by
decomposing the total star formation history to the contribution
from early- and late-type galaxies. The convolution of the split
star formation history and a given lifetime distribution predicts
the fraction of SHBs in each host type. \cite{Zheng06} find that if
the lifetime distribution is a single power law,  $f(\tau) \propto
\tau^{\eta}$, then the observations suggest $\eta \gtrsim 3/2$
implying a typical lifetime of $\sim 10$ Gyr (note that this result
is inconsistent with the recent observations of \citet{Berger06c}
that suggest that $\eta \lesssim 0$).

\cite{Shin06} follow a similar route to that of \cite{Zheng06}, but
rather than using the relative SHB rate in early- and late-type
galaxies, they use the relative rates in cluster and field
early-type galaxies. The advantage of this method over the
early/late ratio is that the star formation histories of early-type
galaxies are simpler. The disadvantage is that this method does not
sample the lifetime distribution below $\sim 1$ Gyr since
short-lived SHBs would not take place in early-type galaxies.
\cite{Shin06} find that the current ratio of cluster to field
early-type galaxies, 2, corresponds to $0<\eta<1$, which is
consistent with earlier results from redshift distribution and
early-type/late-type ratio analysis.

\subsubsection{Intrinsic local rate}
The observed \BATSE local rate is ${\cal R}_{SHB,obs} \sim 10 ~\rm
Gpc^3 ~yr^{-1}$ (\S\ref{SEC: obs rate}). This is clearly a lower
limit on the true intrinsic rate. An upper limit on the SHB rate can
be derived under the most probable assumption that the progenitors
of SHBs are a product of at least one core-collapse supernova (e.g.,
a neutron star) and that the SHB itself is a catastrophic
non-repeating event. In this case the rate of SHBs is bound by the
rate of core-collapse SNe. However, since the typical lifetime of
SHB progenitors seems to be several Gyr, the local SHB rate
corresponds to the rate of core-collapse SNe at a redshift $\approx
0.7$, which is $\sim 5 \times 10^5 \rm ~Gpc^{-3} ~yr^{-1}$
\citep{Dahlen04}. Therefore, the intrinsic local rate, ${\cal
R}_{SHB}\equiv R_{SHB}(z=0)$, must be within the range:
\begin{equation}\label{EQ: SNe rate}
   10 \lesssim  {\cal R}_{SHB} \lesssim 5 \times 10^5 \rm ~Gpc^{-3} ~yr^{-1}.
\end{equation}

Constraining the local rate within this range requires an estimate
of the fraction of undetected bursts. These may be events that are
collimated and point away from the observer, or bursts that are
simply too dim to be detected. The number of bursts that are beamed
away from us can be estimated by the beaming factor (i.e., the
fraction of the total solid angle into which the prompt gamma-rays
are emitted), $f_b^{-1}$. As I discuss in \S\ref{SEC: beaming}, the
value of the beaming factor is reasonably constrained only in two
bursts. In one case (SHB 051221A) it seems to be $\approx 80$
(assuming a `top hat' jet) and in the other (SHB 050724) it is
probably much smaller. Thus it is impossible at this stage to
confidently estimate the beaming factor, but current observations
suggest that $1 \ll f_b \lesssim 100$. The correction for undetected
dim bursts depends on the luminosity function and most strongly on
its lower cutoff, $L_{min}$ (defined so $\phi(L)=0$ for
$L<L_{min}$). Assuming the single power-law luminosity function that
is found to be consistent with the data ($\phi(L) \propto
L^{-2}~~~L>L_{min}$) the local rate is \citep{NakarGalYamFox06}:
\begin{equation}\label{EQ: R_SHB_Z0}
    {\cal R}_{SHB} \approx 40 f_b^{-1} \left(\frac{L_{min}}{10^{49}
    \rm{erg/s}}\right)^{-1}
    \rm ~Gpc^{-3} ~yr^{-1}.
\end{equation}

\subsection{Coalescence of a compact binary}\label{SEC: Binary merger}
The coalescence of two neutron stars was recognized as a potential
GRB progenitor already two decades ago. This possibility was briefly
mentioned in \cite{Blinnikov84}, \cite{Paczynski86},
\cite{Goodman86} and \cite{Goodman87}, and discussed in detail for
the first time by \cite{Eichler89} and later by \cite{Narayan92} and
many others. \cite{Paczynski91}, \cite{Narayan92} and
\cite{Mochkovitch93} discussed similar coalescence models of a
neutron star (NS) and a stellar mass black hole (BH) as a possible
progenitor. Binary mergers take place because of orbital energy and
angular momentum loss to gravitational-wave radiation, as was
observationally confirmed \citep{Taylor82}. These models are natural
candidates since such events must take place at a reasonable rate
(see below) and the amount of gravitational energy that is liberated
during the coalescence ($\sim 10^{54}$ erg) is large enough, so a
small fraction of it can generate the GRB. Additionally, in the
`standard' version of the merger model, the outcome of the
coalescence is the same central engine as in long GRB models - a
disk accreting onto a black hole. This similarity between the
engines naturally explains the great similarity between the
observational properties of the two phenomena. In both scenarios the
duration of the burst is determined by the lifetime of the disk. The
difference is that in binary mergers the disk is expected to be
consumed within a fraction of a second, while in the `collapsar'
model for long GRBs,  in-falling matter from the collapsing star
feeds the disk for much longer, enabling the production of a long
duration GRB.

\subsubsection{The ``central engine''}\label{SEC: central engine}

{\it I.  An accretion driven engine}\newline In the most popular
central engine scenario a binary merger results in the formation of
a disk around a black hole (newly born in the NS-NS case). Accreting
black holes are known to produce relativistic jets in other systems
such as active galactic nuclei and microquasars, however the
accretion rates that are required in order to power GRBs are higher
by many orders of magnitude, implying different physical conditions.
The models of the formation and operation of this type of SHB engine
are discussed below.

{\it I.I  Forming an accretion disk in a NS-NS merger}\newline An
extensive effort was dedicated to numerical simulations of NS-NS
mergers in the context of the GRB central engine. Starting from
3-dimensional Newtonian simulations with a polytropic equation of
state (EOS) and no neutrino effects \citep{Davies94}, the
simulations evolved to include increasingly more realistic EOS,
neutrino modeling and  asymmetric scenarios. These simulations were
carried out mainly by two groups using different numerical schemes -
Lagrangian Smooth Particle Hydrodynamics (SPH)
\citep{Rosswog99,Rosswog00,Rosswog02a} and Eulerian Piecewise
Parabolic Method (PPM)
\citep{Ruffert96,Ruffert97,Ruffert98,Ruffert99,Ruffert01}. The
outcomes of these simulations are rather similar. A central
quasi-axissymmetric object is formed within a few orbital periods
after the initial contact. This central object is surrounded by a
thick disk of material at a radius of tens of km with densities of
$\sim 10^{11}-10^{12} \rm ~gr ~cm^{-3}$ and temperatures of $\sim
1-10$ MeV. The mass of the disk depends somewhat on the initial
conditions (e.g., spin, mass ratio) but is typically in the range of
$0.03-0.3 M_\odot$ for a system with an initial mass of $\approx 3
M_\odot$. A narrow funnel along the axis of symmetry is relatively
baryonic free and may serve as a potential site for launching a GRB
jet \citep[although it is unclear if this funnel is clean
enough;][]{Ruffert98,Ruffert99}.

In recent years several groups carried out approximate
\citep[e.g.,][]{Oechslin02} and fully \citep[e.g.,][]{Shibata00}
general relativistic (GR) simulations of double neutron star merger.
In the context of SHBs these simulations show that the general
relativistic picture may be more complicated than the Newtonian one.
\cite{Oechslin06} use a relativistic SPH code with conformally flat
approximation of the Einstein field equations and a realistic EOS.
For a range of initial conditions they find results that are rather
similar to those of Newtonian simulations - a central object with a
surrounding disk at a radius of tens of km with a mass
$0.06-0.26M_\odot$. Shibata and collaborators
\citep{Shibata05,Shibata06b} carried out a full general relativity
simulation with a hybrid EOS that mimics a realistic stiff nuclear
EOS. They find that depending on the initial mass of the system, the
merger can proceed in two ways: (i) If the total mass is below some
threshold ($\lesssim 2.6 M_\odot$ depending on the EOS) a
hypermassive neutron star is formed at first \citep[a NS that is
supported by differential rotation; ][]{Baumgarte00}. If this
hypermassive neutron star collapses to a black hole after a
sufficient delay so it can transport angular momentum to the
surrounding matter, then the mass of the remaining disk is $\gtrsim
0.01 M_\odot$, independent of the initial NS mass ratio. Adding
magnetic fields, \cite{Shibata06} carry out a full GR
magneto-hydrodynamic simulation of the final stages of the merger
starting with a hypermassive NS as initial conditions \citep[see
also ][]{Duez06}. They find that magnetic breaking and
magnetorotational instability transfer angular momentum very
efficiently and as a result a disk of $\sim 0.05 M_\odot$ is formed
by the time that the NS collapses to a BH. (ii) On the other hand,
if the total mass of the system is above the threshold, the central
object collapses promptly (in less than $1$ ms) into a black hole,
taking with it most of the mass and leaving a very small disk. A
massive enough disk to be a possible SHB central engine ($\sim
0.01M_\odot$) is left only in cases where the mass of the two
progenitor neutron stars is significantly different (mass ratio
$\lesssim 0.8$), contrary to the results of \cite{Oechslin06}.
\cite{Shibata06b} suggest that the most likely reason for this
difference is the approximations made by \cite{Oechslin06}, that do
not take into account gravitational waves radiation reaction. In the
simulation by \cite{Shibata06b}, GW radiation is one of the main
channels through which the central object loses angular momentum and
ultimately collapses.

{\it I.II  Forming an accretion disk in a NS-BH merger}\\
The outcome of a NS-BH merger is less clear. The most important
initial condition affecting the coalescence evolution is the mass
ratio $q \equiv M_{NS}/M_{BH}$. If $q \ll 0.1$ the tidal disruption
radius of the NS is within the innermost stable circular orbit, for
any BH spin, and the NS plunges into the BH without leaving any
residual disk
\citep[e.g.,][]{Vallisneri00,Miller05,Rosswog05,Faber06b}. As I
describe below, the fate of systems with $q \gtrsim 0.1$ seems more
promising as SHB progenitor, but the outcome of this case is not
entirely known yet.

Numerous numerical simulations of NS-BH coalescence were carried out
using Newtonian potentials. The main result of these simulations is
that the formation of the disk depends strongly on the NS equation
of state (EOS). In the case of soft equation of state (polytropes
with $\Gamma \leqslant 2.5$ or Lattimer and Swesty EOS;
\citealt{Lattimer91}) the NS is tidally disrupted during its first
approach, forming a massive disk \citep[$\sim 0.3 M_\odot$;
][]{Janka99,Lee99b,Lee00,Lee01}. A more complex evolution is found
in the case of a stiff EOS (Shen EOS [\citealt{Shen98}] or a
polytrope with $\Gamma=3$;
\citealt{Kluzniak98,Lee99a,Lee00,Rosswog04}), as seems to be
required by the highest measured pulsar mass $M =2.1\pm 0.2 M_\odot$
\citep{Nice05}. These simulations show that a ``stiff'' NS is not
disrupted during the first episode of mass transfer. Instead, in
some cases, it bounces back into an eccentric orbit and the mass
transfer resumes only after additional angular momentum is lost to
gravitational wave radiation. The result is an episodic mass
transfer process that lasts much longer than the duration of the
simulations (massive disks are not formed before the simulations
end). Based on these results, \cite{Davies05} developed a
semi-analytic model of intermittent mass transfer, finding that the
NS is disrupted after $\sim 1$ s, leaving a $\sim 0.1 M_\odot$ disk.

While newtonian simulations seem to give a qualitatively correct
picture of the dynamics in at least some of the NS-NS merger
scenarios, it is not clear at all that they are applicable also in
the case of NS-BH coalescence. An important general relativistic
(GR) feature which is not treated by Newtonian simulations is the
innermost stable circular orbit (ISCO). \cite{Miller05} argues that
the NS tidal radius may be within the ISCO not only in the case of
low $q$, but also in the case of comparable NS and BH masses.  The
reason is that for $q \gtrsim 0.25$ the mass of the NS itself begins
to affect, increasing the the ISCO radius \citep[compared to an
isolated BH; e.g.,][]{Buonanno99}. For $q > 0.5$, even in the case
of a maximally spinning BH, the tidal radius may be within the ISCO.
Moreover, \cite{Miller05} suggests that even in cases where the
tidal radius is slightly outside of the ISCO, GR effects can cause
the NS to plunge directly into the BH without leaving a disk.

The most advanced numerical simulation of a NS-BH merger to date
were presented by \cite{Faber06b,Faber06}, that use the conformal
flatness approximation for GR  for a Schwarzschild\footnote{This
approximation is exact for spherically symmetric potentials and is
less appropriate for a Kerr BH.} BH  and a polytropic EOS with
$\Gamma=2$ for the NS (no simulations were carried out for stiffer
EOS). First, they carry out a simulation with $q=0.1$ and find that
indeed the NS is `swallowed' by the BH without being disrupted. Then
they perform a simulation in which the tidal radius is comparable to
the ISCO (equivalent to q=0.24 for their NS modeling\footnote{As a
result of the conformal flatness approximation \cite{Faber06b} use
q=0.1 while changing the neutron star compactness ($M_{NS}/R_{NS}$)
in order to imitate  a q=0.24 case.}). Here they find that instead
of a direct plunge, angular momentum transfer during tidal
disruption deposits $1/4$  of the NS mass back outside the ISCO.
Half of this mass becomes unbound and leaves the system, while the
rest, $\approx 0.15 M_\odot$, remains bound. Interestingly, only a
fraction of the bound matter forms a hot and dense disk that is
ready to perform as a SHB central engine. The remaining bounded
mass, $\approx 0.05 M_\odot$, is ejected to eccentric orbits and
should fall back toward the BH after more than a second. At the end
of the simulation the polar axis is not significantly polluted by
baryons. These results are promising, however, BH spin is expected
to play an important role in the dynamics of systems with $0.1
\lesssim q \lesssim 0.5$, so the final answer about their fate will
require a full GR simulation.

{\it I.III The accretion rate and the lifetime of the disk}\newline
Regardless of the initial progenitor (NS-NS or NS-BH binary) the SHB
central engine in this model is a hot and dense torus of $0.01-0.3
M_\odot$ that is accreted onto a stellar mass BH. The engine is
active as long as efficient accretion takes place. Assuming that the
burst is powered by accretion, the accretion rate can be estimated
using energy requirements. The observed isotropic equivalent
luminosity of SHBs is $10^{50}-10^{52}$ erg/s. Correcting for
beaming, the true luminosity is probably smaller by about one, or at
most two, orders of magnitude. Taking a reasonable efficiency of
accretion energy conversion into gamma-rays of $\sim 0.01-1\%$ (see
below) implies hyper-accretion rates of $0.01-10 M_\odot/s$. Are
such accretion rates expected for a typical disk that is formed by a
compact binary merger?

The accretion rate and the lifetime of the disk in this scenario was
first estimated analytically and semi-analytically
\citep{Popham99,Narayan01,DiMatteo02}. The amount of energy emitted
during the burst implies that the disk must be accreted efficiently
(i.e, most of the disk mass must be accreted and not expelled as a
wind). Efficient accretion requires efficient cooling, which at
these densities can be achived only by neutrinos. For this to occur,
the temperatures and the densities in the disk should be high
enough, which in turn implies that the radius of the disk should be
small enough. \cite{Narayan01} present an analytic approximation
that is valid assuming that the disk is optically thin to neutrinos.
They find that in order for efficient neutrino cooling to take
place, the radius in which the disk is deposited must satisfy:
\begin{equation}\label{EQ: Rmax}
    R_d < 44 R_s \left(\frac{\alpha}{0.1}\right)^{-2/7} \left(\frac{M_{BH}}{3M_\odot}\right)^{-1}
    \left(\frac{M_d}{0.1M_\odot}\right)^{3/7},
\end{equation}
where $R_s$ is the Schwarzschild radius of a central BH with mass
$M_{BH}$, $M_d$ is the disk mass and $\alpha$ is dimensionless
viscosity parameter \citep{Shakura73}. As discussed above, the disk
that is apparently formed in NS-NS and NS-BH mergers satisfies this
criterion\footnote{Mergers of a BH with less compact objects such as
white dwarfs or helium stars do not satisfy this criterion and
therefore are unlikely to produce long or short GRBs
\citep{Narayan01}.}. For a disk which is within this radius
\cite{Narayan01} find an accretion rate  of:
\begin{equation}\label{EQ: Mdot}
    \dot{M}_{acc} = 0.6  \left(\frac{\alpha}{0.1}\right) \left(\frac{M_{BH}}{3M_\odot}\right)^{-\frac{13}{7}}
    \left(\frac{M_d}{0.1M_\odot}\right)^{\frac{9}{7}}\left(\frac{R_d}{10R_s}\right)^{-\frac{3}{2}} ~M_\odot/s~,
\end{equation}
and a corresponding disk lifetime of
\begin{equation}\label{EQ: t_acc}
    t_{acc} = 0.2 \left(\frac{\alpha}{0.1}\right)^{-1} \left(\frac{M_{BH}}{3M_\odot}\right)^\frac{13}{7}
    \left(\frac{M_d}{0.1M_\odot}\right)^{-\frac{2}{7}}\left(\frac{R_d}{10R_s}\right)^\frac{3}{2}
    ~s.
\end{equation}
Note that the accretion time is set by the viscous time scale and is
linear with $1/\alpha$, which is the least constrained parameter in
the problem. These equations assume that the disk is optically thin
to its own neutrino emission, which for the expected accretion rates
is valid for $R \gtrsim 10 R_s$ \citep{DiMatteo02}. A semi-analytic
extension of this model to the optically thick regime was carried
out by \cite{DiMatteo02} and a further semi-analytic calculations of
the the disk properties evolution in time is presented by
\cite{Janiuk04}. Considerations of the strong magnetic fields that
might be present in hyper-accreting disk as well as observable
quantum electrodynamical effects arising from supercritical fields
are discussed by \citet{Kohri02}. Finally, General Relativistic
effects of Kerr metric on the structure of the accretion disk are
calculated by \citet{Chen07}.

A number of detailed numerical simulations in 2D
\citep{Lee02,Lee04,Lee05b} and 3D \citep{Setiawan04} follow the
evolution of the disk during its accretion. The latest simulations
include detailed equation of state, approximate neutrino treatment
and a viscosity described by an $\alpha$-law. These simulations
provide insights about the neutrino luminosity of the disk (see
below) and they are in rough agreement with analytic calculations of
the accretion rate and duration.

As discussed in \S\ref{SEC: prompt theory}, the burst duration is
most likely comparable to, \citep[e.g.,][]{Kobayashi97}, or at most
larger by an order of magnitude than \citep{Janka06,Aloy06}, the
duration of the engine activity.  Hence, for a typical value of
$\alpha = 0.1$ the accretion rate as well as the burst duration
nicely agree with those required for SHBs. However, late engine
activity, which is suggested by the observations (see \S\ref{SEC:
aft variability} and \S\ref{SEC: Xtail theory}), is not naturally
explained in this model.

Several processes have been suggested to explain prolonged accretion
that lasts longer than the viscous time scale.
\citealt{vanPuttenOstriker01} (see also
\citealt{vanPutten01,vanPutten02,vanPutten03,vanPutten05}) suggest a
configuration in which the accretion is temporarily suspended due to
a ``magnetic wall'' around a rapidly spinning black hole. This model
was originally proposed for long GRBs, but it may be viable during
the post-merger conditions of a compact binary, and explain
long-lasting emission. In this model, the jet is energized by the
spin of the BH and the accretion is suspended as long as the BH is
rapidly spinning (tens of seconds for a $2.5 M_\odot$ BH and a $0.1
M_\odot$ disk). \cite{Proga06} suggest a different scenario in which
magnetic flux can be accumulated near the black hole and behave as
an alternating barrier, that allows for intermittent accretion,
thereby increasing the accretion time. \cite{Perna06} suggest that
instabilities in the disk may cause fragmentation of the outer parts
of the disk, producing separate blobs which slowly inspiral toward
the black hole, producing late engine activity. Yet another
possibility is that during the time that the disk is formed, some of
the material is ejected into an eccentric, but bound, orbit and
falls back into the BH after the disk was accreted. Such behavior is
seen in the NS-BH simulation of \cite{Faber06b} and may take place
also during NS-NS coalescence, where some simulations show that
$\sim 10^{-2} M_\odot$ are ejected from the system
\citep[e.g.,][]{Rosswog99,Ruffert01}.

{\it I.IV Launching the jet}\newline Two classes of processes were
suggested to extract the energy of an accretion disk-black hole
system and launch a relativistic jet: neutrino-anti neutrino
annihilation and magnetically driven mechanisms.

{\it I.IV.I Neutrino driven jet}\newline In neutrino-cooled
accretion disks, a significant fraction of the  disk gravitational
energy is converted during the accretion into neutrino flux. This
energy is available to launch a jet through $\nu\bar{\nu}$
annihilation and subsequent pair production. The possible importance
of neutrino annihilation as the energy source of GRB jets was
recognized by many authors
\cite[e.g.,][]{Goodman87,Eichler89,Narayan92,Meszaros92,Mochkovitch93,Mochkovitch95,Witt94,Jaroszynski93,Jaroszynski96}.
While neutrinos are generated in various sites during the merger
\citep[e.g.,][]{Salmonso01}, the most promising process for energy
extraction is by neutrinos emitted from the  cooling torus. This
emission lasts long enough to power the jet and is originates far
enough from the black hole so it is not strongly affected by
gravitational capture or redshift of the black hole \citep[see ][for
a discussion of gravitational effects]{Asano00,Asano01}. Moreover,
the geometry of the torus implies that the most efficient
annihilation sites are along the rotation axis, which also suffer
the least from baryon pollution. Therefore, most numerical
simulations of these systems
\citep{Ruffert97,Ruffert98,Ruffert99,Fryer99b,Rosswog02b,Rosswog03a,Rosswog03b,Rosswog03c,Lee02,Lee04,Lee05b,Setiawan04},
as well as semi-analytic calculations
\citep{Popham99,DiMatteo02,Janiuk04}, were used to estimate the
energy deposited by neutrinos emitted from the disk.

The energy extraction efficiency is defined as the fraction of the
rest mass energy of accreted matter that is converted into pairs by
$\nu\bar{\nu}$ annihilation $\epsilon_{\nu\bar{\nu}} \equiv
L_{\nu\bar{\nu}}/\dot{M}c^2$ (most of the annihilation takes place
along the rotation axis, where baryonic pollution is low). This
efficiency is a combination of the fraction of the rest mass energy
that is emitted by neutrinos, $f_{\dot{M}\rightarrow L\nu} \equiv
L_\nu/\dot{M}c^2$ , and the fraction of neutrino energy that
annihilates into $e^+e^-$ pairs, $f_{L\nu\rightarrow L\nu\bar{\nu}}
\equiv L_{\nu\bar{\nu}}/L_\nu$. The total efficiency depends on the
accretion rate. At low accretion rates ($\dot{M} < 0.1 \rm
M_\odot/s$) the neutrino luminosity drops fast and the annihilation
rate (which is proportional to the luminosity square) drops even
faster. Therefore, at these accretion rates the total efficiency is
very low, $\ll 10^{-4}$ \citep{Popham99,Setiawan04}. On the other
hand, disks with accretion rates that are too high ($\dot{M} \gg 0.1
\rm M_\odot/s$) are optically thick, and most of the generated
neutrinos are advected into the BH, thereby reducing
$f_{\dot{M}\rightarrow L\nu}$ while $f_{L\nu\rightarrow
L\nu\bar{\nu}}$ remains roughly constant \citep{DiMatteo02}. Maximal
efficiency is obtained at $\dot{M} \sim 1 \rm M_\odot/s$ were
$f_{\dot{M}\rightarrow L\nu} \sim 0.01-0.1$ and $f_{L\nu\rightarrow
L\nu\bar{\nu}} \sim 10^{-3}-10^{-2}$ \citep[e.g.,][]{DiMatteo02}.
Therefore, the total efficiency of neutrino annihilation under
optimal conditions is $\epsilon_{\nu\bar{\nu}} \sim 10^{-4}$. In the
context of SHB central engine this optimal accretion rate is also
the expected one, implying that the total energy that can be
expected from this process is $\sim 10^{49} (M_d/0.1M_\odot)$ erg
where $M_d$ is the disk mass. This energy is enough to power
luminous SHBs (e.g., SHB 051221) only if their flow is narrowly
collimated\footnote{Possible mechanical collimation processes are
discussed and simulated by \cite{Rosswog02b} and by \cite{Janka06}}
and less luminous SHBs (e.g., SHB 050509B) also if they are not
narrowly beamed. Current limits on the total energy of SHBs (Table 2) 
are still compatible with this model, but future observations may
require higher energy. \citet{Ruiz05b} suggest the existence a hot
corana that modifies the spectrum of the emitted neutrinos as a way
to significantly increase $f_{L\nu\rightarrow L\nu\bar{\nu}}$.

{\it I.IV.II Magnetically driven jet}\newline An alternative method
to launch relativistic jets from the BH-accretion disk system is via
strong electromagnetic fields
\cite[e.g.,][]{Narayan92,Levinson93,Thompson94,Ruffert97,Meszaros97b,LeeBrown00,Brown00b,Rosswog02b,Rosswog03c,Daigne02a,Lyutikov06}.
Even if the initial magnetic field is low, it is expected to play an
important role due to amplification by the magneto-rotational
instability \cite[for review of MHD accretion disks see
][]{Balbus98}. An extensive numerical effort is invested in systems
of magnetized accretion onto BH, with rapid progress from
simulations that use newtonian and pseudo-Newtonian potentials
\cite[e.g.,][]{Hawley00,Hawley02a,Hawley02b,Armitage01,Armitage03,Proga03,Proga03b,Machida03,Kato04}
to GR simulations using a Kerr metric in two dimensions
\citep{McKinney04,McKinney05,McKinney06} and in three dimensions
\citep{DeVilliers03,DeVilliers05,Hirose04,Krolik05,Hawley06}.
Currently, these simulations do not include neutrino physics and use
simple equation of state (a constant adiabatic index of $4/3$ or
$5/3$ is assumed) and therefore the results may not be directly
applicable to the case of SHBs. Nevertheless, magnetic fields most
likely play important, and maybe the major, role in jet launching.
Different GRMHD simulations show a qualitatively  similar picture.
An accretion flow is generated within the disk by magnetic viscosity
that results from amplification of the magnetic field by the
magnetorotational instability \citep{Balbus91} even if it is
initially weak. The accretion results in a strong wind of magnetized
plasma which is ejected from the system along the boundaries of the
centrifugally evacuated funnel, and a strong Poynting flux jet is
observed within the funnel. The baryonic load within the
Poynting-flux-dominated funnel is low enough to allow a terminal
Lorentz factor $>100$ \citep{McKinney06}.

Recent GRMHD simulations find that the energy of Poynting flux jets
is directly related the spin of the BH and is in general agreement
with the predictions of \citealt{Blandford77}\footnote{For a
discussion of the similarities and the differences between the
simulations and the BZ mechanism see \cite{Hawley06}.}\citep[][see
also
\citealt{Komissarov04,Komissarov05}]{McKinney04,McKinney05,Hawley06}.
Energy output in Poynting flux jets increases sharply with the spin
of the BH, and this process can be far more efficient than a
neutrino-driven jet. \cite{Hawley06} find efficiencies
$\epsilon_{em} = 3\cdot 10^{-4},6\cdot 10^{-3},0.04 ~\&~0.2$ for
spin parameters $a/M = 0,0.5,0.9~\&~0.99$ correspondingly
($\epsilon_{em} \equiv L_{em}/\dot{M}c^2$ where $L_{em}$ is the
luminosity of the Poynting flux jet and $a/M \equiv Jc/GM_{BH}^2$
where $J$ is the BH angular momentum and $G$ is the gravitational
constant). \cite{McKinney06} simulates an accretion on a BH with a
spinning parameter $a/M =0.9375$ and finds an outflow efficiency of
$\epsilon_{em} \approx 0.05$ where $10\%$ of the jet luminosity is
spread roughly uniformly within an opening angle of $0.1$ rad around
the polar axis. Taking an accretion disk of $ 0.1 M_\odot$, this
$\epsilon_{em}$ corresponds to a total emitted energy of $\approx
10^{52}$ erg out of which $10^{51}$ erg are within 0.1 rad of the
jet axis. An observer with a viewing angle within $0.1$ rad of the
jet axis will infer a total isotropic equivalent energy of $\approx
10^{53}$ erg. Therefore, this process, if applicable, can easily
account for the energetics and the Lorentz factors observed in SHBs.


{\it II. Hypermagnetized neutron stars}\newline An interesting
alternative GRB engine is a compact object with an ultra-high
magnetic field ($\gg 10^{15}$ G). Here, it is the energy of the
magnetic field and/or the star rotation that drives a relativistic
outflow, and there is no need for an accompanying accretion disk.
Such a magnetized object can be a transient hypermassive NS that is
formed by the coalescence of two neutron stars
\citep{Kluzniak98,Rosswog02a,Rosswog02b,Rosswog03c,Shibata06,Duez06}.
\cite{Price06} carry out an MHD simulation of the merger of two
neutron stars with initial magnetic fields of $10^{12}$ G. They find
that on very short timescales ($\sim 1$ ms) a magnetic field
$>10^{15}$ G is generated by the Kelvin-Helmholtz instability along
the shear interface that forms where the two neutron stars come into
contact. This field strength is clearly limited by the resolution of
the simulations and therefore the actual generated field  is most
likely much higher. An upper limit on the field strength is the
equipartition level $\sim 10^{17}$ G. If the hypermagnetized
hypermassive neutron star is stable for $\sim 0.1$ s, as suggested
by the results of \cite{Shibata06b}, then the energy in the magnetic
field can drive a relativistic outflow in various ways.
\cite{Price06} suggest that bubbles of equipartition magnetic field
can become buoyant and float up through the star surface producing a
relativistic flow \citep{Kluzniak98,Lyutikov03b,Dai06}. If a large
scale field is built up then an alternative launching mechanism is
the is the magneto-centrifugal slinging
\citep{Usov92,Thompson04,Thompson05}. The emission time scale is the
spin-down time of the star (assuming that it does not collapse
first), which is $\sim 0.1$ s in the case of a $10^{17}$ G magnetic
field (the initial spin of the hypermassive NS is $\sim 1$ ms).
Since the energy source of this engine is the magnetic field and/or
the rotation  of the transient neutron star ($\sim 10^{53}$ erg), it
can in principle produce very energetic bursts with more than
$10^{51}$ erg. Rapidly rotating hyper-magnetized neutron star (i.e.,
msec magnetar) may be produced also by other progenitor systems and
it is discussed in some more detail in \S\ref{SEC: Magnetar}

\subsubsection{The lifetime of compact binaries and their
merger rate}

The merger rate and typical lifetime of NS-NS binaries are estimated
in two ways, based on  observed systems in our Galaxy and using
theoretical population synthesis. In the case of NS-BH binaries only
the latter method is used, as no observed systems of this kind are
currently known\footnote{The fact that no NS-BH systems are observed
is not surprising given that binary detection requires a recycled
pulsar, and that the birth rate of BH-recycled pulsar binaries is
expected to be very low \citep{Pfahl05}.}. These aspects of compact
binaries are covered in detail by other papers in this volume and
therefore I will briefly review here only the main results of each
method.

{\it I. Constraints from observed systems in the Milky Way}

This method is based on counting observed NS-NS binaries in our
Galaxy, and then deducing the Galactic merger rate by correcting for
the completeness of the surveys and the estimated lifetime of each
system (from the second supernova till the coalescence). The
cosmological local rate is then derived by extrapolating the
Galactic rate to a cosmological volume. The main advantage of this
methods is that it is based on solid observations. As such it
provides a robust lower limit on the total merger rate. The main
disadvantage of this method is that it is possible that there are
binary populations that cannot be detected by current surveys. For
example currently detection of NS-NS binaries depends on at least
one of the companions being a recycled pulsar. Therefore, it is
insensitive to formation channels that do not recycle any of the
pulsars, if such exist. The extrapolation from the Galactic rate to
the cosmological one is also not trivial since it is not clear what
is the right indicator which should be used for such an
extrapolation (e.g., blue light?). Finally, the sample is small $-$
there are only three NS-NS binaries that are used in order to
estimate the Galactic rate - PSR~B1913+16, PSR~B1534+12,
PSR~J0737$-$3039\footnote{Other observed NS-NS systems will not
merge within the Hubble time and therefore are irrelevant for this
method. The only exception, PSR~B2127+11C, is excluded because of
its association with a globular cluster
\citep{Anderson90,Prince91}.}.

The first calculations of the Galactic merger rate
\citep{Phinney91,Narayan91} were of the order of
$10^{-6}-10^{-5}$\,\peryear\,,  where the uncertainty was dominated
by uncertain beaming correction and survey selection effects. More
detailed calculations that followed obtained similar values
\citep{Curran95,vandenHeuvel96,Arzoumanian99}. \citet{Kalogera01}
carried out detailed investigation of the uncertainties in these
rate estimates and found that the uncertainty in the correction for
undetected faint systems can span over two orders of magnitude. With
two NS-NS systems observed , PSR~B1913+16 and PSR~B1534+12, they
estimate a rate of $\sim  10^{-6}$  \peryear\ when only the beaming
corrections is included. Considering the correction for faint
undetected pulsars, they estimate that the rate can be as high as
$5\times 10^{-4}$\,\peryear . Later, \cite{Kalogera03} developed a
method to assign statistical significance to these estimates. They
find a merger rate of $8^{+9}_{-5} \times 10^{-6}$\,\peryear\
($68\%$ confidence). The subsequent discovery of the relativistic
binary pulsar PSR~J0737$-$3039 \citep{Burgay03} increased the number
of systems used for rate estimation to three.  The corresponding
range of Galactic merger rates becomes between $1.7\times 10^{-5}$
to $2.9\times 10^{-4}$\,\peryear\ at 95\% confidence
\citep{Kalogera04}, increasing the most likely estimated rate by a
factor of 6-7. This significant revision in the rate is due to the
unique properties of PSR~J0737$-$3039 (short lifetime, 2.4hr period
orbit and different beam profile) that make it difficult to detect.

The extrapolation of the galactic rate to the local cosmological
rate is usually done assuming that the merger rate is proportional
to the blue stellar luminosity. \cite{Phinney91} obtains a
conversion factor of $~10^{7} ~\rm Gpc^{-3}$. Using a similar
conversion factor the most updated estimate of the local universe
merger rate is $200-3000 ~\rm Gpc^{-3} yr^{-1}$ \citep{Kalogera04}.

The merger rate derived based on the observed Galactic sources is
dominated by PSR~J0737$-$3039, which is short-lived ($\approx 100$
Myr). Therefore, this method predicts that the lifetime distribution
of NS-NS binaries is dominated by rather short lived-systems ($\ll
1$ Gyr). The fact that there are observed systems with lifetimes
that are comparable the Hubble time or longer, implies that the
observed systems lifetime distribution is broad.

{\it II. Constraints from population syntheses}

An alternative approach to evaluate the NS-NS merger rate -- and the
only current mean to evaluate the NS-BH merger rate -- is via
population synthesis
\citep[e.g.,][]{Tutukov94,Brown95,Lipunov95,Portegies96,Portegies98,Bethe98,Fryer98,Fryer99,Bloom99,Belczynski99,Brown00a,Belczynski01,Belczynski02c,Belczynski02a,Belczynski02b,Perna02,OShaughness05a,OShaughness05b,deFreitas06,Dewi06}.
The evolution of a binary is followed as a function of the initial
binary properties (e.g., the zero age main sequence [ZAMS] mass of
the two companions and the initial orbital separation). The
advantage of this method is that it can be used to explore binary
systems that cannot be observed, such as non-recycled NS-NS or NS-BH
binaries. The main disadvantage of the method is that the
uncertainties involved are substantial (e.g., the supernova kick
distribution, common envelope evolution, etc.). As a result, the
rate estimates span over two orders of magnitude, which include the
range inferred from observed Galactic binaries.

An interesting outcome of this research is the suggestion by
\cite{Belczynski01} that a formation channel of very short-lived,
practically unobservable, NS-NS binaries exists.  Including this
channel in their most updated population synthesis code,
\cite{Belczynski05,Belczynski06} find that the distribution of the
time that the binary spends as two neutron stars, from the second
supernova until coalescence, is bimodal with one peak at $\sim 10^5$
yr and another at $\sim 10^{10}$ yr. When the time between the
formation of main sequence stars and the two supernovae is added,
the lifetime distribution (from star-formation till death),
$f(\tau)$, becomes flat in log space ($f(\tau) \propto \tau^{-1}$ in
the notation of \S\ref{SEC: lifetime}) with a cutoff below $10$ Myr.
They obtain a rather similar lifetime distribution also for NS-BH
binaries.

An alternative formation channel of NS-NS binaries is by exchange
interactions in globular clusters (as in the case of PSR~B2127+11C
that is associated with M15). The number of  NS-NS binaries that are
formed by this process is expected to be significantly smaller than
the number of binaries that are formed through binary evolution
\citep[e.g.,][]{Phinney91}. \cite{Grindlay06} estimate a Galactic
merger rate in globular clusters of $\sim 4 \cdot 10^{-8}$\,\peryear
 which corresponds to a cosmological local rate of $\sim 2 \rm~
Gpc^{-3}~yr^{-1}$. This rate is indeed much smaller than the
estimates based on the observed systems that are not associated with
globular clusters. \cite{Hopman06} evaluate the lifetime
distribution function of binaries that form in globular clusters and
find that it is dominated by old systems with an average life time
of $\approx 6$ Gyr.

Estimates of the rate of NS-BH mergers vary by orders of magnitude,
and so does the ratio between the NS-NS merger rate and the NS-BH
merger rate. Two major open questions that determine if a NS
collapses into BH during the binary evolution, and thus strongly
affect the ratio between NS-NS and NS-BH binaries, is the accretion
rate of a NS during common envelope phase and the maximal mass of a
NS. The interesting work by \cite{Bethe98} suggests that the mergers
of low-mass BH and NS should be more common than NS-NS mergers
\citep[see also ][]{Chevalier93,Brown95,Wettig96,Bethe05b,Bethe05},
since the first neutron star cannot survive a common envelope phase
and it collapses into a BH following hypercritical accretion from
the companion envelope. They suggest that the only way to avoid this
outcome is by having two companions with similar masses (ZAMS masses
no more than $4\%$ apart) so they get into the helium burning phase
at a similar time, thereby avoiding vigorous mass accretion during a
common envelope phase. Using a Salpeter initial mass function
\citep{Salpeter55} they find that the rate of NS-BH mergers is $20$
times the rate of NS-NS mergers. This ratio is reduced to a factor
of $5$ when a flat initial mass function (constant $dn/dM$) is
considered \citep{Lee06}. The mean lifetime of a NS-BH binary in
this scenario is $\approx 5$ Gyr and the predicted rate (normalized
to the supernova rate at a redshift of $\approx 0.7$) is $\sim 10^4
~\rm Gpc^{-3}~yr^{-1}$ (Bethe, Brown \& Lee, 2007, this volume).

Finally, population synthesis can be used to put an upper limit on
the local rate of NS-NS and NS-BH mergers, by evaluating the
fraction of binaries that survive both supernovae. This fraction
depends mostly on the distribution of the supernova kick velocity,
which is not well constrained. Several works show that this fraction
of surviving binaries is unlikely to exceed $2\%$
\citep[e.g.,][]{Pfahl02,Lipunov97}. Taking this fraction from the
local rate of core-collapse SNe \citep{Cappellaro99} implies an
upper limit of $\sim 1000 ~\rm Gpc^{-3} ~yr^{-1}$ on the merger rate
of short-lived binaries ($\tau \lesssim 1$ Gyr). An upper limit of
$10^4 ~\rm Gpc^{-3} ~yr^{-1}$ on the local merger rate of long-lived
binaries($\tau \gtrsim 4$ Gyr) is obtained by relating it to the
rate of core-collapse SNe at redshift 0.7 \citep{Dahlen04}.
Therefore, if SHBs are mergers of compact binaries, then from Eq.
\ref{EQ: SNe rate}, the SHB-merger rate must be in the range:
\begin{equation}\label{EQ: R_SHB=merger}
    10 \lesssim  {\cal R}_{SHB=merger} \lesssim 10^4 \rm
~Gpc^{-3} ~yr^{-1}.
\end{equation}

\subsubsection{Offsets from host galaxies and external medium densities}\label{SEC: binary offset}
An important clue about the nature of the progenitors is the
location of the bursts with respect to their host galaxies. The
offset of the merger site of a NS-NS or NS-BH binary from its birth
place was explored by several groups
\citep{Fryer99,Bloom99,Perna02,Belczynski06}. This location depends
on the natal kick of the progenitor system and the time between the
kick and the merger, as well as the galactic gravitational
potential. In the case of compact binaries the natal kick is
expected to be between several tens to several hundreds km/s
\citep[e.g.,][]{Hobbs05} and different models of binary formation
channels predict a wide duration range between the second SN and the
merger \citep[e.g.,][]{Belczynski06}. A binary with an inspiral time
of several Gyr born with a kick of several hundred km/s, can merge
after traveling $1$ Mpc, while a binary that merges within $1$ Myr
will always end up near its birth place. Similarly, binaries with
moderate kick velocity are expected to be bound to large galaxies
but not to small ones, and therefore offsets from larger galaxies
are expected to be smaller. Thus, in principle, mergers can take
place at any distance ($\lesssim 1$ Mpc) from their hosts, and the
predicted merger distribution depends on the specific binary model
considered. \cite{Belczynski06} find that for their model of bimodal
inspiral time distribution, the offset from highly star-forming
galaxies is very small (since the merger rate is dominated by the
short inspiral time population). The offset from elliptical
galaxies, in which only long-lived binaries merge, is expected to be
large. The typical offset from a small [giant] elliptical is about
100 [10] kpc.

The circum-burst gas density depends of course on the offset. If the
merger site is outside of a field host, in the inter-galactic medium
(IGM), the density is expected to be $\sim 10^{-6} ~\rm cm^{-3}$.
If, on the other hand, the merger takes place within its host galaxy
or in the intra-cluster medium of a galaxy cluster (ICM) the density
is expected to be $> 10^{-3} ~\rm cm^{-3}$ . Since the brightness of
the afterglow depends on the external density (\S\ref{SEC: afterglow
theory}), afterglows of bursts with large offsets should be fainter.
For example, \cite{Belczynski06} present the distribution of
external densities of merger sites predicted by their model for
different types of hosts. They find that the lowest densities are
expected for binaries that are born in small early-type galaxies,
while the highest densities are found for those that are born in
large star-forming galaxies.

If compact binary mergers lead to SHBs then this behavior predicts a
strong selection effect that biases the observed redshift and
host-type distributions. The reason is that it is very hard to
detect the afterglow of a burst that takes place at a large distance
from its host and even if an afterglow of such a burst is detected,
it is unlikely that a host association will be secure, and therefore
the SHB redshift, can be determined. As a result, a bias is expected
favoring bursts with short-lived progenitors, implying higher
redshifts and late-type hosts. Similarly, a bias is expected in
favor early-type cluster galaxies over early-type field galaxies.

\subsubsection{Comparison with the observations}
The most rudimentary predictions of NS-NS and NS-BH mergers fit the
SHB observations well: (i) These mergers must take place at a rate
that is comparable to the SHB rate (Eq. \ref{EQ: SNe rate}). (ii)
The gravitational binding energy that is liberated in these mergers
is more than enough to power the observed prompt emission and
afterglow. Moreover, extensive numerical and theoretical work
suggests several very plausible processes to channel enough of this
energy into a relativistic outflow. (iii) The duration of SHBs is
explained in a rather natural way in these models, while the engine
size is compact enough to allow for the observed rapid variability
during the prompt emission. Following recent observations, several
papers have shown that the energy and the timescales of specific
events with known redshifts can be explained within the framework of
the merger model \citep{Lee05,Oechslin06}. (iv) The most popular
engine model in this scenario, hyper-accretion onto a black hole, is
similar to the engine suggested for long GRBs, naturally explaining
the similarity between the two phenomena. (v) SHBs do not trace the
star formation, and they take place in both early- and late-type
galaxies, as expected for compact binary mergers. (vi) Many SHBs are
observed at low redshifts ($\sim 0.2$) while there are indications
of a population of SHBs at higher redshift ($\sim 1$). This result
indicates on a wide lifetime distribution, as some papers predict
for compact binary mergers \cite[e.g.,][]{Belczynski06}. (vii)
Observations suggest that some of the bursts take place in very low
densities and at large offsets from their hosts, as predicted for a
binary population with strong natal kicks and long lifetimes.

A more detailed comparison of this model to the observations reveals
that there are also some conflicts and issues that should still be
resolved between the model and the observations. The growing amount
of evidence suggesting that the central engine is active for a
duration that is much longer than the duration of the initial hard
prompt gamma-ray emission is a major challenge to the merger model
(e.g., the late flare in SHB 050709). Here, the natural engine
activity time scale that the merger model predicts ($\lesssim 1$ s)
is an obstacle. Nevertheless, several extensions and modifications
of the model that may explain such activity have already been
suggested. An additional discrepancy between the NS-NS merger model
and the observations is the inferred lifetime distribution of the
merging systems. While the current observations point toward a
lifetime distribution that is dominated by several Gyr old
progenitors, the observed NS-NS systems in our Galaxy represent a
population with a typical lifetime of $\sim 100$ Myr. Note that the
poorly constrained lifetime distribution of NS-BH binaries is
consistent with the observations of SHBs. Both the observed SHB
sample and the observed NS-NS sample are very small, and therefore
this discrepancy may be resolved with any of the two lifetime
estimates, or both, being inaccurate. However, if supported by
future SHB observations and assuming that Galactic binaries are
representative of the cosmological population, lifetime
considerations may rule out NS-NS coalescence and leave only NS-BH
mergers as a viable SHB progenitor system.

To conclude, given the very promising rudimentary comparison of
compact binary merger model predictions and observations, this
remains the leading progenitor model. However, it is important to
remember that all supporting arguments are indirect and none of them
are even close to being as conclusive as the detection of SN spectra
in the afterglows of nearby long GRBs. Therefore, it is important,
especially given the fact that there are some apparent discrepancies
between this model and SHB observations, to explore the possibility
that SHBs are produced by entirely different systems.

\subsection{Other progenitor models}\label{SEC: Other models}

\subsubsection{Accretion induced collapse}\label{SEC: AIC}
An accretion induced collapse (AIC) of a rapidly rotating neutron
star to a black hole was suggested by several authors as the source
of GRBs \citep{Vietri98,Vietri99,MacFadyen05,DermerAtoyan06a}. The
GRB central engine in this scenario is similar to the one expected
in a compact binary merger, the rapid accretion of a disk onto the
newly formed BH. Here, the source of the disk is the NS material
with the highest angular momentum, on the equator, while the BH is
formed by material from the collapsing NS poles.

\cite{MacFadyen05} propose this progenitor model to explain the
$\sim 100$ s X-ray tail observed in some SHBs (see \S\ref{SEC: X-ray
tail}). They suggest that the collapse is initiated by accretion
from a close and less compact companion and that the late X-ray
emission is produced by the interaction of relativistic ejecta with
the companion. Note that this scenario predicts a single smooth
pulse and cannot explain variable X-ray tails \citep[which might
have been already observed;][]{Norris06} or several afterglow
flares, as observed following SHB 050724. \cite{DermerAtoyan06a}
further suggest that neutrino radiation from the NS collapse may
heat and disrupt the companion, sending some of its mass toward the
young BH, leading to late-time flares. They also find that the rate
of such events can fit the SHB rate.

The main uncertainty in this model is the mass of the formed disk.
Semi-analytic calculations and general relativistic simulations of
the collapse into a black hole of a uniformly rotating star spinning
at the mass-shedding limit, show that the mass of the remaining disk
depends strongly on the equation of state
\citep{Cook94b,Cook94a,Shibata00b,Shibata02,Shibata03,Shapiro04}.
These calculations use a polytropic equation of state and they show
that assuming a very soft EOS, $\Gamma - 4/3 \ll 1$,  a massive disk
may form \citep{Shibata02}. However, a star with $\Gamma >1.5$
collapses directly into a BH leaving practically no disk, $M_d <
10^{-3} M_\odot$ \citep{Shibata03,Shapiro04}. The reason is that the
angular momentum of the material on the equator is not high enough
in order to support it from collapsing. It is possible that if there
are strong magnetic fields that help with angular momentum transfer
towards the equator the outcome is different. Since neutron stars
are expected to be much stiffer than a polytrope with $\Gamma =
1.5$, the current results suggest that unless strong magnetic field
significantly alter the collapse outcome, AIC of a NS cannot be a
GRB progenitor.

\subsubsection{Magnetars}\label{SEC: Magnetar}

Central engine models that involve highly-magnetized  neutron stars
(magnetars) with magnetic fields that are similar to or larger than
those observed in our Galaxy\footnote{see \cite{Woods06} for a
review of Galactic magnetars}
\citep[][]{Duncan92,Paczynski92,kouveliotou98}, come in two flavors.
One is a newly born millisecond magnetar and the second is a version
of SGR giant flares, in which case the magnetar already span-down
significantly and is rotating slowly. A millisecond magnetar may be
formed during the merger of two neutron stars as a transient, which
collapses into a black hole during its spin-down,  (as discussed in
\S\ref{SEC: Binary merger}). A span down magnetars in old stellar
population (as required by \S\ref{SEC: lifetime}) must be therefore
formed by a different progenitor system. A milliseconed magnetar
that span-down without collapsing may be formed for example by a
merger of two white dwarfs \citep[e.g.,][]{Saio85,Levan06}, or an
accretion-induced collapse of a white dwarf
\citep[e.g.,][]{Nomoto91}.

A millisecond magnetar was first suggested as the central engine of
GRBs by \citet[][see also \citealt{Thompson04,Thompson05}]{Usov92}.
The energy source in this model is  rotational energy, $\sim 5 \cdot
10^{52}$ erg, which is channeled into a relativistic outflow by
magnetic luminosity. Assuming magnetic dipole emission, and that
gravitational radiation can be neglected, the luminosity is
\citep{Usov92}:
\begin{equation}\label{EQ: magnetar luminosit}
    L_{md} \approx  2.5 \times 10^{52} B_{16}^{2}\Omega_4^4 \rm ~erg/s,
\end{equation}
where B is the magnetic field on the neutron star surface and
$\Omega$ is the magnetar angular velocity. The spin down time is:
\begin{equation}\label{EQ: Tmagnetar}
    T_{md} \approx  2 B_{16}^{-2}\Omega_4^{-2} \rm s.
\end{equation}
Therefore,  SHB energy considerations require $B \gtrsim 10^{15}$ G
while time considerations require either $B \gtrsim 10^{16}$ G or
termination of the magnetic emission before the entire rotational
energy is emitted (e.g., by collapse of the magnetar into a BH).

A millisecond magnetar with a lower magnetic field ($\sim
10^{14}-10^{15}$) is unlikely to produce the SHB itself, but it may
affect the afterglow \citep{Dai98a,Dai98b,Wang01,Zhang01a,Dai04}. If
somehow the end-product after the SHB is launched is such a magnetar
then the rotational energy powers an outflow over $10^2-10^4$ s and
this energy is injected into the external shock that produces the
afterglow. This energy injection was suggested as an explanation of
late flares and light curve flattening observed in SHBs 050724 and
051221A \citep{Gao06,Fan06b}.

The second scenario is of an older magnetar that is producing the
SHB in the same way as it produces SGR giant flares. Here the energy
source is the magnetic field. If the field of the magnetar is
similar to those inferred for Galactic objects ($\sim 10^{15}$) then
it cannot produce a burst with more than $\sim 10^{47}$ erg.
\cite{Dar05b} suggests that giant flares, such as the one observed
from SGR 1806-20, are beamed in such a way that for some observers
the isotropic-equivalent energy appears to be  $\sim 10^{50}$ erg,
thereby explaining the whole SHB population as similar extragalactic
giant flares. While this model is very economic, it fails to produce
the observed SHB afterglows, which are brighter by many orders of
magnitude than the unbeamed afterglow of the giant flare from SGR
1806-20 \citep{Cameron05,Gaensler05}. An alternative intriguing
possibility is that there are old magnetars with external magnetic
fields of $10^{17}$ G (T. Thompson, private communication). Giant
flares from such magnetars can be energetic enough to produce SHBs.
Such magnetars may be the end product of a merger of two white
dwarfs \citep{Levan06}, and their birthrate and lifetime may be such
that it is not surprising that none are observed in our Galaxy. Note
that in this model is currently the only one where the same source
may produce more than one SHB, enabling, in principle, repeated
detection of SHBs from the same source.

\subsubsection{Quark stars}
Quark stars, if they exist, may be an entirely different energy
source for both long and short GRBs. There are many models in which
GRB progenitors involve a quark star
\cite[e.g.,][]{Schramm92,Ma96,Cheng96,FryerWoosley98,Dai98b,Dar99,Bombaci00,Ouyed02a,Ouyed02b,Berezhiani03,Bombaci04,Bombaci05,Paczynski05,Ouyed05,Staff06,Menezes06,Drago06}.
In these models the burst is initiated by a nuclear phase transition
in the composition of some, or all, of the star material, usually
the conversion of  hadrons to quarks. This phase transition is also,
usually, the main source of the burst energy. The duration of the
bursts in different models is determined by different mechanisms.
For example, \cite{Menezes06} suggest that the duration in which an
entire neutron star is converted into a strange quark star,
triggered by the formation of seed strange quark matter, is between
$1$ ms and $1$ s, and may lead to a SHB. \cite{Ouyed02a} suggest
that both long and short GRBs are produced by hot quark stars where
the difference between the two types of bursts is the progenitor
mass. They find that the structure of the star is different above
and below a critical mass and that the phase transition in hot quark
stars below this critical mass takes $\sim 1$ s while for a more
massive star it takes $\sim 80$ s. \cite{Ouyed02b} suggest a model
in which the engine is activated by the accretion onto a quark star.
In this scenario, similar to typical GRB accretion models, the
difference between long and short GRBs is in the duration of the
accretion. \cite{Dar05b} suggest that SHBs, as well as soft
gamma-ray repeaters and their giant flares, originate from
hyper-stars, i.e. neutron stars where a considerable fraction of
their neutrons have converted to hyperons. The energy source is a
continuous episodic phase transition from neutron matter to hyper
matter in the inner star layers.

The outcome of most quark star scenarios is the  release of a large
amount of energy in the form of radiation within a short time on the
surface of the star. At this point these models merge with the
loaded pair-radiation fireball model that leads to a relativistic
outflow.

\subsubsection{Type Ia SN} \cite{Dar03} suggested that SHBs are
associated with Type Ia SNe. The prompt emission and the afterglow
in this model are produced by ``cannon balls'' \citep[for a review
of the cannon ball model see ][]{Dar04}. The stringent limits on SN
emission, which are a factor of $\sim 1000$ below that of a typical
SN Ia, together with the fact that the variance on SN Ia peak
magnitudes is small, disfavor this model.

\section{Gravitational waves from SHBs}\label{SEC: GW}

If SHBs are generated by NS-NS or NS-BH mergers then they are the
electromagnetic counterparts of the most accessible
gravitational-wave (GW) sources for ground-based GW observatories.
As such, SHBs received a lot of attention from the GW community. The
local SHB rate might have direct implications for the detection rate
of GW telescopes, and more importantly searching for GW signals in
association with SHBs may increase the sensitivity, and thus the
detection rate, of these telescopes. Even if SHBs have a different
origin they still may be powerful GW sources, although in this case
the prospects for detecting their GW signals are not as promising.

\subsection{NS-NS or NS-BH coalescence}
Mergers of compact binaries are expected to be the first
cosmological source of GW detected by ground-based observatories.
The exact nature of the signal expected and the prospects for its
detection are discussed in numerous papers \citep[see ][for a
comprehensive review]{Cutler02}. In principle the mergers are
expected to have three different phases of GW emission: the chirp
inspiral signal, the coalescence signal, and the signal from the
ringdown of the remaining BH. The frequency of the typical signal
during the final inspiral stages of NS-NS and NS-stellar mass BH
systems passes conveniently through the most sensitive frequency
range of GW observatories such as LIGO\footnote{The Laser
Interferometer Gravitational-Wave Observatory;
\href{http://www.ligo.caltech.edu/}{http://www.ligo.caltech.edu/}}
and
VIRGO\footnote{\href{http://wwwcascina.virgo.infn.it/}{http://wwwcascina.virgo.infn.it/}}
($\sim 100-500$) Hz, making it the main target for detection. At its
current sensitivity LIGO-I can detect a NS-NS merger up to an
average distance of about $15$ Mpc, and a merger of a NS with a $10
M_\odot$ BH up to about $30$ Mpc. Beginning at the end of 2005
LIGO-I started its S5 science run, in which a full year of data is
expected to be collected at this sensitivity (which is comparable to
LIGO design sensitivity over almost the entire frequency range). The
VIRGO design sensitivity is similar, and it is expected to be
reached in the near future.

If SHBs are generated by compact binary mergers then their local
rate is given by Eq. \ref{EQ: R_SHB=merger}, implying a detection
rate by current observatories of $1[10] \times 10^{-4} \lesssim
{\cal R}_{GW} \lesssim 0.1[1] \rm ~yr^{-1}$ for NS-NS [NS-$10
M_\odot$BH] mergers. Therefore, in the most optimistic scenario,
where SHBs are narrowly beamed and/or there are many dim undetected
SHBs, LIGO-I might detect GW signals from undetected SHBs. In all
other cases such a detection is unlikely. This GW detection rate is
for a blind search within the data for signal from mergers. If the
search is preformed only during some short time window (say 1 min)
around the time in which electromagnetic emission from SHBs is
observed, the sensitivity increases (a search for a significant
signal within data with longer duration requires higher signal to
noise ratio). \cite{Kochanek93} estimate that a search preformed
only when SHBs are detected can increase the range of LIGO-I and
VIRGO by a factor of $\approx 1.5$. If SHBs are preferentially
beamed perpendicular to the binary orbital plane the range is
increased by an additional factor of\footnote{The average range is
obtained by averaging over all the possible orientations. The signal
from a merger that is observed face-on is larger by a factor of
$\sqrt{5}/2$.} $\approx 1.5$. Therefore, a SHB-coincident search for
GW signals may increase the detection range to $\approx 30[60]
~\rm{Mpc}$. The detection rate of \swift is about $10$ SHBs per year
and that of \hete and other IPN spacecraft is similar. If several
percent of these bursts are within $50-100$ Mpc, as suggested by
\cite{Tanvir05}, then a coincident detection may be obtained with
current facilities. Note that if SHBs are beamed so their GW-SHB
coincident detection range is significantly increased then their
true rate (after beaming correction) within $15$ Mpc is at least
comparable to the observed SHB rate within $30$ Mpc. However, given
that even the most optimistic scenarios predict a rate that is
smaller than one merger event per year at these distances, a
coincident search for SHB and a GW signal may make the difference
between a detection and a non-detection by current GW observatories.

Simultaneous detection of the inspiral GW signal from a compact
merger and a SHB will provide conclusive evidence that SHBs
originate from compact mergers and would improve our understanding
of both merger physics and SHBs significantly. GWs can provide a
unique view of the formation, and possibly the operation, of the
inner engine powering the burst, which are difficult to observe via
any other method. A coincident detection is also most valuable in
order to determine the cosmological parameters \citep{Dalal06}, as
the GW signal of a binary merger enables an accurate determination
of the luminosity distance, while the electromagnetic signal may
provide an accurate measurement of the redshift. If GW signal from a
merger is detected without a coincident SHB, an association may
still be secured by the detection of an orphan afterglow -
afterglows that are not associated with prompt emission
\citep[e.g.,][]{Rhoads97,Granot02b,Nakar02d,Totani02,NakarOnAxis03,Levinson02}.
However, orphan afterglows should be searched for quickly, within
days or weeks after the detection of the GW signal. The reason is,
that the nearby SHBs, which might be detected by GW observatories,
are most likely at the low end of the luminosity function ($\lesssim
10^{47} erg$), simply because these are the most frequent, and
therefore their afterglow is not expected to be detectable many days
after the burst (even though they are nearby). Moreover, the
transition to Newtonian blast wave and quasi spherical emission
takes place days-weeks after the burst if its energy is low.

Next generation observatories, which are planned to become
operational in the first half of the next decade, are expected to be
ten times more sensitive. For example advanced-LIGO (LIGO-II)  is
designed to detect NS-NS [NS - $10 M_\odot$ BH] up to about
$300[650]$ Mpc. As discussed above, simultaneous detections will
increase LIGO-II range by a factor of 1.5-2.5 to $\approx 0.6[1.3]
~\rm{Gpc}$ . So far, \hete observed one burst at a distance $<700$
Mpc while {\it Swift} detected at least $2$ additional  SHBs at a
distance $\lesssim 1$ Gpc. The GBM detector on {\it
GLAST}\footnote{\href{http://f64.nsstc.nasa.gov/gbm/}{http://f64.nsstc.nasa.gov/gbm/}}
is expected to have a threshold that is similar to BATSE and more
than half-sky field of view, and thus it is expected to detect at
least $5$ SHBs within a distance of $500$ Mpc every year. Therefore,
if SHBs are compact binary mergers, and assuming that an efficient
GRB detector will be operational simultaneously with LIGO-II, a
coincident electromagnetic and GW detection is guaranteed. The
benefit of the simultaneous operation of LIGO-II and an efficient
GRB detector goes beyond the high likelihood to observe simultaneous
SHBs and mergers if they are associated - it will also enable to
disprove the association if no simultaneous detections are observed.

\subsection{Other processes that radiate gravitational waves }

SHBs may emit GWs also if the progenitor is not a compact binary
merger. Any progenitor that involves the collapse of a rotating
compact object to a black hole (e.g., the collapse of a rotating
neutron star triggered by accretion; \S\ref{SEC: AIC}) will produce
gravitational waves  \citep[e.g., ][]{Stark85}. The amplitude of
these waves is highly uncertain. Moreover, the absence of an
accurate signal templates will reduce thier detectability. Such GW
signals would most likely not be detected even by LIGO-II at
distances much greater than $10$ Mpc \citep[][, and references
therein]{Kokkotas05}.

If the central engine is an accretion disk - BH system then disk
instabilities may break the axissymmetry and lead to GW emission
\citep[e.g.,][]{vanPutten01,vanPutten02,vanPutten03,Kobayashi03,Araya04,vanPutten05,Bromberg06}.
The strength of this signal is uncertain and it depends on the
evolution of the disk, which is different from one model to the
other. Optimistic estimates suggest that it might be detectable by
LIGO-II to a distance of $\sim 100$ Mpc
\citep{Kobayashi03,Bromberg06}.

An additional source of GW is the acceleration of relativistic jets
\citep[e.g.,][]{Segalis01,Sago04,Piran05}. Unlike electromagnetic
emission, GW radiation in this case is not collimated. However, it
is expected to be too weak for detection by LIGO-I and even with
LIGO-II the detection range is not expected to exceed $10$ Mpc by
much.

\section*{Acknowledgements}
I am indebt to A. Gal-Yam for numerous discussions and invaluable
remarks that considerably improved this review. I am also especially
grateful to E. O. Ofek and J. Granot for detailed valuable comments.
It is my pleasure to thank K. Belczynski, E. Berger, G. E. Brown, D.
N. Burrows, S. B. Cenko, Z. Dai, D. B. Fox, D. A. Frail, N. Gehrels,
G. Ghirlanda, D. Guetta, D. Z. Holz, S. R. Kulkarni, P. Kumar, C.
Kouveliotou, W. H. Lee, Z. Li, R. Perna, E. S. Phinney, T. Piran, R.
D. Preece, E. Ramirez-Ruiz, R. Sari, A. M. Soderberg, E. Waxman and
D. Wei for helpful discussions and comments. I thank J. Racusin and
D. N. Burrows for preparing Figure \ref{FIG: XRT_lightcurves} and to
S. D. Barthelmy for providing the X-ray light curve of SHB 050724. I
am grateful to R. Quimby for creating the GRBlog, which I used
frequently during the preparation of this review. This work was
partially supported by a senior research fellowship from the Sherman
Fairchild Foundation and by NASA NNH05ZDA001N grant.

\newpage

\end{document}